\documentclass[twocolumn]{aastex62}

\usepackage{amsmath}
\usepackage{graphicx}   % Include figure files
\usepackage{bm}         % bold math
\usepackage{natbib}
\usepackage{etoolbox}
\usepackage{rotating}
\usepackage{array}
\usepackage{booktabs}
\usepackage{amssymb}

\makeatletter
\patchcmd{\NAT@citex}
  {\@citea\NAT@hyper@{%
     \NAT@nmfmt{\NAT@nm}%
\hyper@natlinkbreak{\NAT@aysep\NAT@spacechar}{\@citeb\@extra@b@citeb}%
     \NAT@date}}
  {\@citea\NAT@nmfmt{\NAT@nm}%
   \NAT@aysep\NAT@spacechar\NAT@hyper@{\NAT@date}}{}{}

\patchcmd{\NAT@citex}
  {\@citea\NAT@hyper@{%
     \NAT@nmfmt{\NAT@nm}%
\hyper@natlinkbreak{\NAT@spacechar\NAT@@open\if*#1*\else#1\NAT@spacechar\fi}%
       {\@citeb\@extra@b@citeb}%
     \NAT@date}}
  {\@citea\NAT@nmfmt{\NAT@nm}%
\NAT@spacechar\NAT@@open\if*#1*\else#1\NAT@spacechar\fi\NAT@hyper@{\NAT@date}}
  {}{}
\makeatother

\makeatletter
\DeclareRobustCommand{\textsupsub}[2]{{%
  \m@th\ensuremath{%
    ^{\mbox{\fontsize\sf@size\z@#1}}%
    _{\mbox{\fontsize\sf@size\z@#2}}%
  }%
}}
\makeatother

\newcommand{\lsun}{\mbox{\,$L_\odot$}}
\newcommand{\msun}{\mbox{\,$M_\odot$}}

\newcommand{\kms}{\mbox{\,km\,s$^{-1}$}}
\newcommand{\ghz}{\mbox{\,GHz}}
\newcommand{\mhz}{\mbox{\,MHz}}
\newcommand{\khz}{\mbox{\,kHz}}

\newcommand{\lbol}{\mbox{$L_{\rm bol}$}}
\newcommand{\tbol}{\mbox{$T_{\rm bol}$}}
\newcommand{\tbc}{\mbox{$T_{\rm b, cont}$}}
\newcommand{\ee}[1]{\mbox{${} \times 10^{#1}$}}% scientific number format
\newcommand{\jyb}{\mbox{Jy\,beam$^{-1}$}}
\newcommand{\cc}{\mbox{cm$^{-3}$}}

\newcommand{\J}{\mbox{$J$}}
\newcommand{\K}{\mbox{$K$}}
\newcommand{\N}{\mbox{$N$}}
\newcommand{\F}{\mbox{$F$}}
\newcommand{\Ka}{\mbox{$K_\text{a}$}}
\newcommand{\Kc}{\mbox{$K_\text{c}$}}

\newcommand{\rt}{\mbox{$\rightarrow$}}
\newcommand{\unc}[2]{\mbox{$^{+#1}_{-#2}$}}

\newcommand{\htcn}{\mbox{H$^{13}$CN}}
\newcommand{\methylformate}{\mbox{CH$_{3}$OCHO}}
\newcommand{\methylformatev}{\mbox{CH$_{3}$OCHO\,$v=1$}}
\newcommand{\methanol}{\mbox{CH$_{3}$OH}}
\newcommand{\tmethanol}{\mbox{$^{13}$CH$_{3}$OH}}
\newcommand{\etmethanol}{\mbox{CH$_{3}^{18}$OH}}
\newcommand{\dmethanol}{\mbox{CH$_{2}$DOH}}

\newcommand{\dimethylether}{\mbox{CH$_{3}$OCH$_{3}$}}
\newcommand{\acetone}{\mbox{CH$_{3}$COCH$_{3}$}}
\newcommand{\ethanol}{\mbox{C$_{2}$H$_{5}$OH}}
\newcommand{\acetaldehyde}{\mbox{CH$_{3}$CHO}}
\newcommand{\ethylcyanide}{\mbox{CH$_{3}$CH$_{2}$CN}}

\newcommand{\methylcyanide}{\mbox{CH$_{3}$CN}}
\newcommand{\dmethylcyanide}{\mbox{CH$_{2}$DCN}}

\newcommand{\sosigma}{\mbox{SO\,$^{3}\Sigma$}}
\newcommand{\tfso}{\mbox{$^{34}$SO}}
\newcommand{\sotwo}{\mbox{SO$_{2}$}}
\newcommand{\cctht}{\mbox{c-C$_{3}$H$_{2}$}}
\newcommand{\glycolaldehyde}{\mbox{CH$_{2}$OHCHO}}
\newcommand{\formamide}{\mbox{NH$_{2}$CHO}}
\newcommand{\thcooh}{\mbox{$t$-HCOOH}}
\newcommand{\cch}{\mbox{C$_2$H}}

\shorttitle{PEACHES}
\shortauthors{Yang et al.}

\begin{document}

\title{The Perseus ALMA Chemistry Survey (PEACHES). I. The Complex Organic Molecules in Perseus Embedded Protostars}

\author{Yao-Lun Yang}
\affiliation{Department of Astronomy, University of Virginia, Charlottesville, VA 22904-4235, USA}
\affiliation{RIKEN Cluster for Pioneering Research, Wako-shi, Saitama, 351-0106, Japan}
\author{Nami Sakai}
\affiliation{RIKEN Cluster for Pioneering Research, Wako-shi, Saitama, 351-0106, Japan}
\author{Yichen Zhang}
\affiliation{RIKEN Cluster for Pioneering Research, Wako-shi, Saitama, 351-0106, Japan}
\author{Nadia M. Murillo}
\affiliation{RIKEN Cluster for Pioneering Research, Wako-shi, Saitama, 351-0106, Japan}
\author{Ziwei E. Zhang}
\affiliation{RIKEN Cluster for Pioneering Research, Wako-shi, Saitama, 351-0106, Japan}
\author{Aya E. Higuchi}
\affiliation{National Astronomical Observatory of Japan, Osawa, Mitaka, Tokyo 181-8588, Japan}
\author{Shaoshan Zeng}
\affiliation{RIKEN Cluster for Pioneering Research, Wako-shi, Saitama, 351-0106, Japan}
\author{Ana L\'{o}pez-Sepulcre}
\affiliation{Univ. Grenoble Alpes, CNRS, Institut de Plan\'{e}tologie et d'Astrophysique de Grenoble (IPAG), F-38000 Grenoble, France}
\affiliation{Institut de Radioastronomie Millim\'{e}trique (IRAM), 300 rue de la Piscine, 38406 Saint-Martin-D'H\`{e}res, France}
\author{Satoshi Yamamoto}
\affiliation{Department of Physics, The University of Tokyo, 7-3-1, Hongo, Bunkyo-ku, Tokyo 113-0033, Japan}
\author{Bertrand Lefloch}
\affiliation{Univ. Grenoble Alpes, CNRS, Institut de Plan\'{e}tologie et d'Astrophysique de Grenoble (IPAG), F-38000 Grenoble, France}
\author{Mathilde Bouvier}
\affiliation{Univ. Grenoble Alpes, CNRS, Institut de Plan\'{e}tologie et d'Astrophysique de Grenoble (IPAG), F-38000 Grenoble, France}
\author{Cecilia Ceccarelli}
\affiliation{Univ. Grenoble Alpes, CNRS, Institut de Plan\'{e}tologie et d'Astrophysique de Grenoble (IPAG), F-38000 Grenoble, France}
\author{Tomoya Hirota}
\affiliation{National Astronomical Observatory of Japan, Osawa, Mitaka, Tokyo 181-8588, Japan}
\author{Muneaki Imai}
\affiliation{Department of Physics, The University of Tokyo, 7-3-1, Hongo, Bunkyo-ku, Tokyo 113-0033, Japan}
\author{Yoko Oya}
\affiliation{Department of Physics, The University of Tokyo, 7-3-1, Hongo, Bunkyo-ku, Tokyo 113-0033, Japan}
\author{Takeshi Sakai}
\affiliation{Department of Communication Engineering and Informatics, Graduate School of Informatics and Engineering, \\The University of Electro-Communications, Chofugaoka, Chofu, Tokyo 182-8585, Japan}
\author{Yoshimasa Watanabe}
\affiliation{Shibaura Institute of Technology, 3-9-14 Shibaura, Minato-ku, Tokyo 108-8548, Japan}

\correspondingauthor{Yao-Lun Yang}
\email{yaolunyang.astro@gmail.com}

\begin{abstract}
  To date, about two dozen low-mass embedded protostars exhibit rich spectra with lines of complex organic molecule (COM).  These protostars seem to possess different enrichment in COMs.  However, the statistics of COM abundance in low-mass protostars are limited by the scarcity of observations.  This study introduces the Perseus ALMA Chemistry Survey (PEACHES), which aims at unbiasedly characterizing the chemistry of COMs toward the embedded (Class 0/I) protostars in the Perseus molecular cloud.  Of 50 embedded protostars surveyed, 58\%\ of them have emission from COMs.  A 56\%, 32\%, and 40\%\ of the protostars have \methanol, \methylformate, and N-bearing COMs, respectively.  The detectability of COMs depends neither on the averaged continuum brightness temperature, a proxy of the H$_2$ column density, nor on the bolometric luminosity and the bolometric temperature.  For the protostars with detected COMs, \methanol\ has a tight correlation with \methylcyanide, spanning more than two orders of magnitude in column densities normalized by the continuum brightness temperature, suggesting a chemical relation between \methanol\ and \methylcyanide\ and a large chemical diversity in the PEACHES samples at the same time.  A similar trend with more scatter is also found between all identified COMs, hinting at a common chemistry for the sources with COMs.  The correlation between COMs is insensitive to the protostellar properties, such as the bolometric luminosity and the bolometric temperature.  The abundance of larger COMs (\methylformate\ and \dimethylether) relative to that of smaller COMs (\methanol\ and \methylcyanide) increases with the inferred gas column density, hinting at an efficient production of complex species in denser envelopes.

\end{abstract}

\section{Introduction}

Planet formation may start during the embedded phase of star formation \citep{2020AA...640A..19T}.  In the scenario in which planets form from the embedded disks, resulting in substructures, the chemistry of embedded disks may play a significant role for the chemical composition of the forming planets.  In the past two decades, observations show the emission of complex molecules toward the center of several protostellar cores.  From the astronomical point of view, complex molecules are usually defined as a species that contains six or more atoms \citep{2009ARAA..47..427H}.  All detected complex molecules are organic.  It can be saturated \citep[e.g., ][ and the ALMA PILS Survey]{2003ApJ...593L..51C,2007AA...463..601B,2016AA...595A.117J,2017ApJ...850..176C,2017ApJ...843...27L} or unsaturated \citep[e.g., carbon-chain molecules: ][]{2013ChRv..113.8981S,2014Natur.507...78S,2018ApJ...863...88L}.  The saturated organic molecules, often called complex organic molecules (COMs) or interstellar COMs (iCOMs), have single covalent bonds between carbon atoms, making them rich in hydrogen, while the unsaturated organic molecules contain double or triple bonds between carbon atoms, making them poor in hydrogen.  Most of the COMs appear in the inner warm envelope \citep[e.g., ][]{2004ApJ...617L..69B} and/or the surface of the embedded disk where the temperature is warm \citep[$\gtrsim$ 100 K; e.g., ][]{2017ApJ...843...27L}, while COMs in some sources, such as L1157, are linked to molecular shocks \citep[e.g.,][]{2001AA...372..899B,2017MNRAS.469L..73L,2020AA...635A..17C}.  The connection between a disk-like structure and the emission of COMs has been elusive \citep{2020AA...635A.198B}.  However, if the disk formation can inherit COMs from the embedded envelope \citep[e.g., ][]{2014MNRAS.445..913D}, the chemistry of COMs at the embedded phase of star formation may implicate future developments of organics on the planets.

For low-mass protostars, gas-phase COMs typically emit from the warm inner protostellar envelope ($T\gtrsim100$\,K), which corresponds to $\lesssim$ 100\,au for typical low-mass protostars \citep{2004ApJ...617L..69B,ceccarelli2007extreme}.  Thus, COMs serve as a tracer of the inner envelope where a disk may be forming \citep{2013ChRv..113.8961A}, measuring the chemistry and dynamics of embedded disks \citep{2014Natur.507...78S,2017ApJ...843...27L}.  Other processes, such as jets and outflows, can also produce the emission of COMs by sputtering and/or shock chemistry \citep[e.g., ][]{2008ApJ...681L..21A,2017MNRAS.469L..73L}.  Furthermore, accretion shocks occurred when the gas falls onto the disk may enhance the desorption of COMs \citep{2016ApJ...824...88O,2017ApJ...839...47M,2018AA...617A..89C}.  However, \citet{2020AA...635A.198B} showed no clear correlation between the COM emission and the occurrence of disk-like structures possibly due to specific conditions required for efficient desorption via accretion shocks \citep{2017ApJ...839...47M} or insufficient sensitivity.  Outbursting protostars, such as FU Orionis objects, can temporarily increase the COM abundance with elevated temperature \citep{2018ApJ...864L..23V,2019NatAs...3..314L}.  In the more evolved stage, COMs have only been detected in a few disks \citep{2015Natur.520..198O,2016ApJ...823L..10W,2018ApJ...862L...2F,2020AA...642L...7P}.

While recent observations show abundant COMs in several embedded protostars, the probability of the existence of COMs and their relationship to the star formation process yet remain to be understood.  Several protostars are rich in long carbon-chain molecules but have few COMs, such as L1527 \citep{2010ApJ...722.1633S} and IRAS 15398$-$3359 \citep{2009ApJ...697..769S,2014ApJ...795..152O,2018ApJ...864L..25O}.  In comparison, interferometric observations detected both COMs and carbon-chain molecules in several protostars, such as IRAS 16293$-$2422 \citep{2017AA...597A..40J,2018AA...617A.120M}, L483 \citep{2017ApJ...837..174O}, and B335 \citep{2016ApJ...830L..37I}.  Single-dish surveys showed no correlation between COMs and long carbon-chain molecules \citep{2016ApJ...833..125G,2018ApJS..236...52H}.  Interestingly, carbon-chain molecules tend to distribute at a larger spatial scale than that of COMs \citep{2016ApJ...830L..37I,2017ApJ...837..174O,2020AA...636A..19B}.  \citet{2020ApJ...897..110A} demonstrated that the conditions of prestellar cores can result in a larger spatial extent of carbon-chain molecules that coexist with COMs or deficient of carbon-chain molecules in COM-rich sources.  Moreover, dust opacity may obscure the emission of COMs in submillimeter wavelengths, introducing additional uncertainty to our understanding of COMs \citep{2020ApJ...896L...3D}.  Therefore the chemical pathways of complex molecules at the embedded protostars remain ill-constrained, highlighting the need of unbiased chemistry surveys to characterize the statistics of COMs \citep[e.g., ][]{2011AA...532A..23C,2018MNRAS.477.4792L}.

The Perseus molecular cloud is one of the most active nearby star-forming regions, which extends $\sim$10\,pc on the sky.  Many studies assume a distance of 235\,pc for Perseus based on the maser observations toward NGC 1333 SVS 13 \citep{2008PASJ...60...37H}.  Recently, \citet{2018ApJ...865...73O} estimated a distance of 321$\pm$10\,pc for IC 348 using VLBA and a distance of 293$\pm$22\,pc for NGC 1333 using Gaia parallaxes.  \citet{2020AA...633A..51Z} combined the Gaia parallaxes and photometric data with a Bayesian framework to revise the distances toward several sightlines of Perseus, resulting in distances ranging from 234 to 331\,pc.

In Perseus molecular cloud, infrared and submillimeter surveys reveal more than 400 young stellar objects as well as $\sim$100 dense cores, which contains $\sim$50 Class 0 and I protostars (\citealt{2005AA...440..151H,2008ApJ...683..822J,2013AJ....145...94D}; hereafter the embedded protostars).  The Perseus molecular cloud contains star-forming regions in a wide range of environments.  The majority of protostars in Perseus are associated with the two clusters, NGC\,1333 and IC\,348.  NGC\,1333, which has many active outflows \citep{2000AA...361..671K,2013ApJ...774...22P}, was thought to be younger than IC 348 based on the higher extinction and abundance of protostars; however, \citet{2016ApJ...827...52L} showed a similar age for two regions based on the evolutionary modeling of young stars and brown dwarfs.  The embedded protostars associated with IC\,348 lie at the southwest of the open cluster, IC\,348, near a prominent outflow, HH\,211 \citep[e.g., ][]{2009ApJ...699.1584L}.  Most of the other embedded protostars at Perseus are related to L1448, L1455, Barnard 1 (B1), and Barnard 5 (B5).  L1448 has active outflows that may regulate the ongoing star formation \citep{2010MNRAS.408.1516C}; The low abundance of Class 0 protostars in L1455 suggests a more evolved protostellar population than that in other regions \citep{2007AA...468.1009H}; B1 exhibits rich spectra of deuterated species \citep{2005ApJ...620..308M,2005AA...438..585R}.  B5 has a dominant embedded protostar, IRS1, along with three gravitationally bound gas condensation at a separation greater than 1000 au \citep{2015Natur.518..213P}.  Furthermore, the protostellar multiplicity in Perseus has been extensively studied by the VLA Nascent Disk and Multiplicity Survey (VANDAM) of Perseus protostars, showing a multiplicity fraction of 0.57$\pm$0.09 and 0.23$\pm$0.08 for Class 0 and I protostars, respectively \citep{2016ApJ...818...73T,2018ApJ...867...43T,2018ApJ...866..161S,2018ApJS..238...19T}.  \citet{2016AA...592A..56M} further characterized the spectral energy distributions of all sources identified in the VANDAM survey to derive protostellar properties, such as bolometric luminosities (\lbol), bolometric temperatures (\tbol), and evolutionary stage classification.  Several surveys using CO isotopologues and NH$_3$ probe chemical variations and kinematics at the cloud scale \citep[e.g., ][]{2008ApJ...679..481P,2017ApJ...843...63F}.  Thus, the Perseus molecular cloud provides an ideal test bed for chemistry in embedded protostars by surveying the protostars within each region and across the entire cloud.

\citet{2018ApJS..236...52H} presented a pilot survey of the chemistry in the Perseus embedded protostars with Nobeyama 45m telescope and IRAM 30m telescope, which surveyed all Class 0/I protostars \citep{2007AA...468.1009H} that have \lbol\ greater than 1\,\lsun\ (0.7\,\lsun for protostars in B1 and B5) and the envelope mass greater than 1\,\msun.  This pilot survey probes the molecules such as C$_2$H, c-C$_3$H$_2$, and \methanol.  The majority of the sources have emission of both \methanol\ and carbon-chain molecules.  They suggested a possible correlation between the location of sources within the clouds and the ratio of \methanol/\cch\, indicating the environmental effect on the chemistry.  However, single-dish observations of \methanol\ are contaminated by the contribution of the photodissociation region of the molecular clouds hosting the protostar \citep{2020AA...636A..19B}.  In addition, chemical surveys with single-dish telescopes are often less sensitive to detect the COMs in the vicinity of the protostars.  To trace those COMs , high spatial resolution from interferometric observations is crucial.  Based on the NOEMA observations of 26 embedded protostars in several molecular clouds, \citet{2020AA...635A.198B} detected \methanol\ and other COMs in about a half of them.  Interestingly, \methanol\ and \methylcyanide\ show a strong correlation, while their chemical link remains unclear.  They also investigated the origins of COMs and revealed apparent chemical difference among binary systems in their survey.  Internal luminosity is the most impacting parameter for the COM chemical composition, while the existence of a disk-like structure has no obvious impact to the COM emission.  A combined analysis of a few protostars from four different star-forming regions (Perseus, Serpens, Ophiuchus, and Orion) showed no significant correlation between the occurrence of COMs and the \lbol\ \citep{2020AA...639A..87V}.  Furthermore, the abundance of a few O-bearing COMs to \methanol\ are similar for the sources in different regions.  They suggested an inheritance scenario where the evolution in the prestellar phase dominates the chemistry.  Despite the growing sample of COM detections in embedded protostars, the statistics of COM abundance remain unconstrained due to the biases from the source selection, limited resolution, and small sample sizes.  To unbiasedly understand the chemistry, we conducted the Perseus ALMA Chemistry Survey (PEACHES) that probes the complex chemistry toward Perseus embedded protostars.  The source selection followed the same criteria as in \citet{2018ApJS..236...52H} with a few pointing modifications using the results from the VANDAM of Perseus protostars \citep{2016ApJ...818...73T}.  The PEACHES survey also probed the outflows of embedded protostars \citep[e.g., NGC1333 IRAS 4C: ][]{2018ApJ...864...76Z}.

Section\,\ref{sec:observations} describes the details of our ALMA observations.  Section\,\ref{sec:analyses} presents the identification of protostellar sources and the methodology of spectral extraction, line identification, and modeling.  Section\,\ref{sec:results} shows the detection statistics of COMs and their correlations.  Section\,\ref{sec:discussion} discusses the implications of the observed COM abundance on the current understanding of chemistry of COMs at protostellar sources.  Finally, Section\,\ref{sec:conclusions} summarizes the findings of this study.

\section{Observations}
\label{sec:observations}
% Observation details

\begin{deluxetable*}{cccccc}[htbp!]
  \tabletypesize{\scriptsize}
  \tablecaption{ALMA Projects for PEACHES \label{tbl:obs_projects}}
  \tablewidth{\textwidth}
  \tablehead{\colhead{Project Code} & \colhead{Regions} & \colhead{Calibrators (amplitude, bandpass, phase)} & \colhead{Baselines} & \colhead{$\theta_\text{beam}$\tablenotemark{a}} & \colhead{$\theta_\text{MRS}$\tablenotemark{b}}}
  \startdata
  2016.1.01501.S\tablenotemark{c} & L1448, L1455, NGC 1333 & J0238$+$1636, J0237$+$2848, J0336$+$3218 & 15--919 m & $\sim$0\farcs{6}$\times$0\farcs{4} & $\sim$6\farcs{3} \\
  2017.1.01462.S\tablenotemark{d} & B1, IC 348, B5 & J0237$+$2848, J0237$+$2848, J0336$+$3218 & 15--1231 m & $\sim$0\farcs{5}$\times$0\farcs{3} & $\sim$4\farcs{3} \\
  \enddata
  \tablenotetext{a}{The beam size varies slighly from source to source.  Table\,\ref{tbl:continuum} gives the synthesized beam size for each source.}
  \vspace{-1em}\tablenotetext{b}{Maximum recoverable scale.}
  \vspace{-1em}\tablenotetext{c}{Observations were performed on 2016 November 16, 19, 26, 29, and 30}
  \vspace{-1em}\tablenotetext{d}{Observations were performed on 2018 September 10 and 12}
  \vspace{-4em}
\end{deluxetable*}

The PEACHES observations were conducted in two ALMA projects (2016.1.01501.S and 2017.1.01462.S; PI: N. Sakai), surveying 37 fields toward the Perseus molecular cloud.  Each project surveyed different regions in the Perseus molecular cloud.  Table\,\ref{tbl:obs_projects} lists the basic information for each project, while Table\,\ref{tbl:source_list} shows basic information of the sources observed in the PEACHES survey.  The ALMA correlator was configured to have 13 spectral windows at Band 6.  The windows have 12 narrow spectral windows with 480 channels and a wide spw with 980 channels.  The narrow spectral windows were tuned to observe specific molecular species, such as SiO, \cch, CS, \methanol, and \methylcyanide, with a spectral resolution of 122\khz\ ($\sim$0.15\kms), while the wide spectral window was scheduled to observe the continuum with a resolution of 0.976\mhz\ ($\sim$1.2\kms).  In later imaging processes, we combined the two spectral windows potentially affected by the broad SiO emission, resulting in a total of 12 spectral windows.  The frequency setups for the two ALMA projects are largely identical, except for the wide continuum window, which shifts by $\sim$380\mhz.  Table\,\ref{tbl:spw} lists the frequency ranges for each spectral window and the corresponding channel widths.

We used CASA \citep{2007ASPC..376..127M} for standard calibration and imaging of the continuum and spectral lines with \texttt{tclean}.  Self-calibration is not applied in this study because some of the samples have weak continuum emission.  Because of the rich spectra, we manually flagged the lines for the continuum imaging.  The images were cleaned down to 0.022\,\jyb, except for the continuum spectral windows, whose images were cleaned down to 0.008\,\jyb\ due to their lower spectral resolution.  The line imaging used the ``multiscale'' deconvolver with a robust parameter of 0.5 because the targeted emission traces different spatial scales (e.g., SiO for outflows and COMs for the inner envelope/disk).  Last, we applied the primary beam correction to the image cubes.  The synthesized beam is about 0\farcs{6}$\times$0\farcs{4} averaged across all spectral windows.  In this study, we assume a distance of 300\,pc for the entire Perseus cloud, resulting in a synthesized beam of 180\,au$\times$120\,au.

\begin{deluxetable*}{cccccc}[htbp!]
    \tabletypesize{\scriptsize}
    \tablecaption{PEACHES Sample \label{tbl:source_list}}
    \tablewidth{\textwidth}
    \tablehead{\colhead{Source} & \colhead{Common Names} & \colhead{R.A. (J2000)} & \colhead{Decl. (J2000)} & 
    \colhead{$v_\text{lsr}$} & \colhead{Reference ($v_\text{lsr}$)} \\
    \colhead{} & \colhead{} & \colhead{(hh:mm:ss)} & \colhead{(dd:mm:ss)} & 
    \colhead{(km s$^{-1}$)} & \colhead{}}
    \startdata
    Per-emb 22 B   &                & 03:25:22.35    & 30:45:13.11    & 4.3 & S19   \\
    Per-emb 22 A   &                & 03:25:22.41    & 30:45:13.26    & 4.3 & S19   \\
    L1448 NW       & L1448 IRS 3C   & 03:25:35.67    & 30:45:34.16    & 4.2 & H18   \\
    Per-emb 33 B/C &                & 03:25:36.32    & 30:45:15.19    & 5.3 & S19   \\
    Per-emb 33 A   &                & 03:25:36.38    & 30:45:14.72    & 5.3 & S19   \\
    L1448 IRS 3A   &                & 03:25:36.50    & 30:45:21.90    & 4.6 & H18   \\
    Per-emb 26     &                & 03:25:38.88    & 30:44:05.28    & 5.4 & S19   \\
    Per-emb 42     &                & 03:25:39.14    & 30:43:57.90    & 5.8 & S19   \\
    Per-emb 25     & IRAS 03235$+$3004 & 03:26:37.51    & 30:15:27.81    & 5.5 & S18   \\
    Per-emb 17     & L1455 IRS 1, IRAS 03245$+$3002 & 03:27:39.11    & 30:13:02.96    & 6.0 & S19   \\
    Per-emb 20     & L1455 IRS 4    & 03:27:43.28    & 30:12:28.88    & 5.3 & S19   \\
    L1455 IRS 2    &                & 03:27:47.69    & 30:12:04.33    & 5.1 & H18   \\
    Per-emb 35 A   & NGC 1333 IRAS 1 & 03:28:37.10    & 31:13:30.77    & 7.4 & this study\\
    Per-emb 35 B   & NGC 1333 IRAS 1 & 03:28:37.22    & 31:13:31.74    & 7.3 & this study\\
    Per-emb 27     & NGC 1333 IRAS 2A & 03:28:55.57    & 31:14:36.97    & 6.5 & this study\\
    EDJ2009-172    &                & 03:28:56.65    & 31:18:35.43    & \nodata & \nodata\\
    Per-emb 36     & NGC 1333 IRAS 2B & 03:28:57.37    & 31:14:15.77    & 6.9 & S19   \\
    Per-emb 54     & NGC 1333 IRAS 6 & 03:29:01.55    & 31:20:20.49    & 7.9 & S19   \\
    SVS 13B        & NGC 1333 SVS 13B & 03:29:03.08    & 31:15:51.73    & 8.5 & S19   \\
    SVS 13A2       & VLA 3          & 03:29:03.39    & 31:16:01.58    & 8.4 & S18   \\
    Per-emb 44     & NGC 1333 SVS 13A & 03:29:03.76    & 31:16:03.70    & 8.7 & S19   \\
    Per-emb 15     &                & 03:29:04.06    & 31:14:46.23    & 6.8 & S19   \\
    Per-emb 50     & IRAS 03260$+$3111 A & 03:29:07.77    & 31:21:57.11    & 9.3 & this study\\
    Per-emb 12 B   & NGC 1333 IRAS 4A2 & 03:29:10.44    & 31:13:32.08    & 6.9 & S19   \\
    Per-emb 12 A   & NGC 1333 IRAS 4A1 & 03:29:10.54    & 31:13:30.93    & 6.9 & S19   \\
    Per-emb 21     & NGC 1333 IRAS 7 SM2 & 03:29:10.67    & 31:18:20.16    & 8.6 & this study\\
    Per-emb 18     & NGC 1333 IRAS 7 SM1 & 03:29:11.27    & 31:18:31.09    & 8.1 & S19   \\
    Per-emb 13     & NGC 1333 IRAS 4B1 & 03:29:12.02    & 31:13:07.99    & 7.1 & S19   \\
    IRAS4B'        & NGC 1333 IRAS 4B2 & 03:29:12.85    & 31:13:06.87    & 7.1 & S19   \\
    Per-emb 14     & NGC 1333 IRAS 4C & 03:29:13.55    & 31:13:58.12    & 7.9 & S19   \\
    EDJ2009-235    &                & 03:29:18.26    & 31:23:19.73    & 7.7 & this study\\
    EDJ2009-237    &                & 03:29:18.74    & 31:23:25.24    & \nodata & \nodata\\
    Per-emb 37     &                & 03:29:18.97    & 31:23:14.28    & 7.5 & this study\\
    Per-emb 60     &                & 03:29:20.05    & 31:24:07.35    & \nodata & \nodata\\
    Per-emb 5      & IRAS 03282$+$3035 & 03:31:20.94    & 30:45:30.24    & 7.3 & S19   \\
    Per-emb 2      & IRAS 03292$+$3039 & 03:32:17.92    & 30:49:47.81    & 7.0 & S19   \\
    Per-emb 10     & B1-d           & 03:33:16.43    & 31:06:52.01    & 6.4 & S19   \\
    Per-emb 40     & B1-a           & 03:33:16.67    & 31:07:54.87    & 7.4 & S19   \\
    Per-emb 29     & B1-c           & 03:33:17.88    & 31:09:31.74    & 6.1 & this study\\
    B1-b N         &                & 03:33:21.21    & 31:07:43.63    & 6.6 & C16   \\
    B1-b S         &                & 03:33:21.36    & 31:07:26.34    & 6.6 & C16   \\
    Per-emb 16     &                & 03:43:50.97    & 32:03:24.12    & 8.8 & S19   \\
    Per-emb 28     &                & 03:43:51.01    & 32:03:08.02    & 8.6 & S19   \\
    Per-emb 1      & HH 211 MMS     & 03:43:56.81    & 32:00:50.16    & 9.4 & S19   \\
    Per-emb 11 B   & IC 348 MMS     & 03:43:56.88    & 32:03:03.08    & 9.0 & S19   \\
    Per-emb 11 A   & IC 348 MMS     & 03:43:57.07    & 32:03:04.76    & 9.0 & S19   \\
    Per-emb 11 C   & IC 348 MMS     & 03:43:57.70    & 32:03:09.82    & 9.0 & S19   \\
    Per-emb 55     & IRAS 03415$+$3152 & 03:44:43.30    & 32:01:31.22    & 12.0 & S19   \\
    Per-emb 8      &                & 03:44:43.98    & 32:01:35.19    & 11.0 & S19   \\
    Per-emb 53     & B5 IRS 1       & 03:47:41.59    & 32:51:43.62    & 10.2 & this study\\
    \enddata
    \tablerefs{C16$=${\citet{2016AA...586A..44C}}; H18$=${\citet{2018ApJS..236...52H}}; 
               S18$=${\citet{2018ApJS..237...22S}}; S19$=${\citet{2019ApJS..245...21S}}.}
\end{deluxetable*}

\begin{table}[htbp!]
  \centering
  \caption{Frequency Setup}
  \label{tbl:spw}
  \scriptsize
  \begin{tabular}{ccc}
    \toprule
    Frequency Range & \multicolumn{2}{c}{Channel Width} \\
    (MHz) & (kHz) & (\kms)  \\
    \midrule
    \multicolumn{3}{c}{2016.1.01501.S} \\
    \midrule
    243483--243542 & 122.070 & 0.150 \\
    243878--243937 & 122.070 & 0.150 \\
    244200--244259 & 122.070 & 0.150 \\
    244898--244957 & 122.070 & 0.149 \\
    246186--247124 & 976.562 & 1.187 \\
    257489--257548 & 122.070 & 0.142 \\
    258218--258276 & 122.070 & 0.142 \\
    259288--259347 & 122.070 & 0.141 \\
    258974--259032 & 122.070 & 0.141 \\
    262046--262104 & 122.070 & 0.140 \\
    260442--260551 & 122.070 & 0.140 \\
    261787--261845 & 122.070 & 0.140 \\
    \midrule
    \multicolumn{3}{c}{2017.1.01462.S} \\
    \midrule
    243502--243561 & 122.070 & 0.150 \\
    243897--243956 & 122.070 & 0.150 \\
    244219--244278 & 122.070 & 0.150 \\
    244917--244975 & 122.070 & 0.149 \\
    245805--246743 & 976.562 & 1.189 \\
    257509--257568 & 122.070 & 0.142 \\
    258238--258296 & 122.070 & 0.142 \\
    259308--259366 & 122.070 & 0.141 \\
    258993--259052 & 122.070 & 0.141 \\
    262066--262124 & 122.070 & 0.140 \\
    260462--260571 & 122.070 & 0.140 \\
    261807--261865 & 122.070 & 0.140 \\
    \bottomrule
  \end{tabular}
\end{table}

\section{Analyses}
\label{sec:analyses}

\subsection{Identification of Young Stellar Objects}
\label{sec:continuum}
In the 37 fields of view, 51 continuum peaks are identified.  We used the CASA task \texttt{imfit} to iteratively fit a 2D Gaussian profile for continuum emission down to 5$\sigma$ of the residual image within the central 70\%\ of the size of the primary beam (20\arcsec).  We measured the noise of the continuum emission ($\sigma$) from the vicinity of the continuum emission.  For the field centered on Per-emb 16, which also covers Per-emb 28, the fitting used a threshold of 4$\sigma$ and extended the mask to the entire primary beam because Per-emb 28 is detected toward the edge of the primary beam, where the noise is higher.  The continuum of the multiple systems were manually fitted to the individual continuum peaks.  The two protostars L1448 IRS 2E and SVS 3 become nondetections in our observations due to their low brightness.  SVS 13C locates at the edge of the primary beam, resulting in a noisy continuum.  Thus, we excluded SVS 13C from this study, making a total of 50 protostars.

\begin{figure*}[htbp!]
  \centering
  \includegraphics[width=\textwidth]{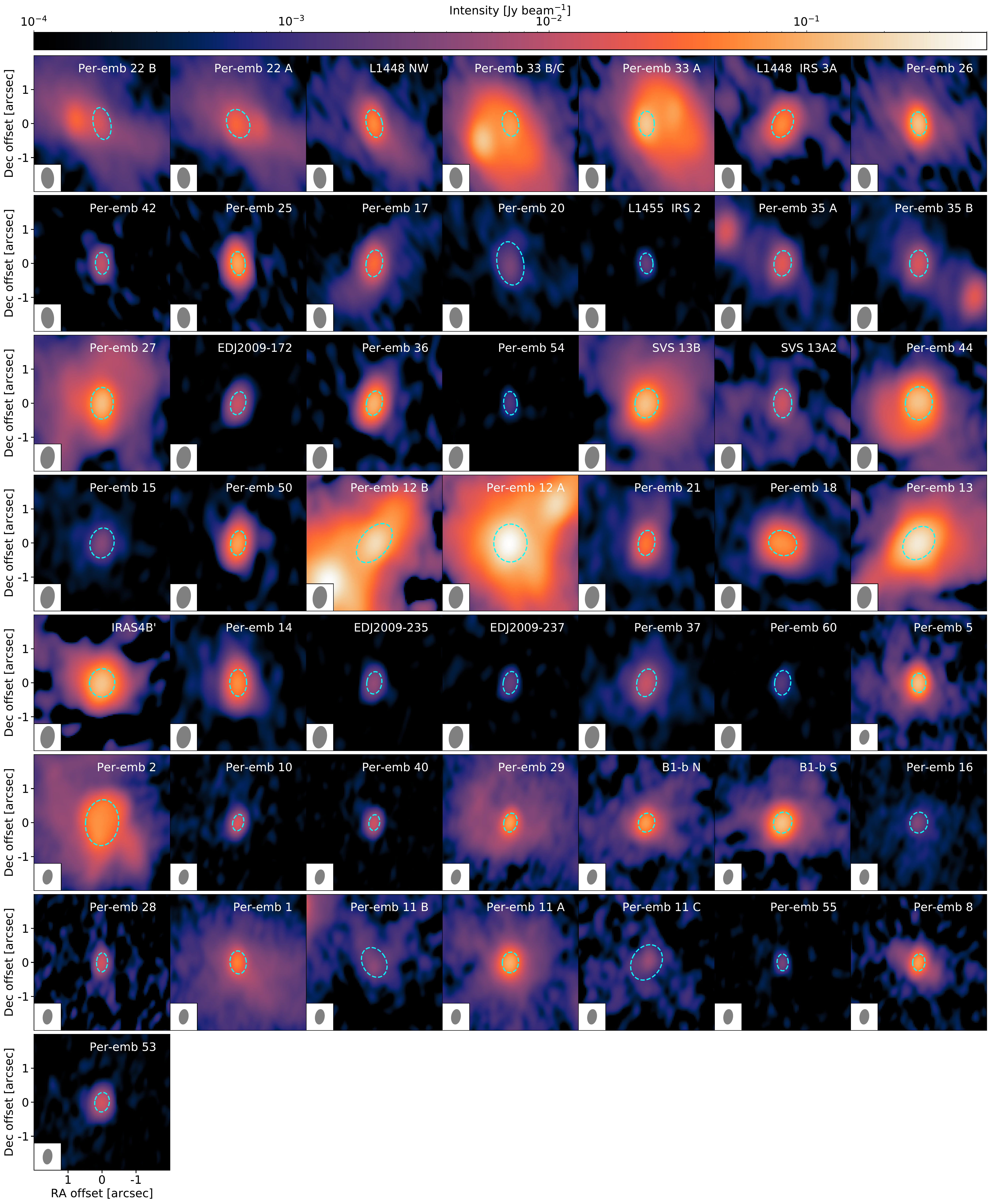}
  \caption{The continuum images of the PEACHES sample.  Each panel is a zoom-in view of each continuum source with a size of 4\arcsec$\times$4\arcsec\ and has the same color scale.  The corresponding beam size is shown in the lower left corner of each panel.  Nondetections toward L1448 IRS\,2E and NGC 1333 SVS 3 are not shown.  EDJ2009-172, EDJ2009-237, and Per-emb 60 have no emission line.  The dashed ellipses illustrate the size of the fitted continuum in which 1D spectra are extracted.}
  \label{fig:continuum}
\end{figure*}

Table\,\ref{tbl:continuum} lists the properties ot the fitted continuum emission, while Figure\,\ref{fig:continuum} shows the continuum emission along with the fitted shapes.  The continuum emission peak appears as a compact circular or elliptical shape.  Some sources show extended continuum emission resembling the shape of outflow cavities, such as Per-emb 22 A and B.  The source velocities collected from the literatures agree with our observations.  Three sources, EDJ2009-237, Per-emb 60, and EDJ2009-172, show no spectral line and no reliable measurement of source velocity in literature; therefore we excluded them from spectral extraction as well as from the line identification and modeling.  However, these three sources still contribute to the total number of sources for characterizing detection statistics.

\begin{deluxetable*}{cccccccc}
    \tabletypesize{\scriptsize}
    \tablecaption{Fitted Continuum \label{tbl:continuum}}
    \tablewidth{\textwidth}
    \tablehead{\colhead{Source} & \colhead{Beam} & \colhead{Convolved size} & \colhead{Deconvolved size} & \colhead{PA} & 
               \colhead{$T_\text{b, cont}$ (K)} & \colhead{Multiplicity\tablenotemark{a}}}
    \startdata
    Per-emb 22 B   & 0\farcs{64}$\times$0\farcs{39} & 0\farcs{95}$\times$0\farcs{51} & 0\farcs{72}$\times$0\farcs{34} & 16\fdg{0}$\pm$11\fdg{0} & 0.92   &     \\
    Per-emb 22 A   & 0\farcs{64}$\times$0\farcs{39} & 0\farcs{86}$\times$0\farcs{65} & 0\farcs{64}$\times$0\farcs{47} & 49\fdg{0}$\pm$18\fdg{0} & 1.71   &     \\
    L1448 NW       & 0\farcs{64}$\times$0\farcs{39} & 0\farcs{83}$\times$0\farcs{47} & 0\farcs{55}$\times$0\farcs{25} & 22\fdg{6}$\pm$5\fdg{5} & 3.15   & binary \\
    Per-emb 33 B/C & 0\farcs{64}$\times$0\farcs{39} & 0\farcs{75}$\times$0\farcs{48} & 0\farcs{42}$\times$0\farcs{28} & 24\fdg{0}$\pm$76\fdg{0} & 5.55   & binary \\
    Per-emb 33 A   & 0\farcs{64}$\times$0\farcs{39} & 0\farcs{73}$\times$0\farcs{45} & 0\farcs{38}$\times$0\farcs{24} & 172\fdg{0}$\pm$19\fdg{0} & 10.33  &     \\
    L1448 IRS 3A   & 0\farcs{64}$\times$0\farcs{39} & 0\farcs{85}$\times$0\farcs{59} & 0\farcs{65}$\times$0\farcs{32} & 135\fdg{4}$\pm$4\fdg{3} & 3.21   &     \\
    Per-emb 26     & 0\farcs{64}$\times$0\farcs{39} & 0\farcs{69}$\times$0\farcs{45} & 0\farcs{31}$\times$0\farcs{23} & 24\fdg{0}$\pm$17\fdg{0} & 8.03   &     \\
    Per-emb 42     & 0\farcs{64}$\times$0\farcs{39} & 0\farcs{64}$\times$0\farcs{39} & 0\farcs{17}$\times$0\farcs{05} & 155\fdg{0}$\pm$28\fdg{0} & 0.66   &     \\
    Per-emb 25     & 0\farcs{64}$\times$0\farcs{39} & 0\farcs{69}$\times$0\farcs{41} & 0\farcs{32}$\times$0\farcs{16} & 10\fdg{5}$\pm$1\fdg{5} & 5.27   &     \\
    Per-emb 17     & 0\farcs{64}$\times$0\farcs{40} & 0\farcs{79}$\times$0\farcs{48} & 0\farcs{53}$\times$0\farcs{24} & 155\fdg{5}$\pm$5\fdg{2} & 2.00   & binary \\
    Per-emb 20     & 0\farcs{64}$\times$0\farcs{40} & 1\farcs{29}$\times$0\farcs{78} & 1\farcs{13}$\times$0\farcs{68} & 12\fdg{1}$\pm$8\fdg{9} & 0.14   &     \\
    L1455 IRS 2    & 0\farcs{64}$\times$0\farcs{40} & 0\farcs{60}$\times$0\farcs{38} & 0\farcs{21}$\times$0\farcs{07}\tablenotemark{b} & \nodata \tablenotemark{b} & 0.13   &     \\
    Per-emb 35 A   & 0\farcs{66}$\times$0\farcs{42} & 0\farcs{75}$\times$0\farcs{51} & 0\farcs{36}$\times$0\farcs{29} & 9\fdg{9}$\pm$82\fdg{0} & 0.93   &     \\
    Per-emb 35 B   & 0\farcs{66}$\times$0\farcs{42} & 0\farcs{78}$\times$0\farcs{53} & 0\farcs{44}$\times$0\farcs{30} & 24\fdg{0}$\pm$25\fdg{0} & 0.75   &     \\
    Per-emb 27     & 0\farcs{66}$\times$0\farcs{42} & 0\farcs{93}$\times$0\farcs{66} & 0\farcs{66}$\times$0\farcs{50} & 9\fdg{2}$\pm$13\fdg{9} & 5.79   & binary \\
    EDJ2009-172    & 0\farcs{66}$\times$0\farcs{42} & 0\farcs{69}$\times$0\farcs{44} & 0\farcs{20}$\times$0\farcs{10} & 133\fdg{0}$\pm$12\fdg{0} & 0.62   &     \\
    Per-emb 36     & 0\farcs{66}$\times$0\farcs{42} & 0\farcs{73}$\times$0\farcs{46} & 0\farcs{31}$\times$0\farcs{19} & 153\fdg{4}$\pm$4\fdg{0} & 5.56   & binary \\
    Per-emb 54     & 0\farcs{66}$\times$0\farcs{42} & 0\farcs{69}$\times$0\farcs{40} & \nodata \tablenotemark{c} & \nodata \tablenotemark{c} & 0.07   &     \\
    SVS 13B        & 0\farcs{66}$\times$0\farcs{42} & 0\farcs{87}$\times$0\farcs{68} & 0\farcs{56}$\times$0\farcs{54} & 168\fdg{0}$\pm$77\fdg{0} & 6.64   &     \\
    SVS 13A2       & 0\farcs{66}$\times$0\farcs{42} & 0\farcs{86}$\times$0\farcs{53} & 0\farcs{55}$\times$0\farcs{31} & 11\fdg{0}$\pm$25\fdg{0} & 0.61   &     \\
    Per-emb 44     & 0\farcs{66}$\times$0\farcs{42} & 0\farcs{98}$\times$0\farcs{79} & 0\farcs{72}$\times$0\farcs{67} & 174\fdg{0}$\pm$74\fdg{0} & 6.84   & binary \\
    Per-emb 15     & 0\farcs{66}$\times$0\farcs{42} & 0\farcs{89}$\times$0\farcs{70} & 0\farcs{60}$\times$0\farcs{56} & 146\fdg{0}$\pm$81\fdg{0} & 0.17   &     \\
    Per-emb 50     & 0\farcs{66}$\times$0\farcs{42} & 0\farcs{73}$\times$0\farcs{44} & 0\farcs{30}$\times$0\farcs{15} & 177\fdg{2}$\pm$2\fdg{1} & 4.13   &     \\
    Per-emb 12 B   & 0\farcs{66}$\times$0\farcs{42} & 1\farcs{33}$\times$0\farcs{81} & 1\farcs{19}$\times$0\farcs{63} & 133\fdg{7}$\pm$7\fdg{3} & 10.04  &     \\
    Per-emb 12 A   & 0\farcs{66}$\times$0\farcs{42} & 1\farcs{11}$\times$0\farcs{98} & 0\farcs{91}$\times$0\farcs{86} & 34\fdg{0}$\pm$82\fdg{0} & 21.85  &     \\
    Per-emb 21     & 0\farcs{66}$\times$0\farcs{42} & 0\farcs{74}$\times$0\farcs{48} & 0\farcs{34}$\times$0\farcs{25} & 168\fdg{0}$\pm$15\fdg{0} & 2.05   &     \\
    Per-emb 18     & 0\farcs{66}$\times$0\farcs{42} & 0\farcs{84}$\times$0\farcs{73} & 0\farcs{73}$\times$0\farcs{30} & 75\fdg{0}$\pm$2\fdg{0} & 3.42   & binary \\
    Per-emb 13     & 0\farcs{66}$\times$0\farcs{42} & 1\farcs{07}$\times$0\farcs{83} & 0\farcs{91}$\times$0\farcs{64} & 122\fdg{1}$\pm$8\fdg{0} & 14.76  &     \\
    IRAS4B'        & 0\farcs{66}$\times$0\farcs{42} & 0\farcs{83}$\times$0\farcs{74} & 0\farcs{61}$\times$0\farcs{51} & 80\fdg{0}$\pm$17\fdg{0} & 7.13   &     \\
    Per-emb 14     & 0\farcs{66}$\times$0\farcs{42} & 0\farcs{79}$\times$0\farcs{50} & 0\farcs{46}$\times$0\farcs{23} & 18\fdg{6}$\pm$3\fdg{3} & 3.05   &     \\
    EDJ2009-235    & 0\farcs{67}$\times$0\farcs{42} & 0\farcs{66}$\times$0\farcs{44} & 0\farcs{19}$\times$0\farcs{13}\tablenotemark{b} & \nodata \tablenotemark{b} & 0.26   &     \\
    EDJ2009-237    & 0\farcs{67}$\times$0\farcs{42} & 0\farcs{67}$\times$0\farcs{42} & 0\farcs{11}$\times$0\farcs{05} & 132\fdg{0}$\pm$55\fdg{0} & 0.12   &     \\
    Per-emb 37     & 0\farcs{67}$\times$0\farcs{42} & 0\farcs{82}$\times$0\farcs{57} & 0\farcs{49}$\times$0\farcs{38} & 155\fdg{0}$\pm$28\fdg{0} & 0.56   &     \\
    Per-emb 60     & 0\farcs{67}$\times$0\farcs{42} & 0\farcs{73}$\times$0\farcs{47} & 0\farcs{31}$\times$0\farcs{20} & 15\fdg{0}$\pm$74\fdg{0} & 0.08   &     \\
    Per-emb 5      & 0\farcs{45}$\times$0\farcs{30} & 0\farcs{56}$\times$0\farcs{41} & 0\farcs{37}$\times$0\farcs{27} & 22\fdg{9}$\pm$5\fdg{1} & 15.29  &     \\
    Per-emb 2      & 0\farcs{45}$\times$0\farcs{30} & 1\farcs{35}$\times$0\farcs{97} & 1\farcs{28}$\times$0\farcs{93} & 175\fdg{6}$\pm$3\fdg{3} & 7.41   & binary \\
    Per-emb 10     & 0\farcs{46}$\times$0\farcs{30} & 0\farcs{49}$\times$0\farcs{32} & 0\farcs{20}$\times$0\farcs{14} & 161\fdg{0}$\pm$19\fdg{0} & 1.82   &     \\
    Per-emb 40     & 0\farcs{46}$\times$0\farcs{30} & 0\farcs{47}$\times$0\farcs{32} & 0\farcs{16}$\times$0\farcs{09} & 38\fdg{0}$\pm$13\fdg{0} & 1.44   & binary \\
    Per-emb 29     & 0\farcs{46}$\times$0\farcs{30} & 0\farcs{56}$\times$0\farcs{39} & 0\farcs{34}$\times$0\farcs{26} & 170\fdg{0}$\pm$34\fdg{0} & 8.41   &     \\
    B1-b N         & 0\farcs{46}$\times$0\farcs{30} & 0\farcs{56}$\times$0\farcs{47} & 0\farcs{38}$\times$0\farcs{33} & 87\fdg{0}$\pm$25\fdg{0} & 7.67   &     \\
    B1-b S         & 0\farcs{46}$\times$0\farcs{30} & 0\farcs{63}$\times$0\farcs{53} & 0\farcs{45}$\times$0\farcs{42} & 108\fdg{0}$\pm$28\fdg{0} & 14.79  &     \\
    Per-emb 16     & 0\farcs{50}$\times$0\farcs{32} & 0\farcs{61}$\times$0\farcs{52} & 0\farcs{44}$\times$0\farcs{41} & 86\fdg{0}$\pm$58\fdg{0} & 0.35   &     \\
    Per-emb 28     & 0\farcs{50}$\times$0\farcs{32} & 0\farcs{56}$\times$0\farcs{32} & 0\farcs{34}$\times$0\farcs{14} & 3\fdg{0}$\pm$6\fdg{9} & 1.52   &     \\
    Per-emb 1      & 0\farcs{49}$\times$0\farcs{32} & 0\farcs{68}$\times$0\farcs{48} & 0\farcs{52}$\times$0\farcs{36} & 20\fdg{0}$\pm$22\fdg{0} & 4.57   &     \\
    Per-emb 11 B   & 0\farcs{50}$\times$0\farcs{33} & 0\farcs{92}$\times$0\farcs{69} & 0\farcs{84}$\times$0\farcs{58} & 39\fdg{0}$\pm$24\fdg{0} & 0.40   &     \\
    Per-emb 11 A   & 0\farcs{50}$\times$0\farcs{33} & 0\farcs{61}$\times$0\farcs{48} & 0\farcs{41}$\times$0\farcs{38} & 14\fdg{0}$\pm$72\fdg{0} & 10.47  &     \\
    Per-emb 11 C   & 0\farcs{50}$\times$0\farcs{33} & 1\farcs{10}$\times$0\farcs{86} & 1\farcs{01}$\times$0\farcs{79} & 141\fdg{0}$\pm$57\fdg{0} & 0.34   &     \\
    Per-emb 55     & 0\farcs{50}$\times$0\farcs{32} & 0\farcs{49}$\times$0\farcs{33} & 0\farcs{23}$\times$0\farcs{10} & 36\fdg{0}$\pm$34\fdg{0} & 0.32   &     \\
    Per-emb 8      & 0\farcs{50}$\times$0\farcs{32} & 0\farcs{49}$\times$0\farcs{36} & 0\farcs{23}$\times$0\farcs{19} & 65\fdg{0}$\pm$24\fdg{0} & 8.51   &     \\
    Per-emb 53     & 0\farcs{51}$\times$0\farcs{33} & 0\farcs{58}$\times$0\farcs{42} & 0\farcs{36}$\times$0\farcs{31} & 150\fdg{0}$\pm$17\fdg{0} & 1.55   &     \\
    \enddata
    \tablenotetext{a}{The ``binary'' label indicates that the continuum source is an unsolved binary in our observations, according to \citet{2016ApJ...818...73T}.}
    \vspace{-1em}\tablenotetext{b}{The deconvolved continuum source is point source no larger than the listed size.}
    \vspace{-1em}\tablenotetext{c}{The continuum cannot be deconvolved because the emission may be resolved in only one direction.}
\end{deluxetable*}

The observations resolved or marginally resolved 94\% (47 of 50) of the continuum sources.  Our sample includes single sources as well as resolved and unresolved multiple systems.  According to the VANDAM survey \citep{2016ApJ...818...73T}, nine close binary systems remain unresolved in the PEACHES survey (Table\,\ref{tbl:continuum}).

\begin{figure}[htbp!]
  \centering
  \includegraphics[width=0.48\textwidth]{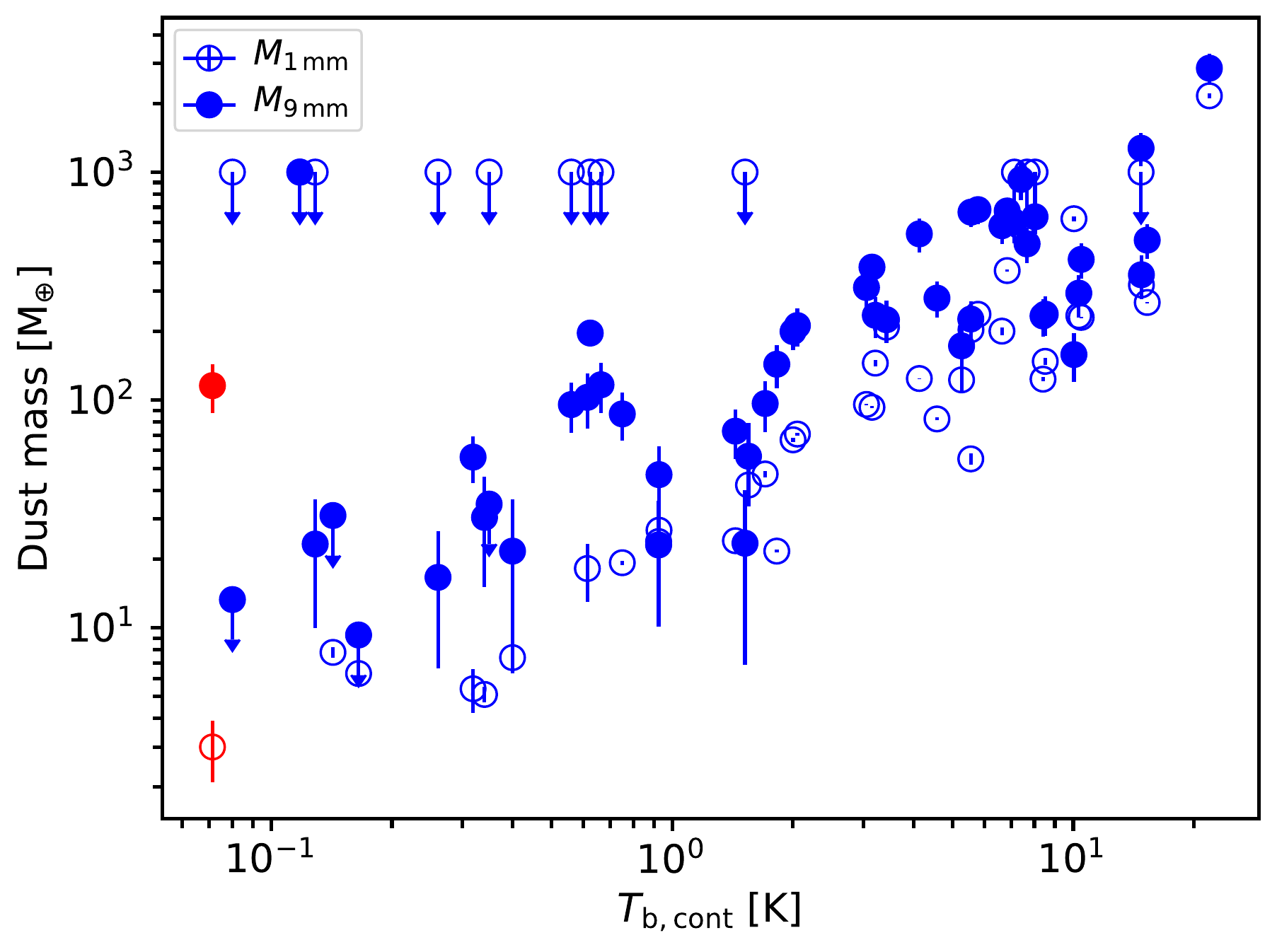}
  \caption{The averaged continuum brightness temperature compared with the masses derived with the 1\,mm and 9\,mm observations in opened and filled circles, respectively \citep{2020AA...640A..19T}.  The red points highlight the mass of Per-emb 54 as an outlier of the linear trend with the 9\,mm masses.}
  \label{fig:Tb_M}
\end{figure}

The mean continuum brightness temperature (\tbc) traces the optically thin dust mass; however, the dust emission from the center of protostars may be marginally optically thick at ALMA Band 6 \citep{2020ApJ...889..172K}, making the derived dust mass a lower limit.  The \tbc\ of the PEACHES sample correlates with the dust mass of the disk-like structure of Perseus protostars derived from the 1 mm ALMA and 9 mm VLA observations \citep{2020AA...640A..19T} except for that of Per-emb 54 (Figure\,\ref{fig:Tb_M}).  The 1 mm masses are systematically lower than the 9 mm masses because the inner disk-like structure becomes optically thick at 1 mm.  Nonetheless, the \tbc\ increases linearly with the 1 mm masses with a similar slope as that with the 9 mm masses, suggesting that the \tbc\ traces the central mass without significant impact of the optical depth.  The 1 mm mass of Per-emb 54 follows the linear trend, indicating that the deviation of its 9 mm mass may be due to excessive free-free emission.

Seven sources in our sample have \tbc\ $>$ 10\,K (Table\,\ref{tbl:continuum}).  Per-emb 12 A has a \tbc\ of 21.9\,K, the brightest among our sample.  In fact, the continuum opacity limits the detectability of COMs toward 4A1, where COMs show no emission or appear in absorption at ALMA Band 7 \citep{2017AA...606A.121L,2019ApJ...872..196S} and are observed in emission at centimeter wavelengths \citep{2020ApJ...896L...3D}.  Toward the other six sources with \tbc\ $>$ 10 K, Per-emb 33 A, Per-emb 12 B, Per-emb 13, Per-emb 5, B1-b S, and Per-emb 11 A (see Section\,\ref{sec:spec_extraction} and figures mentioned there), our observations detected several molecular lines from the central region, suggesting that the continuum opacity has less impact than that for Per-emb 12 A.  Previous observations also suggest less impact from the continuum opacity for Per-emb 13 \citep{2020AA...635A.198B} and B1-b S \citep{2018AA...620A..80M}.  Most of the sources have a mean brightness temperature lower than 10\,K, while spatially resolved embedded disks tend to have a warm temperature, $>20$ K within a radius of 100--150 au \citep{2018AA...615A..83V,2020AA...633A...7V,2020ApJ...901..166V}.  Thus the continuum source is likely to be unresolved and have an insignificant impact on the COMs emission at the angular resolution of this study.  Note that some of the sources, such as Per-emb 18 and Per-emb 21, show a systematical shift of the peaks in multiple species.  This could be understood by the high dust opacity at the continuum peak.  However, correlations between the column density of COMs (Section\,\ref{sec:correlations}) would be less affected because the dust opacity effect could be similar for all the species.  We take the \tbc\ as a tracer of the averaged gas column densities, judging from a reasonable correction in Figure\,\ref{fig:Tb_M}.  Constraining the actual column densities requires higher resolution observations at multiple frequency bands and insights into the dust properties.

\subsection{Spectral Extraction}
\label{sec:spec_extraction}
\begin{figure*}[htbp!]
  \centering
  \includegraphics[width=\textwidth]{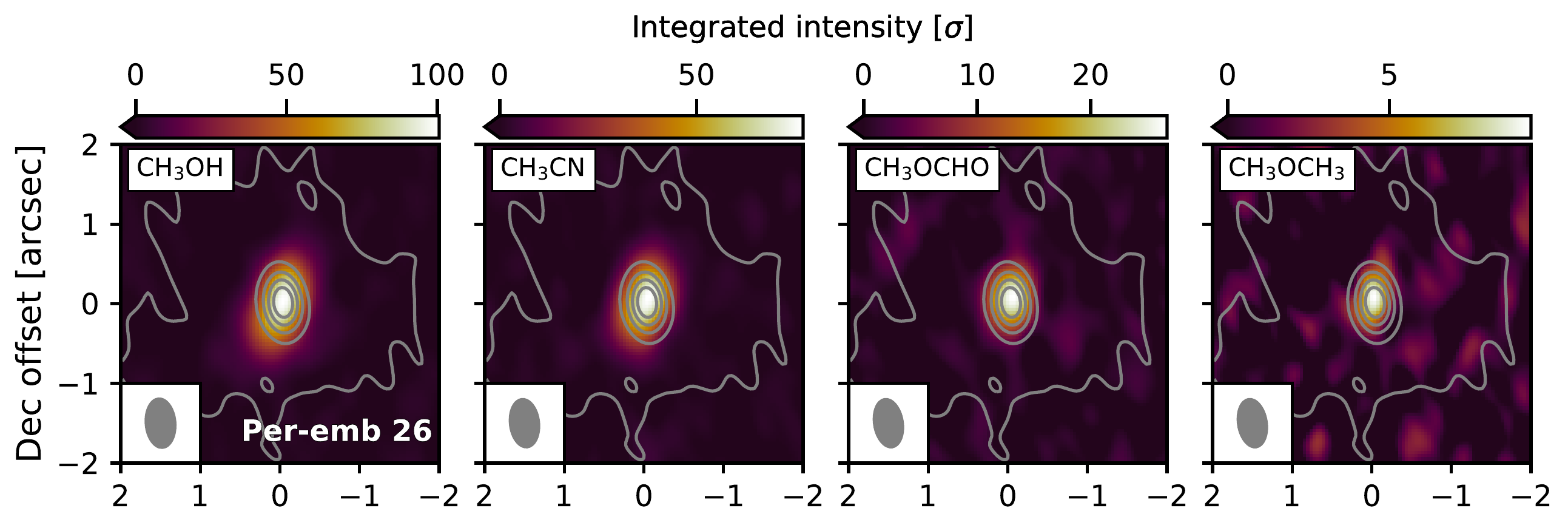}
  \includegraphics[width=\textwidth]{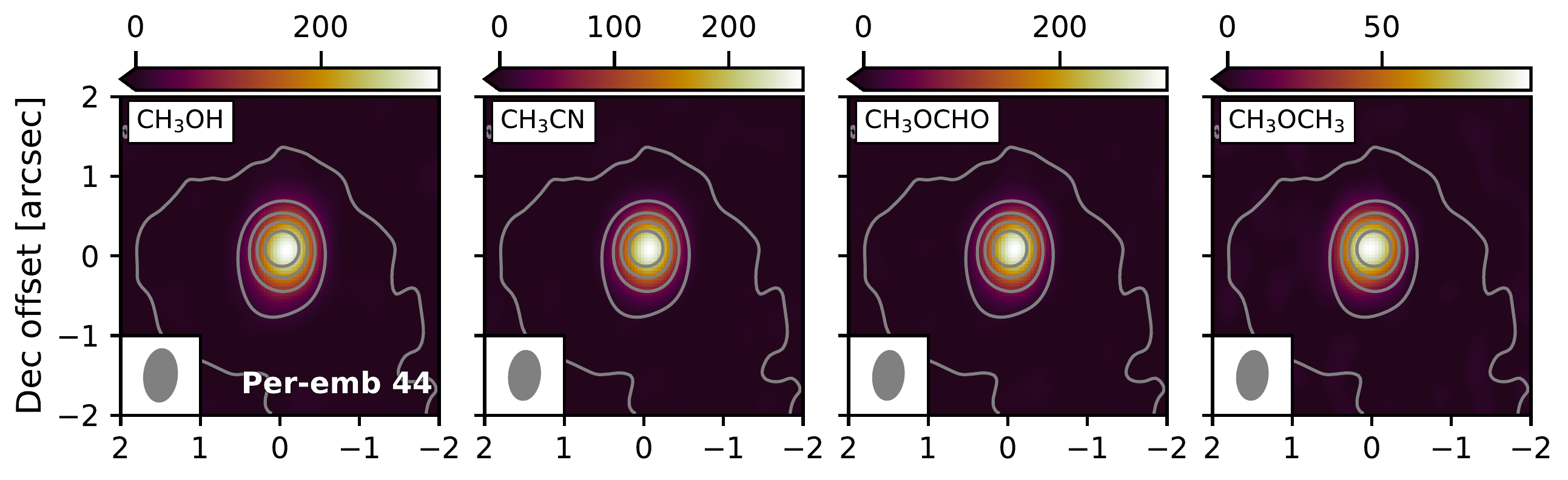}
  \includegraphics[width=\textwidth]{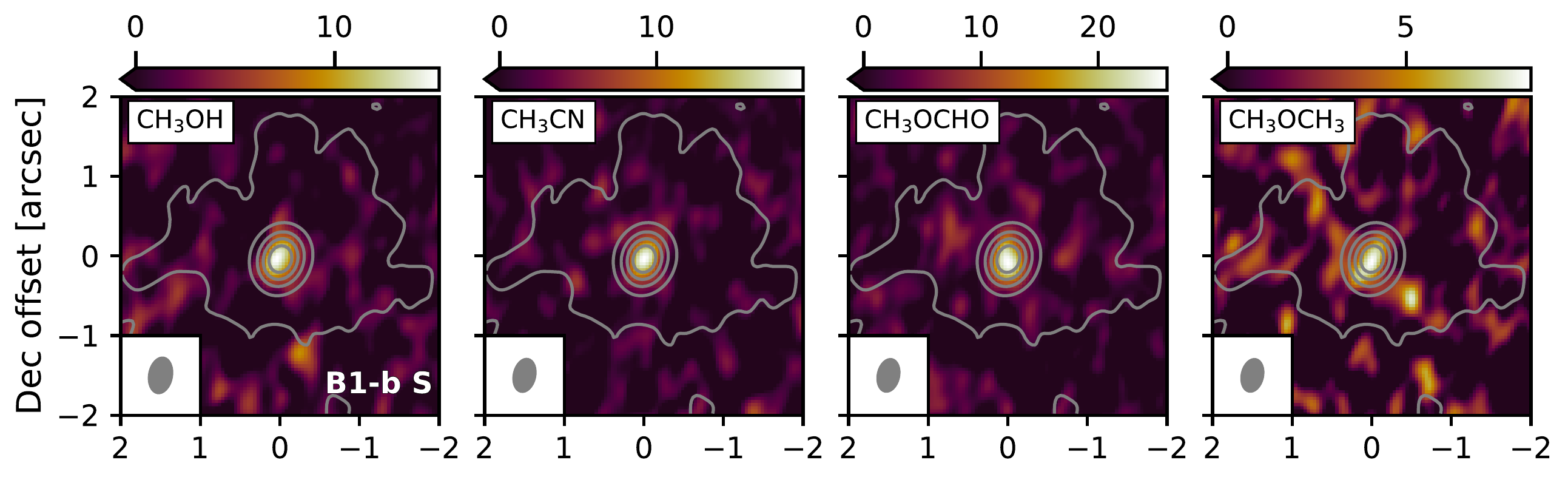}
  \includegraphics[width=\textwidth]{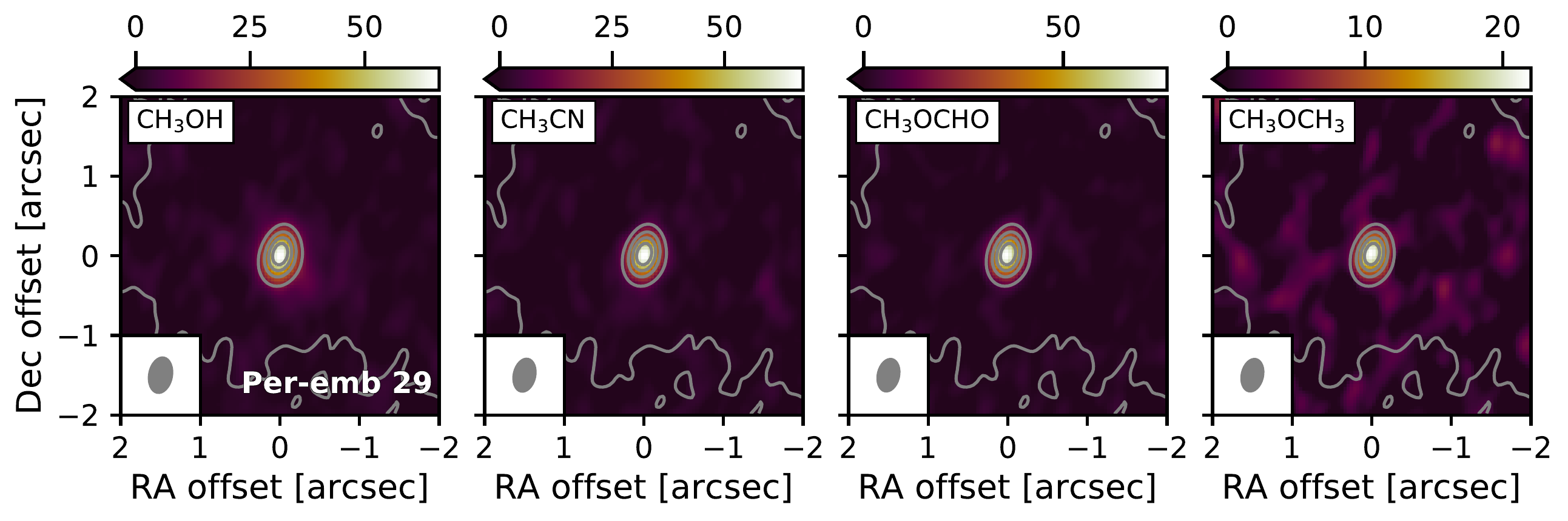}

  \caption{The intensity maps of the most frequently detected COMs, \methanol, \methylcyanide, \methylformate, and \dimethylether\ (from left to right) at selected sources.  Each row shows the emission from Per-emb 26, Per-emb 44, B1-b S, and Per-emb 29 (top to bottom).  The intensity is calculated by integrating over 3\kms\ around the line centroid shown in the unit of the corresponding noise.  The gray contours illustrate the continuum emission, with six contours linearly separated from 3$\sigma$ to the maximum intensity.}
  \label{fig:coms_map_sample}
\end{figure*}

The ALMA image cubes were post-processed to extract 1D spectra for identifying the emission of complex molecules and further analyses.  Given the compact size of COM emission ($\lesssim$100 au) and the spatial resolution of $\sim$0\farcs{5} ($\sim$150\,au at 300\,pc), we focused on the spectra toward the continuum sources to search for the COMs in the inner envelope.  Figure\,\ref{fig:coms_map_sample} shows a representative sample of the COMs emission, while the maps for the rest of the sample are shown in Appendix\,\ref{sec:coms_maps}.  In most of the cases, COM emission is concentrated around the protostars at $\lesssim$300\,au scale.  Three steps of post-processing reduced the image cubes to 1D spectra.  The steps are summarized below.

\begin{itemize}
  \item Extracting spectra:  We used the CASA task \texttt{specflux} to extract the mean flux density within the ellipse that has the same major and minor axes as well as the position angle as the fitted continuum sources (see also the dashed ellipses in Figure\,\ref{fig:continuum}).
  \item Baseline subtraction:  The continuum has been removed before the imaging process; however, the extracted spectra sometimes still show imperfect baselines due to the rich spectra of COMs and broad features.  Thus, we manually selected the frequency ranges for baseline subtraction for each spectral window and each field.
  \item Velocity correction:  Finally, the frequencies of the extracted spectra were corrected according to the source velocities.  We collected the source velocities from the literature as well as from the strong emission lines in our spectra, such as SO and CS.  Table\,\ref{tbl:source_list} lists the adopted source velocities and the corresponding references.
\end{itemize}

\subsection{Line Identification}
\label{sec:identification}
Line identification starts with manual identification and verification for a few sources with rich spectra, including Per-emb 12 B and B1-b S.  Then, we tested the list of identified species to the rest of sources for identification.  We used \textsc{splatalogue}\footnote{\href{http://www.splatalogue.net/}{http://www.splatalogue.net/}} to identify the molecular species, where the molecular data come from the Cologne Database of Molecular Spectroscopy (CDMS; \citealt{2001AA...370L..49M,2005JMoSt.742..215M,2016JMoSp.327...95E}) and the Jet Propulsion Laboratory Millimeter and Submillimeter Spectral Line Catalog (JPL; \citealt{1998JQSRT..60..883P}).  Appendix\,\ref{sec:catalogs} and Table\,\ref{tbl:molcat} summarize the catalogs relevant for each species.  Any tentatively identified species requires verification using synthetic spectra modeled with the \textsc{xclass} \citep{2017AA...598A...7M}, which performs local thermodynamic equilibrium (LTE) radiative transfer calculations using the molecular data from CDMS and JPL.  We describe the details of the \textsc{xclass} modeling in Section\,\ref{sec:modeling}.  An identification needs to satisfy the following criteria:
\begin{itemize}
  \item The predicted strengths in synthetic spectra agree with the observations, assuming a fiducial column density that produces the observed intensity, an excitation temperature of 100\,K, and a source size of 0\farcs{5}.  Identification of a species requires an unblended line detected at a signal-to-noise ratio (S/N) $>$ 3.
  \item For each species, at least one spectral line is not blended with other emission.  The emission of several species, such as HDCO \&\ \tmethanol, \methanol\ \&\ \methylformate, \acetaldehyde\ \&\ \dmethanol, $^{34}$SO \&\ \ethanol, and \dimethylether\ \&\ \dmethylcyanide, are partially blended (blending occurs at a few lines, but other lines remain isolated).  The fittings of these species were performed together to verify their identification.
  \item Identified molecules need to be already found toward young stellar objects, as summarized in \citet{2018ApJS..239...17M}.
\end{itemize}

The identified species and transitions that are detected in at least one of the PEACHES sample are listed in Appendix\,\ref{sec:line_id}.  Only identifiable transitions are listed.  The \textsc{xclass} modeling includes all the transitions for the identified species in our frequency coverage so that we can test the relative strengths of different transitions in the identification process.  Not all emission is identified; we reserve the analysis of the unidentified lines in future studies.

\subsection{Modeling the Spectra of COMs}
\label{sec:modeling}
\begin{figure*}[htbp!]
  \centering
  \includegraphics[width=\textwidth]{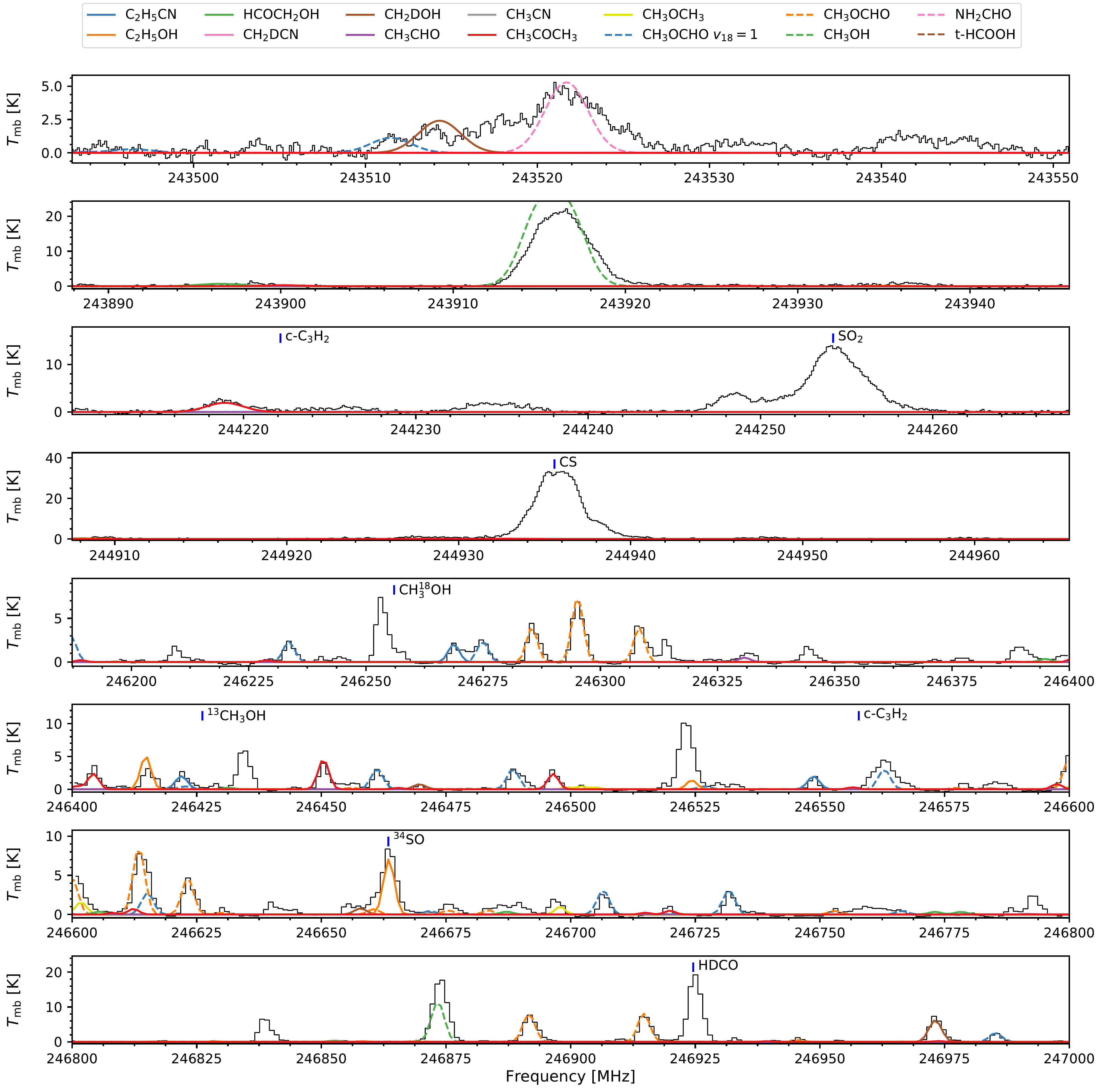}
  \caption{The spectra of Per-emb 44 (SVS 13 A) along with the best-fitting model assuming $T_\text{ex} = 150$\,K.  The fitted spectra for each species are illustrated by colored lines (solid and dashed) with the legends shown at the top of the figure.  The frequencies of identifiable simple organic species, carbon-chain molecules, and the blended COMs emission, such as \tmethanol\ and  \etmethanol, are shown as blue vertical bars with the species listed next to it.}
  \label{fig:svs13a}
\end{figure*}

In addition to verifying the identification, we use the \textsc{xclass} to model the observed spectra to constrain the column densities of COMs.  The \textsc{xclass} solves the radiative transfer equation for an isothermal source in 1D, called detection equation \citep{2004fost.book.....S}.  The source brightness distribution follows a 2D Gaussian distribution, described by its full width at half maximum (FWHM$_\text{COM}$), the column density within FWHM$_\text{COM}$ ($N_\text{COM}$), and the excitation temperature ($T_\text{ex}$).  The model has two additional parameters, the line width ($\Delta \nu$) and velocity offset ($v_\text{off}$), to calculate the spectra.  For each species, the \textsc{xclass} solves the radiative transfer equations independently.  For a few species whose emission is blended at some frequencies, we fit them together in pairs, such as \methanol\ \&\ \methylformate, HDCO \&\ \tmethanol, \acetaldehyde\ \&\ \dmethanol, $^{34}$SO \&\ \ethanol, and \dimethylether\ \&\ \dmethylcyanide.  We detect several simple molecules, such as CS, \htcn, HDCO, SO, and SO$_2$.  However, these lines often exhibit double-peaked or complex line profiles (see Figure\,\ref{fig:svs13a} for an example) due to the self-absorption or contamination from the envelope and outflows, which will be analyzed in future studies.  While our modeling includes these species to account for the weaker emission in the spectral coverage, we excluded these species from the following analysis because of the limitation of the simple LTE model.

\begin{figure}[htbp!]
  \centering
  \includegraphics[width=0.48\textwidth]{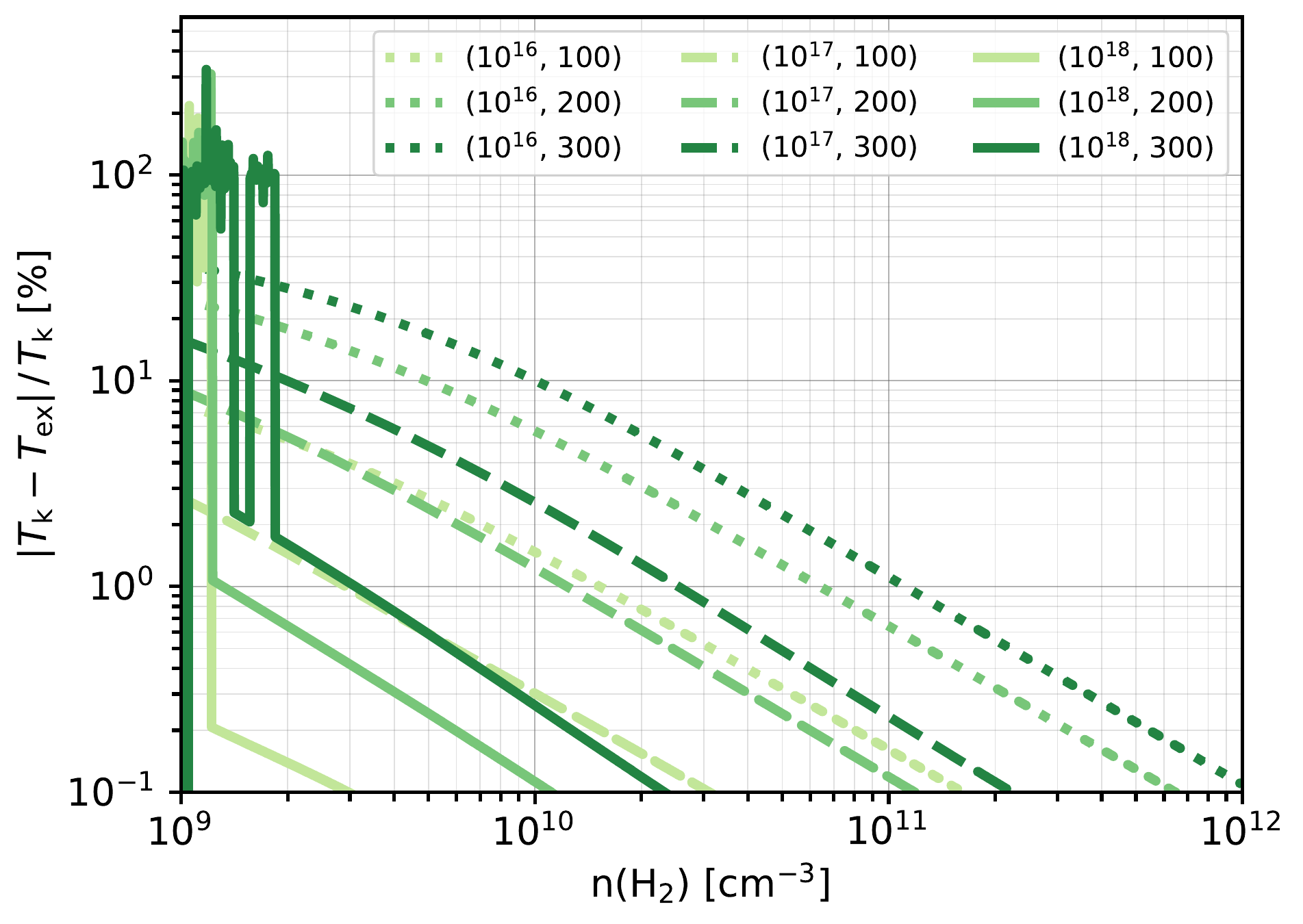}
  \caption{The difference between the excitation temperature ($T_\text{ex}$) and the kinetic temperature ($T_\text{k}$) normalized to $T_\text{k}$ for the \methanol\ $2_{11}\rightarrow 1_{01}$ transition at 261805.68\mhz\ as a function of the number density of H$_2$ at different $T_\text{k}$.  The calculations were performed using RADEX.  The results with \methanol\ column densities of 10$^{16}$, 10$^{17}$, and 10$^{18}$\,cm$^{-2}$ are shown in dotted, dashed, and solid lines, respectively, while the results with $T_\text{k}$ of 100, 200, and 300\,K are shown in decreasing transparency.  The legends indicate the combination of $N_{\rm CH3OH}$ and $T_\text{k}$ as ($N_\text{CH3OH}$, $T_\text{k}$).}
  \label{fig:radex}
\end{figure}

To test the assumption of LTE, we ran 1D non-LTE models to estimate the discrepancy between the kinetic temperature ($T_\text{k}$) and the excitation temperature ($T_\text{ex}$) for the emission of \methanol, which often is the most abundant COM \citep[e.g., ][]{2020AA...635A.198B}.  We followed a similar procedure as by \citet{2016AA...595A.117J} using RADEX, a 1D non-LTE radiative transfer package \citep{2007AA...468..627V}.  The collision rates were taken from the Leiden Atomic and Molecular Database \citep{2005AA...432..369S} based on calculations by \citet{2010MNRAS.406...95R}.  For one of the two most frequently detected \methanol\ transitions at 261805.68\mhz\ in our observations, the $T_\text{k}$ deviates from the $T_\text{ex}$ by more than 10\% at n(H$_2$) $<$ 10$^{10}$ \cc, or the \methanol\ column density $\lesssim$ 10$^{16}$ cm$^{-2}$ (Figure\,\ref{fig:radex}).  The disagreement increases toward lower gas densities where the negative $T_\text{ex}$, indicative of a masing condition, causes the spikes in Figure\,\ref{fig:radex}.  Assuming optically thin continuum emission, we estimated a gas density ranging from $\sim 4\times10^{9}-10^{12}$ \cc\ for the continuum sources in our sample (Appendix\,\ref{sec:n_gas}).  In this range, $T_\text{k}$ agrees with $T_\text{ex}$ within 10--30\%.  Our calculations for the \methanol\ line at 243915.79\mhz\ show a maximum difference smaller than 0.1\%\ for the combinations of parameters.  The 243915.79\mhz\ \methanol\ line has $\Delta K = 0$ so that it is not a maser line, which requires $\Delta K \neq 0$.  Therefore LTE is a valid assumption for modeling the spectra of COMs.  

Our observations cover one transition or a few transitions for each species, making the determination of excitation temperature uncertain.  Thus, instead of choosing an excitation temperature, we optimized the model with five different $T_\text{ex}$, 100, 150, 200, 250, and 300 K.  Then, we took the mean and the range of fitted column density as the best-fit value and the corresponding uncertainty, respectively.  For the COMs with A- and E-species due to the nuclear spins, such as \methanol\ and \methylcyanide, we assumed that the A- and E-species have the same excitation temperature so that they can be modeled together.  These $T_\text{ex}$ cover the typical range of $T_\text{ex}$ for the COMs detected toward embedded protostars \citep[e.g., ][]{2019ECS.....3.1564B}.  We derived a similar range of $T_\text{ex}$ for the species that have multiple detected transitions, such as \methanol\ and \methylformate\ (Appendix\,\ref{sec:Tex}), suggesting that the range of 100--300 K represents the typical $T_\text{ex}$ of COMs in the PEACHES sample.  Therefore we can take the range of the derived column densities as an estimate of the true column density without fine-tuning the model for each source.  To have a robust fitting result, we fixed FWHM$_\text{COM}$ to 0\farcs{5} and assumed $v_\text{off} = 0$.  We also assumed a negligible dust optical depth, making the fitted column density a lower limit if the dust opacity turns out to be substantial.  The synthetic spectra were calculated for the size of continuum emission (Table\,\ref{tbl:source_list}) to include the effect of beam dilution.  We allowed the line width to vary between 1.2 and 3.5\kms\ for a better fitting quality and chose a range of the column density for each molecule according to the strength of the emission.

The fitting function in \textsc{xclass} includes several optimization algorithms that can be used in sequence to reduce biases.  We configured the algorithm chain that starts with the genetic algorithm followed by the Levenberg--Marquardt $\chi^{2}$ minimization.  The genetic algorithm iteratively searches for the best-fitting parameters with generations that evolve like a natural selection, where the better fitting models are modified less over generations.  We set up the genetic algorithm to search for the top two best-fitting models with 30 generations.  Then, the Levenberg--Marquardt $\chi^{2}$ minimization was applied to the two best-fitting models for 20 iterations to determine the best-fitting models.  The genetic algorithm aims to find global minima, and the Levenberg--Marquardt $\chi^2$ minimization further finds the best-fitting models in the global minimums.  The two best-fitting models found by the genetic algorithm are often very similar, suggesting that there is only one minimum.  In the rare cases of two distinct global minima, we selected the model with the lower $\chi^{2}$ values from the two best-fitting models constrained by the Levenberg--Marquardt $\chi^2$ minimization.  We allowed the column density to vary by a few orders of magnitude to ensure that a global minimum was found.

From the five fitted column densities assuming five different $T_\text{ex}$, if a molecule is detected according to the criteria listed in Section\,\ref{sec:identification}, we took the mean column densities as the best-fitting column density, while the range of the column densities indicates the uncertainty.  If a species is considered nondetected (Section\,\ref{sec:identification}), we used the synthetic spectra to derive the upper limit of the column density corresponding to the noise of the spectra.  The fitted column density assumes an FWHM of 0\farcs{5} (150 au).  Figure\,\ref{fig:svs13a} shows an example of the fitted spectra for Per-emb 44 (SVS 13 A), assuming $T_\text{ex} = 150$\,K.  The fitted spectra may under- or overestimate the strengths of some transitions because the assumed excitation temperature may not be the true excitation temperature.

\renewcommand{\thefigure}{\arabic{figure} (Cont.)}
\addtocounter{figure}{-1}
\begin{figure*}[htbp!]
  \centering
  \includegraphics[width=\textwidth]{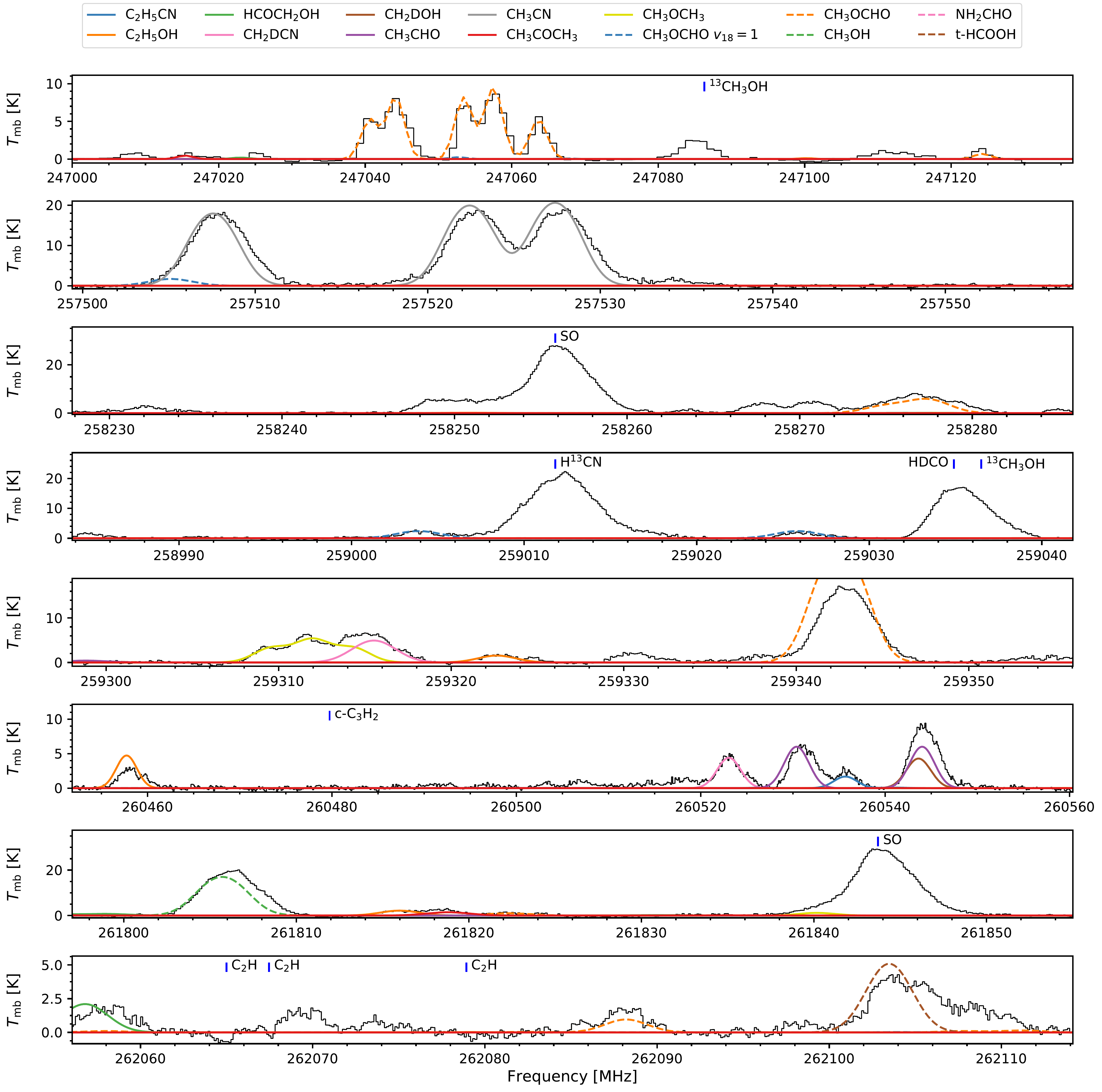}
  \caption{}
\end{figure*}
\renewcommand{\thefigure}{\arabic{figure}}

\begin{figure}[htbp!]
  \centering
  \includegraphics[width=0.47\textwidth]{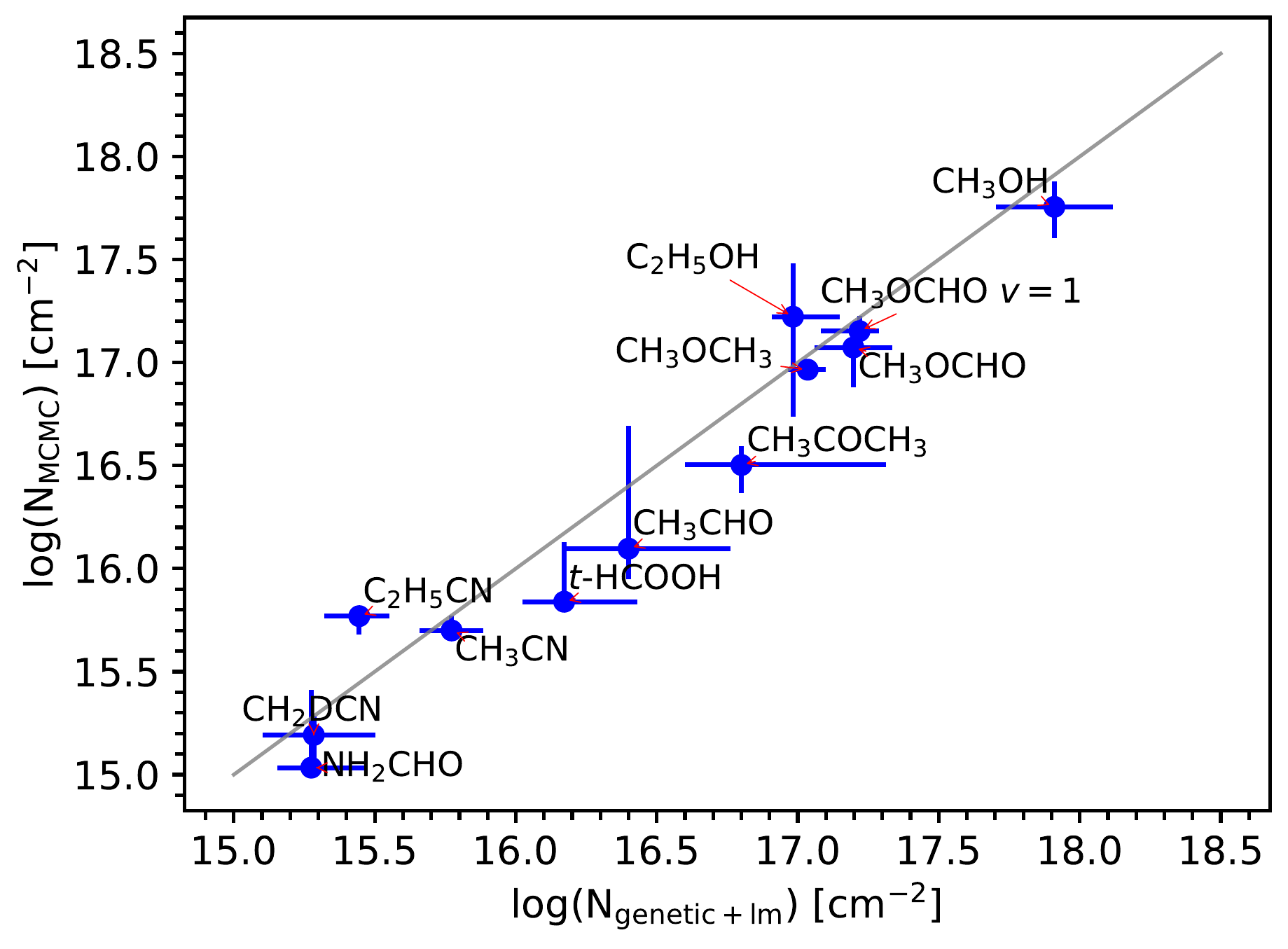}
  \caption{The column densities of Per-emb 44 (SVS 13 A) fit the hybrid algorithm, genetic and Levenberg--Marquardt $\chi^2$ minimization, and the MCMC algorithm.  The molecular species are annotated, and the gray line indicates the equality.}
  \label{fig:genetic_lm_mcmc}
\end{figure}

To benchmark this hybrid optimization process, we compared the fitted column densities with the fitting using the Monte Carlo Markov chain (MCMC) algorithm on the most chemically rich source, Per-emb 44 (SVS 13 A).  The MCMC algorithm uses the affine-invariant MCMC package \textsc{emcee} \citep{2013PASP..125..306F} to sample the parameter space.  In this MCMC optimization, we fit both the excitation temperature and the column density.  For a single-species fitting, we used 100 walkers with 100 iterations after 30 burn-in iterations; for a combined fitting of two species, we used 500 walkers to better sample the parameter space.  The fitting column densities from the hybrid (genetic and Levenberg--Marquardt $\chi^2$ minimizations) method are consistent with that fit with the MCMC optimization (Figure\,\ref{fig:genetic_lm_mcmc}).  Moreover, our hybrid method yields comparable uncertainties to those from the MCMC method, validating our modeling approach.

\section{Results}
\label{sec:results}
\subsection{Detection Statistics}
\begin{figure*}[htbp!]
  \includegraphics[width=\textwidth]{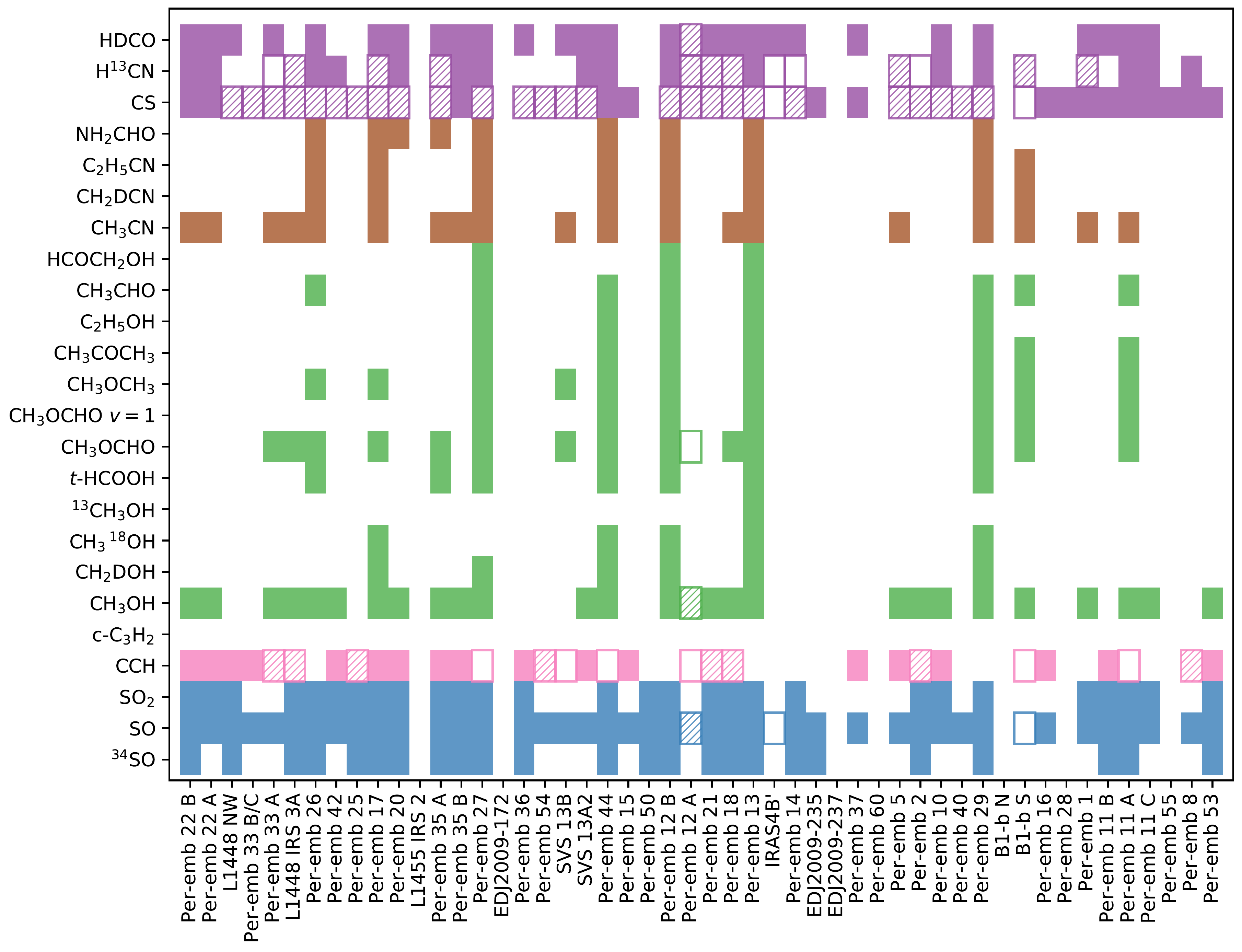}
  \caption{Summary of molecular detections.  The sources are sorted by increasing R.A. from left to right.  The detections are color-coded by the types of species: S-bearing molecules in blue, carbon-chain molecules in pink, O-bearing COMs in green excluding N-bearing molecules, N-bearing COMs in brown, other simple organics in dark purple.  The boxes with solid colors indicate emission and the empty boxes indicate absorption.  The hatched boxes indicate both emission and absorption are seen.}
  \label{fig:stats}
\end{figure*}

We summarize the detection statistics in Figure\,\ref{fig:stats}, which includes COMs, carbon-chain molecules, and simple molecules, such as CS, \htcn, SO, $^{34}$SO, and SO$_{2}$.  To derive the detection fraction, we included the three sources that were excluded from modeling because their source velocity was not reliable (Section\,\ref{sec:continuum}), making a total of 50 sources.  The COMs discussed here are derived from the spectra taken from the area surrounding the continuum peak, as explained in Section\,\ref{sec:continuum}.  We focus on the chemistry of COMs toward the disk-forming regions in this study.  Therefore this study excludes any molecules that are only detected outside the region of continuum emission.  A comparison of the chemical composition in protostellar envelopes (100--1000\,au) requires observations with a larger maximum recoverable scale ($\theta_\text{MRS}$).

The PEACHES sample shows a diverse chemistry of COMs, from no molecule detected (e.g., B1-b N and L1455 IRS 2) to rich spectra of COMs (e.g., Per-emb 13).  Most protostars of the PEACHES sample have simple molecules, such as SO, CS, \htcn, and HDCO, and $\sim 60$\%\ of the sources have SO$_{2}$ and $^{34}$SO.  Emission of \cch\ can be easily identified from the spectra.  However, the \cch\ toward the continuum sources often shows irregular line profiles together with velocity offsets and absorption (Appendix\,\ref{sec:cch}, Figure\,\ref{fig:all_cch}).  Warm environments, such as the outflow cavity wall, easily enhance the abundance of \cch\ because of the elevated abundance of C$^{+}$ \citep[e.g., ][]{2018ApJ...864...76Z,2019ApJ...873L..21I}.  Thus, \cch\ emission tends to extend along with the outflow cavities, which measn that 1D spectra toward the continuum source are no longer representative of the abundance of \cch.  Future studies  will investigate the comparison between unsaturated species and COMs using follow-up observations with a higher dynamic range of the spatial distribution (from $\sim$100 au to several 1000 au scale).  Previous observations with single-dish telescopes have a beam size of $\sim$3000 au \citep[e.g., ][]{2018ApJS..236...52H}.  Thus, observations that fill in the spatial scale from several 100 au to 1000 au are necessary.

Several sources have broad SiO emission, significantly contaminating the emission of \ethylcyanide\ and \acetaldehyde.  In comparison, the emission of COMs has relatively narrow line width of a few \kms.  In the quantitative discussion below, we exclude the spectral windows that are contaminated by the SiO emission.  For a few sources, such as Per-emb 26, we can still distinguish the emission of \ethylcyanide\ and \acetaldehyde\ from the broad SiO emission and include these identification in the detection statistics.

Figure\,\ref{fig:stats} shows the number of COMs detected toward the PEACHES sample.  Twenty-nine (58\%) sources have detected COMs.  \methanol\ is detected in 28 sources (56\%), \methylformate\ is detected in 14 sources (28\%), and N-bearing COMs are detected in 20 sources (40\%).  Compared to the COMs in the CALYPSO survey \citep{2020AA...635A.198B}, the fraction of sources that have methanol, $\sim$50\%, is similar to that for the PEACHES sample.  Furthermore, 30\%\ of the CALYPSO sources have at least three species of COMs, while 28\%\ of the PEACHES sample have at least three species of COMs.  In a smaller sample selected from several molecular clouds, \citet{2020AA...639A..87V} found that 3 of 7 protostars show emission of COMs.

\begin{figure}[htbp!]
  \centering
  \includegraphics[width=0.48\textwidth]{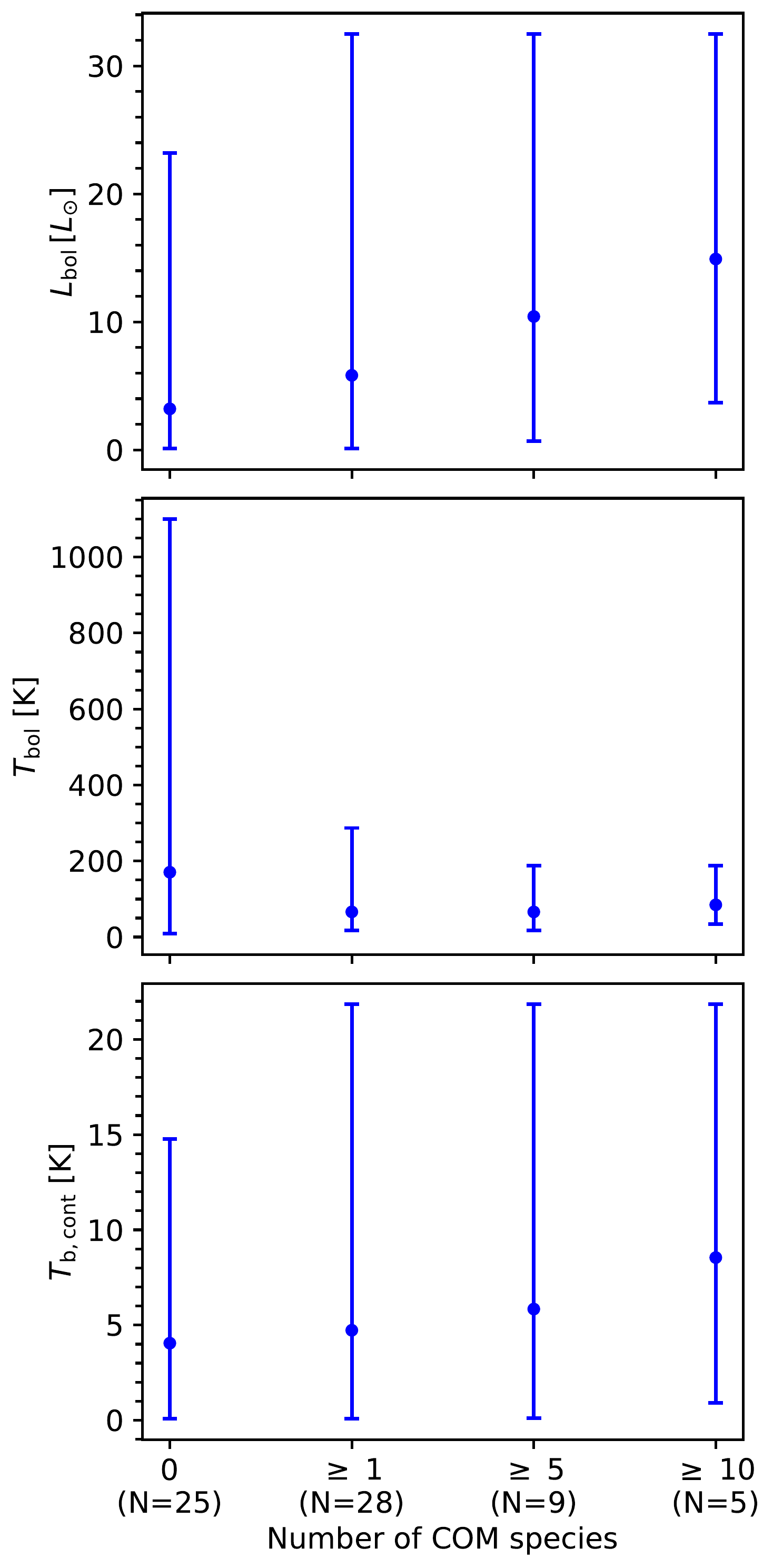}
  \caption{The number of COMs detected toward the PEACHES sample compared with the bolometric luminosity (\lbol), the bolometric temperature (\tbol), and the averaged continuum brightness temperature (\tbc) from top to bottom.  The sample is divided into four groups with detected COM species of 0, $\geq$1, $\geq$5, and $\geq$10, respectively.  For each group, the circles show the mean values of the corresponding properties, while the error bars show the value ranges.  The number of sources in each group is labeled in the parentheses.  The \lbol\ and the \tbol\ are taken from \citet{2016ApJ...818...73T} and \citet{2016AA...592A..56M}.}
  \label{fig:detection_summary}
\end{figure}

To investigate the impact of protostellar properties on the appearance of COMs, we divided the PEACHES sample into four groups according to the number of detected COMs from no COM, at least 1, 5, and 10 species of COMs (Figure\,\ref{fig:detection_summary}).  The compared protostellar properties include the bolometric luminosity (\lbol), the bolometric temperature (\tbol), and the mean continuum brightness temperature (\tbc), where the \tbol\ and \lbol\ are collected from \citet{2016ApJ...818...73T} and \citet{2016AA...592A..56M}.  The \tbol\ traces the evolutionary stage of protostars with higher \tbol\ for more evolved protostars \citep{1993ApJ...413L..47M,1995ApJ...445..377C,2009ApJS..181..321E}; and the \lbol\ serves as a proxy of the size of the warm inner envelope as the accretion luminosity dominates the \lbol\ for Class 0/I protostars \citep{2014prpl.conf..195D}.  The number of detected COMs shows no obvious trend with \lbol, \tbol, and \tbc.  

If the COMs mostly come from thermal desorption, the region with $T>T_\text{desorption}$ may be smaller for low-luminosity sources, resulting in fainter emission of COMs and reducing their detectability.  In contrast to this expectation, the COMs detection shows no obvious trend with \lbol, except that the group with more COMs has a higher minimum \lbol.  However, the range of \lbol\ in each group is much larger than the change in the minimum \lbol; thus, we cannot establish a significant correlation between the appearance of COMs and \lbol.

The high \tbol\ value for the sources without COM detection is driven by the high \tbol\ of EDJ2009-172 ($T_\text{bol}=1100$ K), while the second highest \tbol\ for the group is L1455 IRS 2 with 740 K.  The lower maximum \tbol\ for the sources with more detected COMs indicates that evolved sources may have fewer COMs in the gas phase.  In the Ophiuchus molecular cloud, \citet{2019AA...626A..71A} reported no detection of \methanol\ toward a sample of Class I protostars either.  If we take \tbol\ of 70 K as the boundary between Class 0 and I, our survey shows no obvious difference for the Class 0 and I sources; however, the evolutionary stage classified with \tbol\ may not represent the true evolutionary state of the protostar due to extinction and the effect of inclination \citep{2013AA...551A..98L,2014prpl.conf..195D}.  The detectability of COMs may depend on other factors, such as the chemistry in the envelope scale, instead of the current physical condition.  Thus, the chemo-physical history of protostars may play a dominant role in the chemistry of COMs \citep[e.g., ][]{2008ApJ...682..283G,2013ChRv..113.8961A,2020ApJ...897..110A}.

Several COM-rich protostars in Perseus have been known previously \citep[e.g., ][]{2004ApJ...615..354B,2007AA...463..601B,2006PASJ...58L..15S,2018AA...620A..80M,2019MNRAS.483.1850B}.  Our survey detects more COMs in addition to the known \methanol\ toward Per-emb 35 A and Per-emb 11 A (Figure\,\ref{fig:detection_summary}).  Per-emb 35 A is in a binary system (NGC 1333 IRAS 1) together with Per-emb 35 B.  Source A has emission of \methanol, \methylcyanide, \methylformate, and $t$-HCOOH, while Source B only has weak emission of \methanol, showing an apparent chemical differentiation.  Per-emb 11 A is the brightest source in a wide triple system, where Per-emb 11 B is $\sim$3\arcsec\ away toward the southwest and Per-emb 11 C is $\sim$9\farcs{5} away toward the northeast \citep{2016ApJ...818...73T}.  Per-emb 11 A has rich spectra of COMs, including \methylformate, \dimethylether, \acetone, \acetaldehyde, and \methylcyanide.

Compact emission of COMs is consistent with an origin of thermal desorption at the inner envelope, the so-called hot corinos \citep{2004ASPC..323..195C}.  However, we cannot rule out other origins, such as the local enhancement due to accretion shocks \citep{2016ApJ...824...88O}, because our observations only marginally resolved most of the emission of COMs.  Only L1448 IRS 3A shows extended \methanol\ emission, and B1-b N shows no \methanol\ emission toward the continuum, but peaks at $\sim$1\arcsec\ away from the continuum, which is consistent with the existence of \methanol\ toward the continuum source blocked by opaque continuum \citep{2018AA...620A..80M}.  The enrichment of COMs due to outflows via sputtering \citep{2008ApJ...681L..21A} or photodesorption in the cavity walls \citep{2015MNRAS.451.3836D} dominates at a larger scale ($>$1000 au) and can be time dependent.  It would therefore play a minor role for the COMs that are detected toward the continuum sources.  Excluding L1448 IRS 3A, 28 sources (56\%) are likely to have a hot-corino type chemistry.  However, studies of the origin of COMs require higher spatial resolutions and are beyond the scope of this study.  

\subsection{Column Densities}
From the method described in Section\,\ref{sec:modeling}, we constrained the column densities of COMs along with their uncertainties (Table\,\ref{tbl:Ncol}).  Our modeling pipeline successfully reproduced the spectra of most PEACHES sample, except for Per-emb 17 because of its double-peaked line profile, which we further discuss in Section\,\ref{sec:per17}.  Using the fitted column densities, we discuss the correlation between COMs in Section\,\ref{sec:correlations}.

\subsubsection{The Double-peaked Features of Per-emb 17}
\label{sec:per17}
\begin{figure*}[htbp!]
  \centering
  \includegraphics[width=\textwidth]{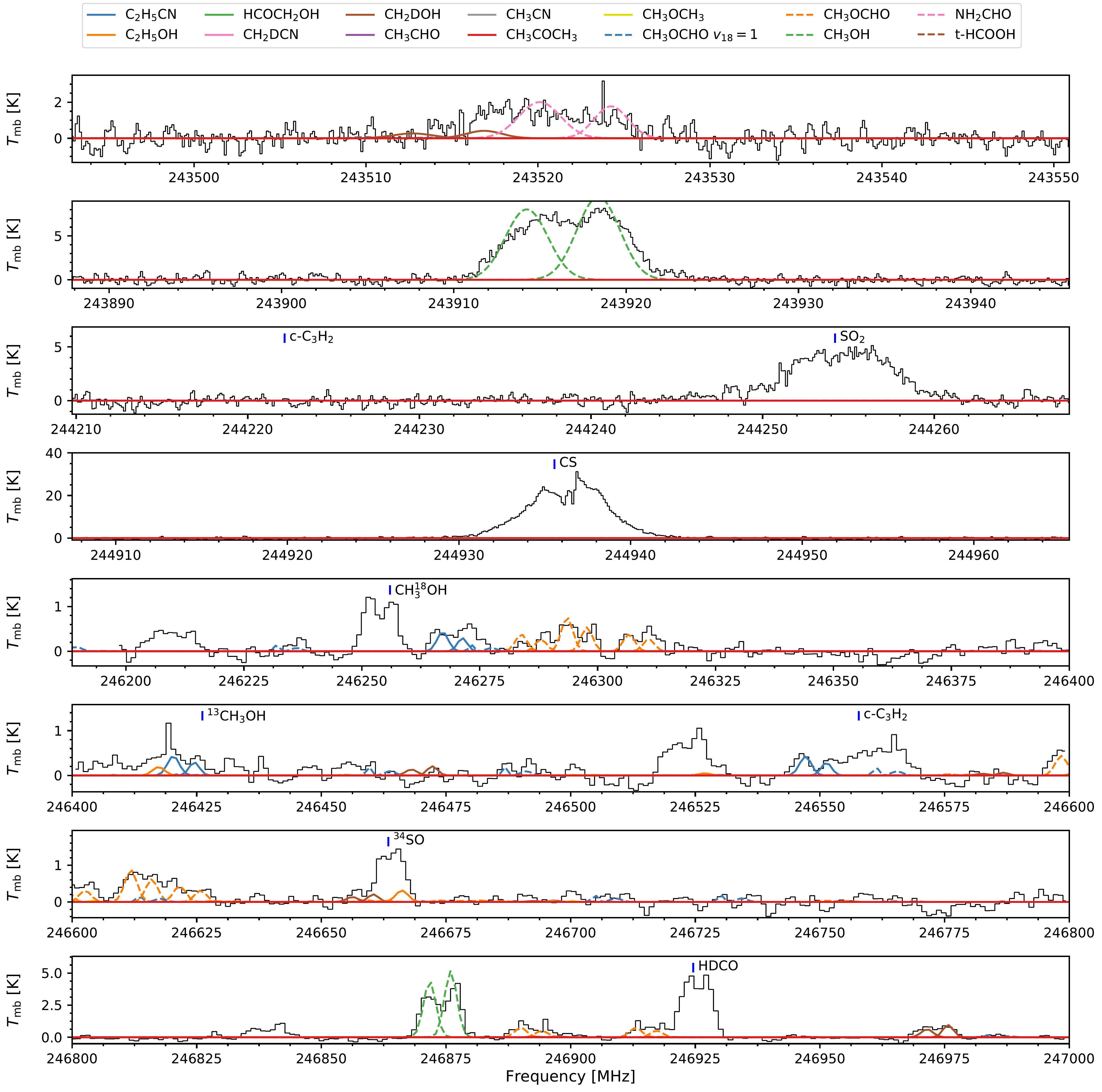}
  \caption{The spectra of Per-emb 17 along with the best-fitting model assuming $T_\text{ex} = 200$\,K.  The legends are similar to the legend in Figure\,\ref{fig:svs13a}, while each species has two velocity components at 1.96\kms\ and $-$3.15\kms.}
  \label{fig:per17}
\end{figure*}

Per-emb 17 is a close binary system with a separation of 0\farcs{278}$\pm$0\farcs{014} (83.3$\pm$4.0 au with a distance of 300 pc; \citealt{2018ApJ...867...43T}).  Our observations show clear double-peaked features for most of the emission (Figure\,\ref{fig:per17}).  The position velocity diagram of the \methanol\ has a bar-shaped morphology, which may come from an unresolved disk \citep[e.g., ][]{2017ApJ...843...27L,2020ApJ...891...61Y}, with asymmetric bright spots (Figure\,\ref{fig:per17_pv}).  However, the nature of these double-peaked features remains unclear because the spaital resolution in our observations is insufficient.  The double-peaked feature challenges our standard modeling pipeline, which assumes a simple Gaussian line profile.  Thus, we modeled the spectra of Per-emb 17 separately from the entire sample.  In \textsc{xclass}, we set up two molecular gas components using the same assumptions as described in Section\,\ref{sec:modeling}.  Then, we configured the fitting to allow both components to vary in their velocity offset from the rest frequency by $\pm$5\kms.  The two velocity components for the emission of COMs appear at 1.96\kms\ and $-$3.15\kms\ on average, while the simple organics appear at slightly smaller velocity offsets.  We reserve detailed analyses of the kinematics of Per-emb 17 for future studies.  The fitting results become less robust with the additional velocity offset parameter because the double-peaked lines are blended.  We therefore performed a two-component fit with fixed velocity offsets at 1.96\kms\ and $-$3.15\kms, which reproduces the observations (Figure\,\ref{fig:per17}).  To minimize the uncertainty due to substantial line blending, we estimated the total column density from the sum of the two components instead of reporting them individually.  Then, we applied the same procedure to derive the mean column densities and upper limits as discussed in Section\,\ref{sec:modeling}.  Line blending still prevents accurate determinations of the column densities for a few species.  In particular, the emission of \dimethylether\ at $\sim$259310\mhz\ may be underestimated or severely contaminated by the emission of \dmethylcyanide, which has another transition at 260523\mhz\ that may be contaminated by SiO emission.

\renewcommand{\thefigure}{\arabic{figure} (Cont.)}
\addtocounter{figure}{-1}
\begin{figure*}[htbp!]
  \centering
  \includegraphics[width=\textwidth]{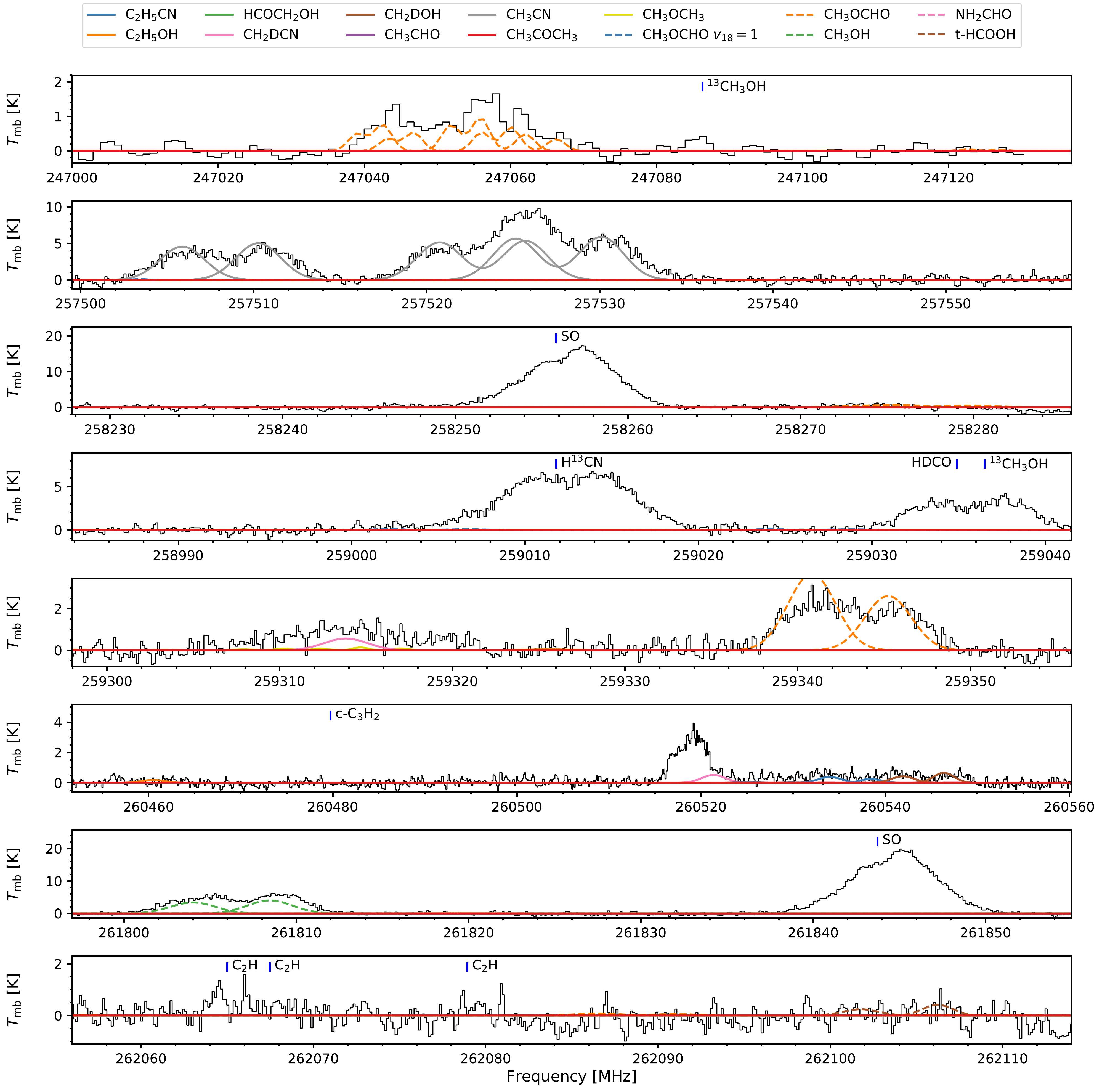}
  \caption{}
\end{figure*}
\renewcommand{\thefigure}{\arabic{figure}}

\begin{figure}[htbp!]
  \centering
  \includegraphics[width=0.48\textwidth]{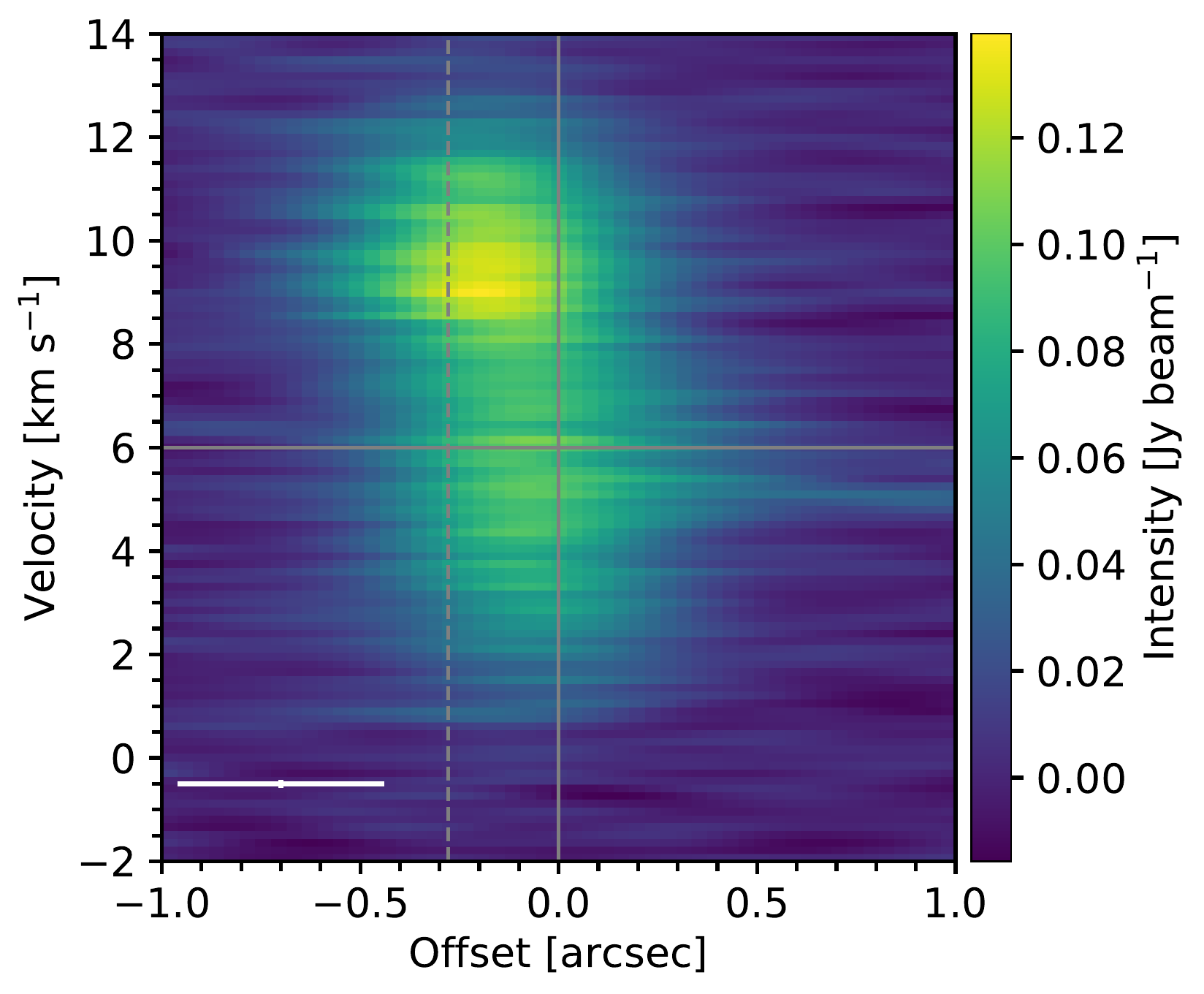}
  \caption{The position-velocity diagram of the \methanol\ emission at 243915.79\mhz\ toward Per-emb 17.  The horizontal solid lines indicate the system velocity of 6.0 \kms.  The vertical solid and dashed lines show the position of Per-emb 17 A and B, respectively.  Per-emb 17 A is the main source.  The white error bars indicate the beam size and the channel width.}
  \label{fig:per17_pv}
\end{figure}

\subsubsection{Correlations of COMs}
\label{sec:correlations}
The chemical evolution of protostars may leave certain patterns in the abundance of molecules as the dynamical evolution determines the density and temperature structures that regulate the chemical reactions.  Thus, the abundance of COMs and their correlations provides critical information for  studying the chemical evolution of embedded protostars.  As described in Section\,\ref{sec:modeling}, we fit the column density and line width with different excitation temperatures, resulting in a range of column densities as their uncertainty.  Thus, we can investigate the correlation between the column density of each species to probe the underlying chemistry.  

\begin{figure}[htbp!]
  \centering
  \includegraphics[width=0.48\textwidth]{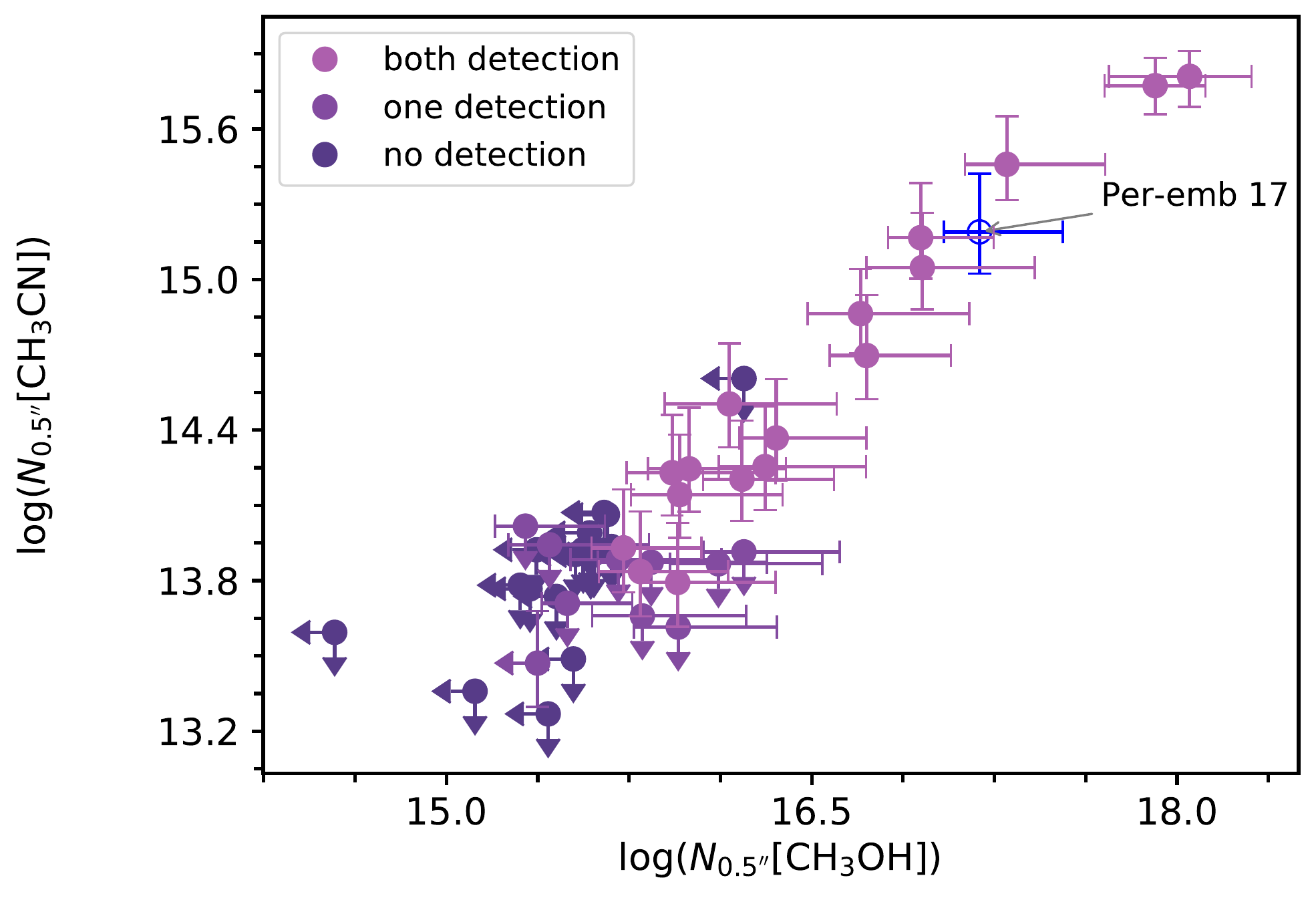}
  \caption{Correlation of the fitted column densities of \methanol\ and \methylcyanide\ from the PEACHES sample.  The data for Per-emb 17, which are estimated separately from the entire sample (Section\,\ref{sec:per17}), are shown as open circles are annotated.}
  \label{fig:ch3oh_ch3cn}
\end{figure}

\begin{figure*}[htbp!]
  \centering
  \includegraphics[width=0.8\textwidth]{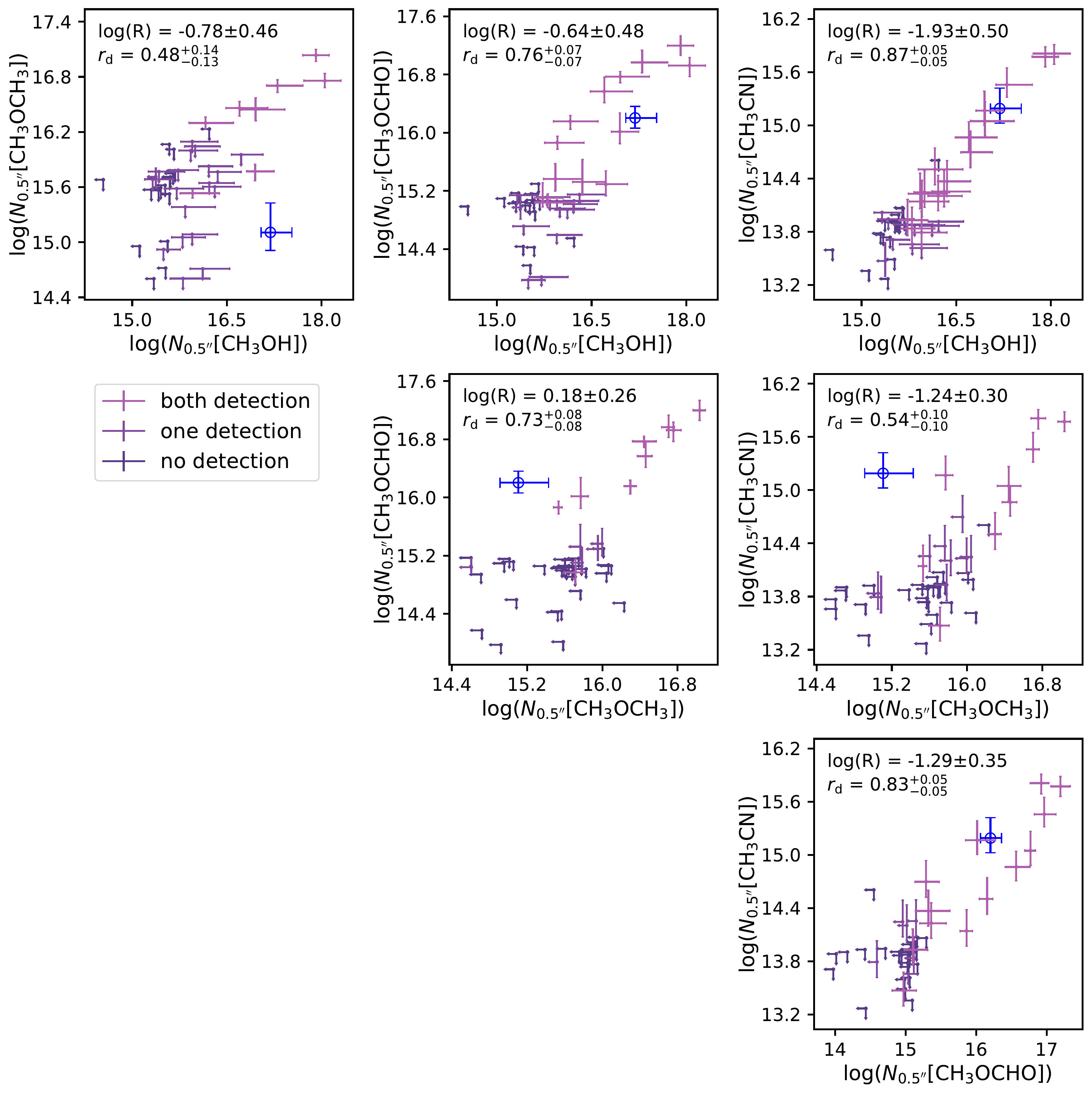}
  \caption{Corner plot of the correlations of the column densities between \methanol, \methylcyanide, \methylformate, and \dimethylether.  The color code follows that of Figure\,\ref{fig:ch3oh_ch3cn}.  The annotated texts indicate the Pearson $r$ ($r_{d}$) and the logarithmic ratio of the two molecules ($N_{y}$/$N_{x}$) for the detection-only sample.  The four most frequently detected COMs are shown in this figure, while other COMs are shown in Figure\,\ref{fig:major_minor}.}
  \label{fig:corner}
\end{figure*}

\begin{figure*}[htbp!]
  \centering
  \includegraphics[width=\textwidth]{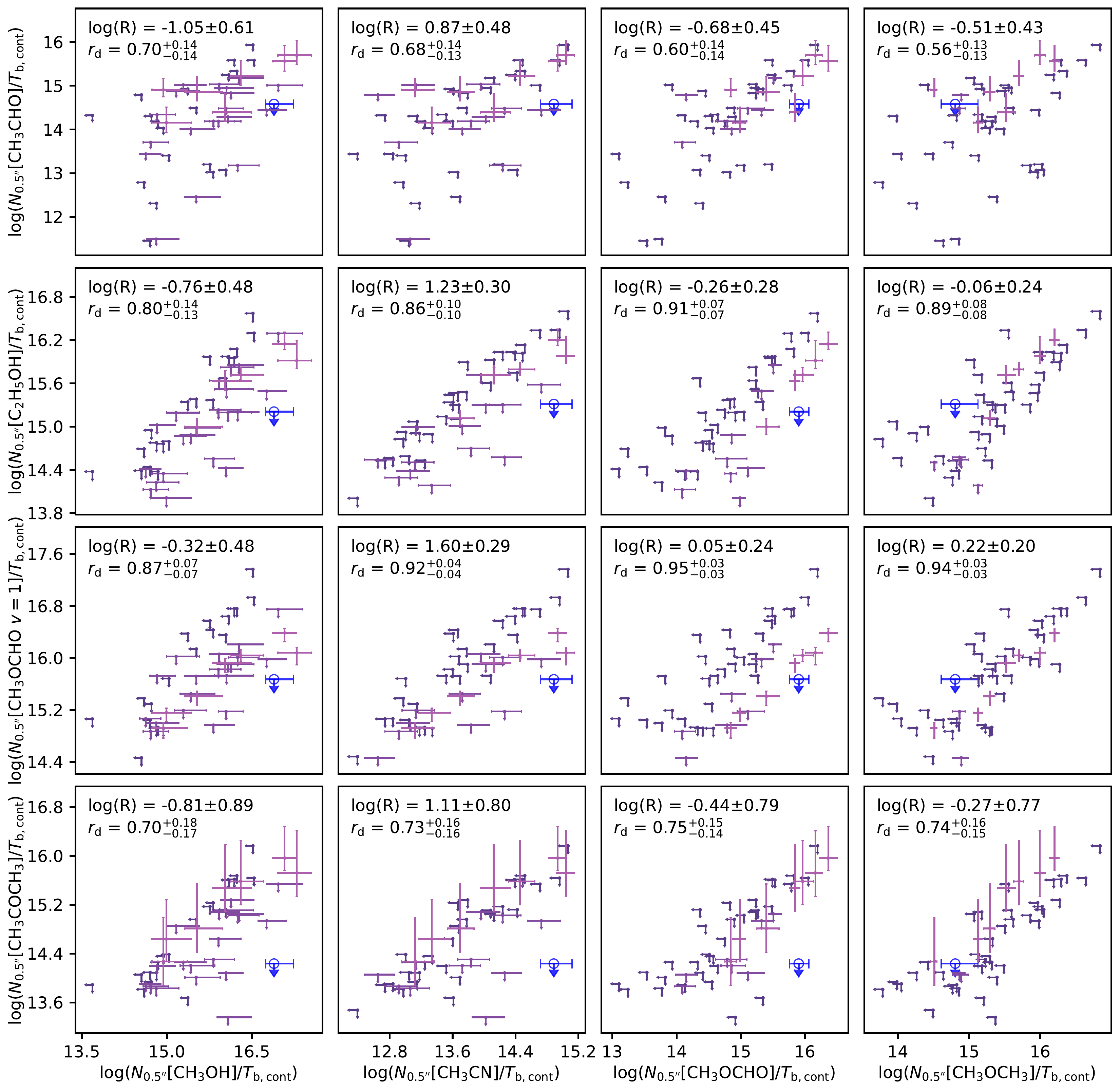}
  \caption{The correlations of the column densities normalized by the continuum brightness temperature between the more abundant COMs, \methanol, \methylcyanide, \methylformate, and \dimethylether, and the less abundant COMs \acetaldehyde, \ethanol, \methylformatev, \acetone, \ethylcyanide, \thcooh\, and \formamide.  The legends are similar to the legends in Figure\,\ref{fig:ch3oh_ch3cn}.}
  \label{fig:major_minor}
\end{figure*}

To quantify the goodness of correlation, we calculated the Pearson correlation coefficient ($r$), which tests the linearity of two variables.  We used the bootstrap method to sample the fitted column densities to calculate the Pearson $r$.  The bootstrap method is an iterative process that resamples the column densities and calculates the Pearson $r$ for the sample that is drawn for each iteration.  We took the correlation between \methanol\ and \methylcyanide\ as an example to demonstrate this bootstrap method (Figure\,\ref{fig:ch3oh_ch3cn}).  For detections, we assumed an asymmetric normal distribution centered on the best-fit column density to account for the asymmetric uncertainty in logarithmic scale.  The distribution on either side of the best-fit value has an equal probability, where the width of each ``half'' normal distribution follows the corresponding uncertainty.  We ran the bootstrap process for 1000 iterations to characterize the distribution of the Pearson $r$.  
Including the nondetections for bootstrapping the correlation coefficient requires assuming that their underlying distribution is consistent with zero.  However, assigning a distribution of the column densities of two species requires assuming that they are covariant, which determines the correlation coefficient.  Thus, we only considered detection for the calculations of the correlation coefficient.
Figure\,\ref{fig:corner} shows the correlations of several COMs that wer selected based on their detection rates as well as the ratios between species, which are discussed in Section\,\ref{sec:ratios}.  The column density of \methanol\ best correlates with that of \methylcyanide\ (see also Figure\,\ref{fig:ch3oh_ch3cn}).  \citet{2020AA...635A.198B} also found a tight correlation between these two molecules from the CALYPSO survey, which has a smaller sample, 26 protostars, in several molecular clouds.  The column densities of \dimethylether\ and \methylformate\ also show a tight correlation.  \citet{2014ApJ...791...29J} also showed a tight correlation between \dimethylether\ and \methylformate\ from different interstellar medium (ISM) sources, ranging from galactic center clouds, hot cores, protostars, cold clouds to comets.  Moreover, \methanol\ correlates with \methylformate\ and \dimethylether, which are the daughter species of \methanol, with a weaker correlation than that with \methylcyanide.  The correlations between \methylcyanide\ and \methylformate\ or \dimethylether\ also show the same behavior.  The decreasing correlation between \methanol\ or \methylcyanide\ and the daughter species of \methanol, \methylformate, and \dimethylether, is consistent with the scenario in which \methanol\ and \methylcyanide\ turns into \methylformate\ and \dimethylether\ along with other COMs as the chemistry evolves \citep[e.g., ][]{2006AA...457..927G}.  The correlations of \methanol, \methylcyanide, \methylformate, and \dimethylether\ with other COMs show positive trends; however, the correlations are driven by a few sources due to the low detection rate of other COMs (Figure\,\ref{fig:major_minor}).

To directly compare the abundance of COMs, we normalized the column densities of COMs by the \tbc, which is a proxy of the gas column density.  The normalized column densities of COMs show similar correlations as that of the column densities (Figure\,\ref{fig:corner_combined}, red markers).  We further normalized the ratio of the column densities to \tbc\ with \lbol\ and \tbol\ to test whether the correlation seen in Figure\,\ref{fig:corner_combined} (red markers) is dominated by the protostellar properties.  After the normalization with \lbol\ and \tbol\ (Figure\,\ref{fig:corner_combined}, blue and black markers), the correlation between \methanol\ and \dimethylether\ weakens significantly, while \methylcyanide\ also shows a considerable lower correlation with \dimethylether.  The correlations between \methylformate\ and \dimethylether\ remain similar with different normalizations.  Per-emb 17 is an outlier with low and uncertain abundance of \dimethylether\ (Section\,\ref{sec:per17}), reducing the correlation strength.  However, the correlation analyses excluding the abundance of Per-emb 17 show similar trends.  In summary, \methylcyanide\ has the best correlation with \methanol, followed by that of \methylformate\ and \dimethylether.

\renewcommand{\thefigure}{\arabic{figure} (Cont.)}
\addtocounter{figure}{-1}
\begin{figure*}[htbp!]
  \centering
  \includegraphics[width=\textwidth]{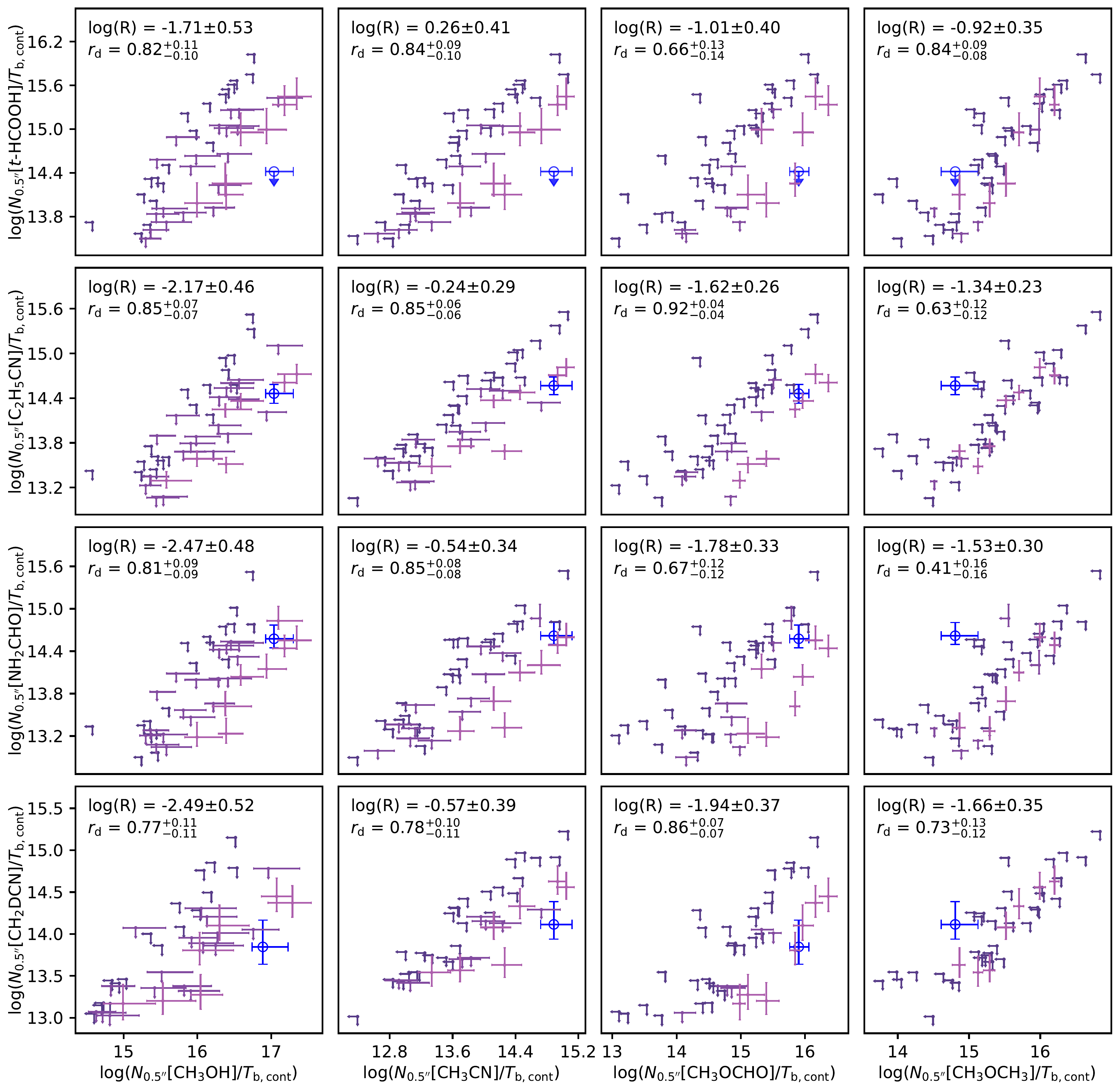}
  \caption{}
\end{figure*}
\renewcommand{\thefigure}{\arabic{figure}}

\begin{figure*}[htbp!]
  \centering
  \includegraphics[width=0.8\textwidth]{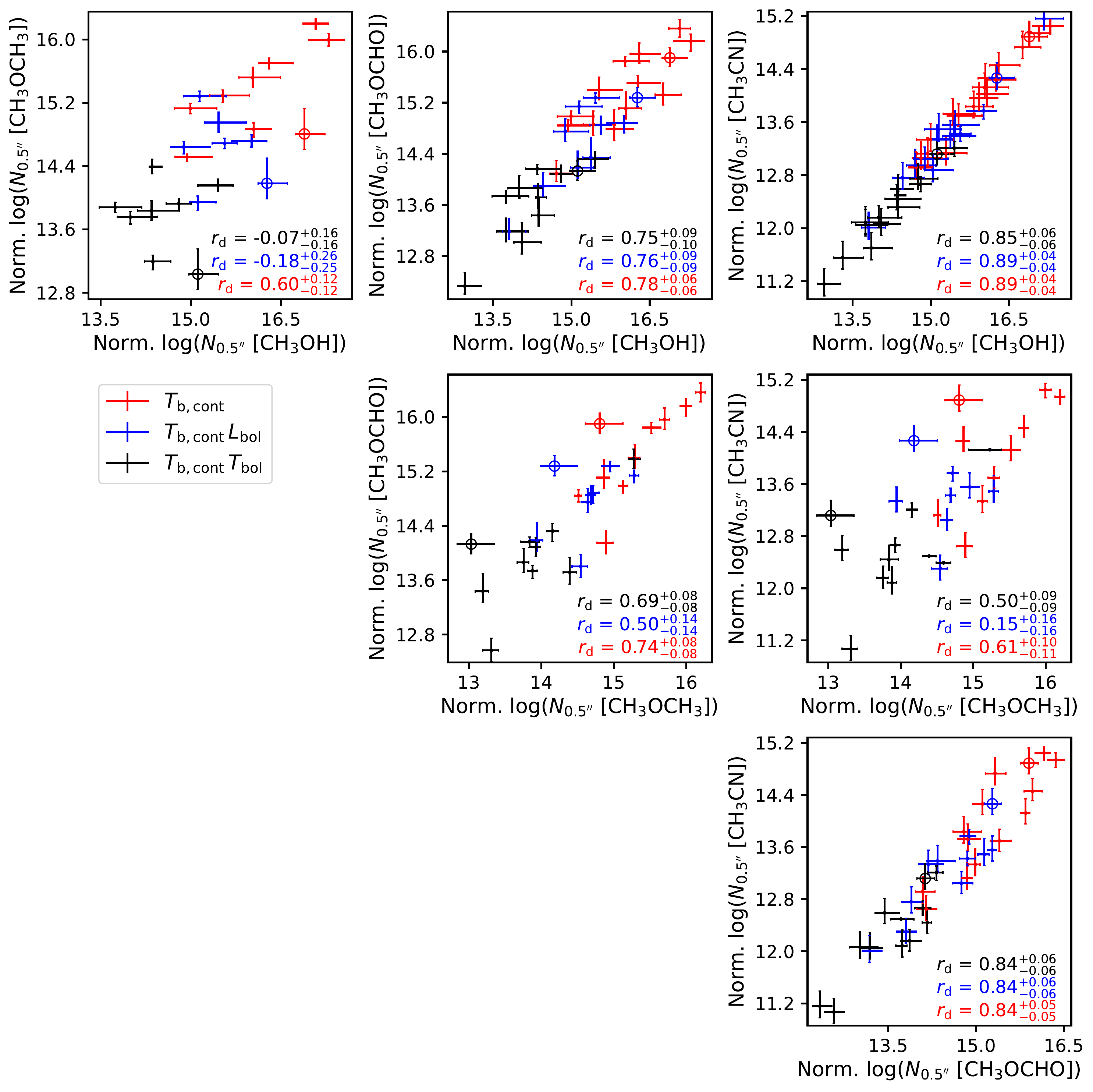}
  \caption{Corner plot of the correlations of the normalized column densities.  The red, blue, and black symbols indicate the column densities normalized by the \tbc, \tbc\lbol, and \tbc\tbol.  We only show the column densities if both molecules are detected.  The Pearson $r$ correlation coefficient for the detections with each normalization is shown in the legend with the corresponding color.  A few close multiple sources, including Per-emb 12 A \&\ B, Per-emb 35 A \&\ B, and Per-emb 11 A \&\ C, are excluded from the normalization of \tbol\ and \lbol\ because their SEDs are poorly determined.}
  \label{fig:corner_combined}
\end{figure*}

\section{Discussion}
\label{sec:discussion}
\subsection{Universal Chemistry among Hot Corinos?}
For the protostars with compact emission of COMs, the so-called hot corinos, the abundance of COMs correlates well between the species.  With the normalization of \tbc, the Pearson $r$ for the correlations between the four most frequently detected species (\methanol, \methylcyanide, \methylformate, and \dimethylether) range from 0.76 to 0.93 (Figure\,\ref{fig:corner}).  The Pearson $r$ for the same four species with all COM species has a median value of 0.86 with a range from 0.57 to 0.95 (Figure\,\ref{fig:major_minor}).  The correlations remain unchanged with the normalizations of \lbol\ and \tbol, suggesting that the correlation that represents the chemistry of COMs may be independent of the evolutionary stage of protostars, as also suggested by \citet{2019MNRAS.483.1850B}.  The source size and beam dilution have limited impact on the derived column densities (see the discussion in Appendix\,\ref{sec:beam_dilution}).  The limited number of detections of the COMs except for the four most frequently detected COMs hinders a further quantification of their correlation strengths.  While the PEACHES sample shows similar abundance ratios between COMs, the absolute column densities of COMs vary by 1--3 orders of magnitude, suggesting a diverse environment in the PEACHES survey.

\begin{figure*}[htbp!]
  \centering
  \includegraphics[width=0.8\textwidth]{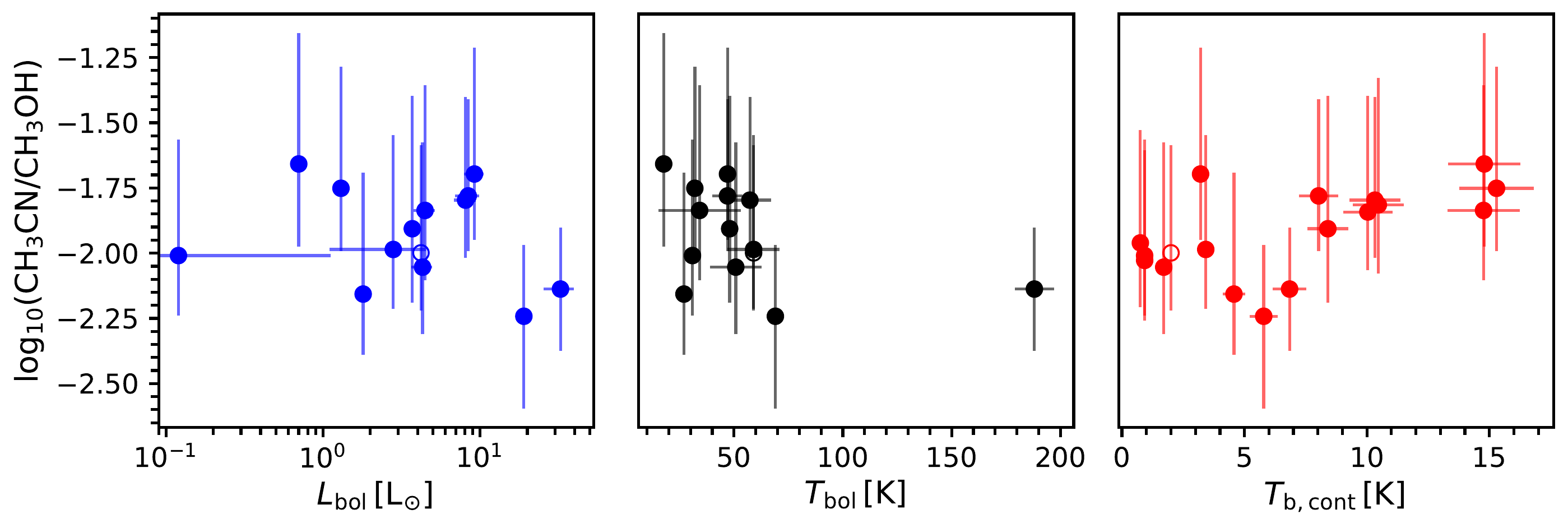}
  \includegraphics[width=0.8\textwidth]{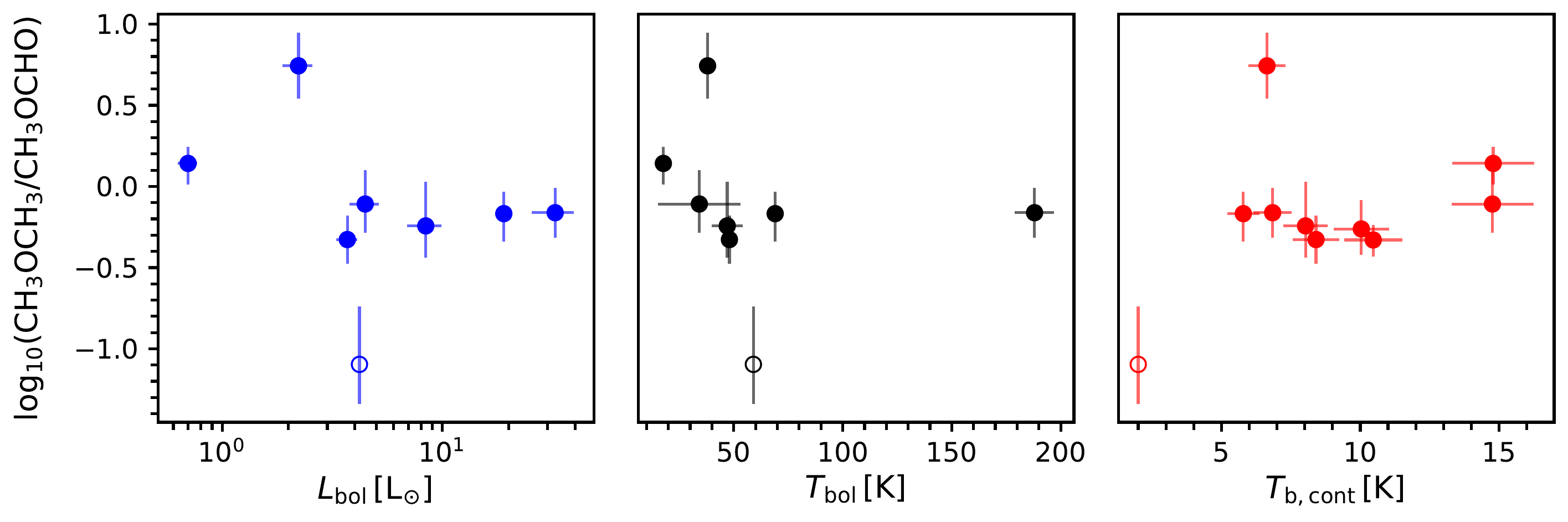}
  \includegraphics[width=0.8\textwidth]{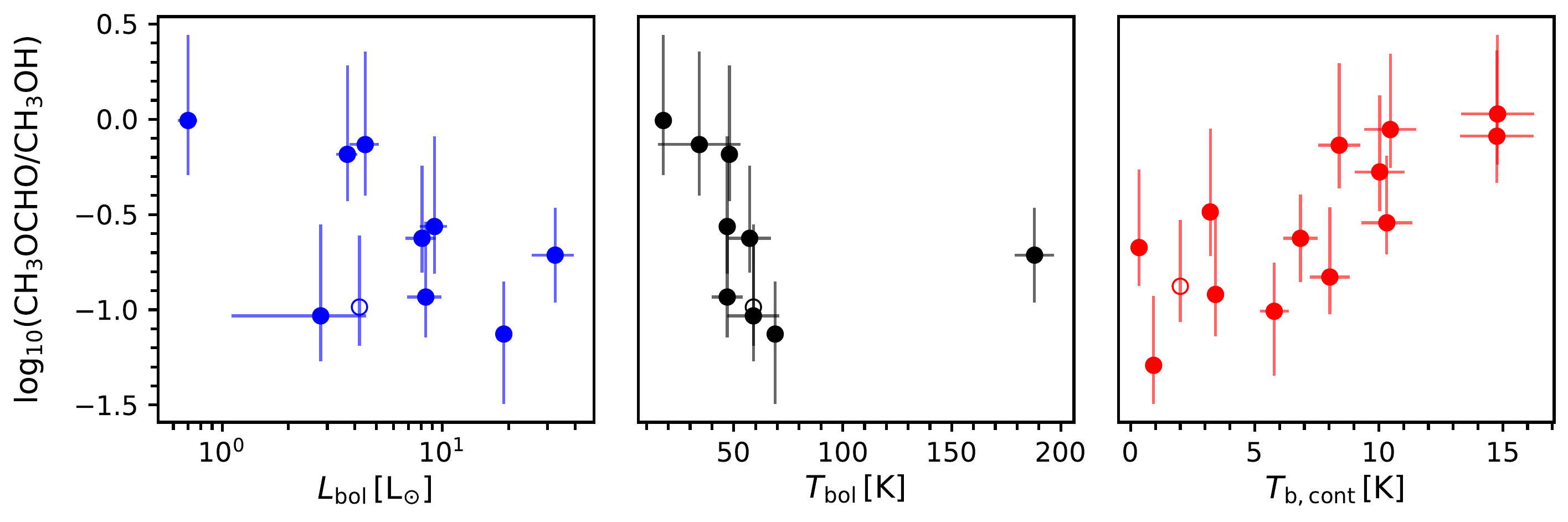}
  \includegraphics[width=0.8\textwidth]{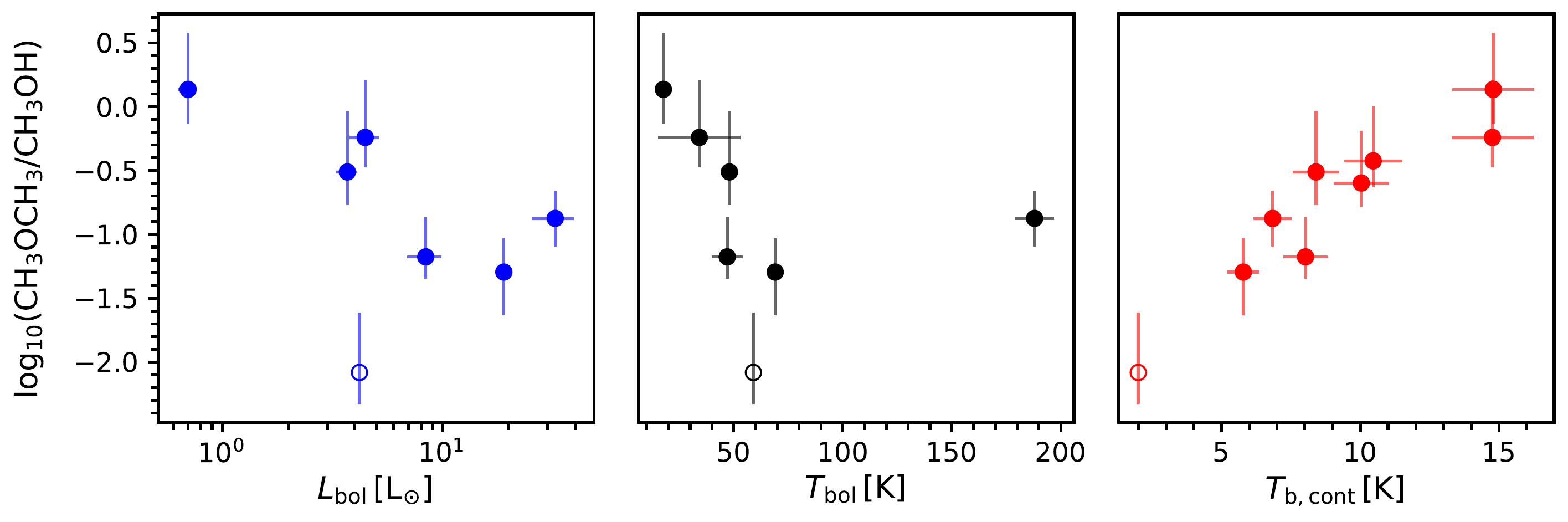}
  \caption{Ratios of well-detected molecules, \methanol, \methylcyanide, \methylformate, and \dimethylether, as functions of \lbol\ (blue), \tbol\ (black), and \tbc\ (red).  The values of Per-emb 17 are shown as open circles, which may have higher systematic uncertainty due to the double-peaked line profile (Section\,\ref{sec:per17}).  The uncertainties of \lbol\ and \tbol\ were taken from \citet{2016ApJ...818...73T} and \citet{2016AA...592A..56M}, where the properties of some sources are unconstrained and are therefore not shown.  The uncertainty on \tbc\ is 10\%.}
  \label{fig:ratios_indicators}
\end{figure*}

To understand the chemical similarity of COMs seen in embedded protostars in different environments,  we further test the effect of protostellar evolution with the abundance ratios of the four most abundant COMs (Figure\,\ref{fig:ratios_indicators}).  Per-emb 17 has a significantly lower column density of \dimethylether\ than the PEACHES sample, resulting in apparent outliers in Figure\,\ref{fig:ratios_indicators} (open circles).  The ratios of \methylcyanide/\methanol\ and \dimethylether/\methylformate, which are molecules with a similar complexity, have little variation as the functions of \lbol, \tbol, and \tbc.  In contrast, the ratios of \methylformate\ or \dimethylether\ to \methanol, which are the molecules with more complexity over the molecules with less complexity, increase with \tbc, whereas the ratios tentatively decrease with \lbol\ and \tbol.  Because of its high abundance, \methanol\ may be more optically thick at a higher gas column density (high \tbc), resulting in an underestimation of the \methanol\ column density, hence the elevated ratios.  The emission of \methylcyanide\ is likely to be optically thin given its low abundance compared to that of \methanol.  In fact, the highest fitted column density of \methylcyanide, $\sim10^{16}$ cm$^{-2}$, and the lowest modeled excitation temperature, 100 K, results in an optical depth of $\sim$0.4.  The \methanol\ becomes optically thick when the column density exceeds 2\ee{17} and 5\ee{17} cm$^{-2}$ at 100 K and 150 K, respectively.  If the temperature is 200 K or higher, \methanol\ remains optically thin when N$\leq 10^{18}$ cm$^{-2}$, while our modeling estimates the highest column density of 1.1\ee{18} cm$^{-2}$ in Per-emb 27.  If the optical depth of \methanol\ plays a major role in the trend between the \methylcyanide/\methanol\ and \dimethylether/\methylformate\ to \tbc, we would expect to see a similar trend in the \methylcyanide/\methanol\ to \tbc, which shows a flat relation instead.  Therefore the optical depth effect is unlikely to dominate the observed trend between the \methylcyanide/\methanol\ and \dimethylether/\methylformate\ to \tbc.  Chemical evolution is an alternative scenario for this trend.  Under the scheme of grain-surface chemistry, \methanol\ primarily forms from the hydrogenation of CH$_3$O or CH$_2$OH on grains, while \methylformate\ and \dimethylether\ primarily form from the radicals associated with H$_2$CO and \methanol\ destruction, which requires $T=20-40$ K so that the radicals become mobile \citep{2006AA...457..927G}.  Thus, the elevated ratios of \methylformate\ and \dimethylether\ to \methanol\ may be a results of a higher abundance of HCO and CH$_3$, a longer time of warm condition at high \tbc\ sources for more efficient formation of \methylformate, or more efficient formation of \methylformate\ and \dimethylether\ in the gas phase after the evaporation/desorption of parent species, such as \methanol\ and CH$_3$O, as suggested by \citet{2015MNRAS.449L..16B}.

\begin{figure}[htbp!]
  \centering
  \includegraphics[width=0.48\textwidth]{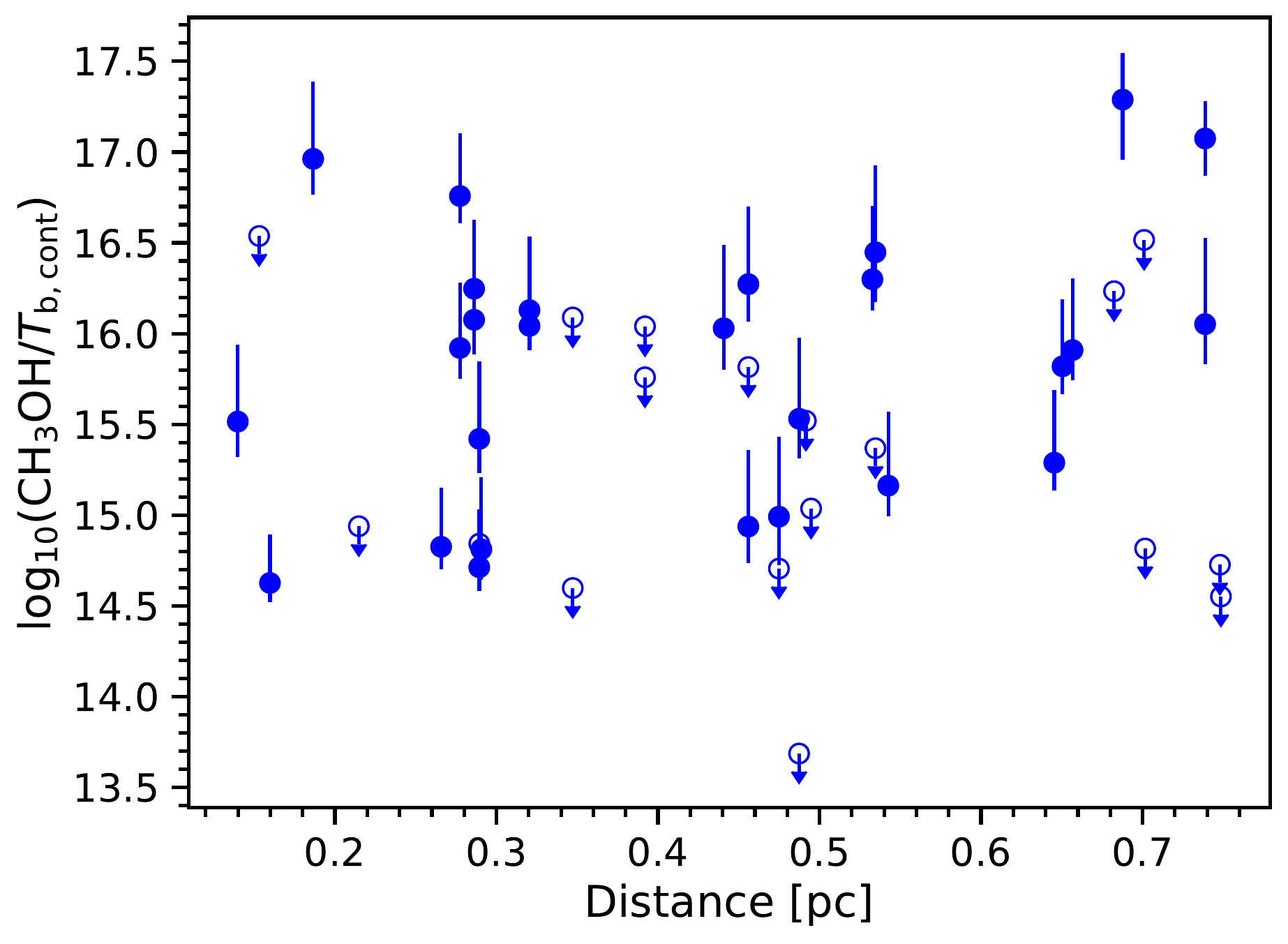}
  \caption{The column densities of \methanol\ toward the PEACHES sample (excluding Per-emb 17) normalized by the averaged continuum brightness temperature as a function of the minimum distance to the 10$\sigma$ contour in the Planck 217\ghz\ observations.  Due to the lower brightness at 217\ghz\ toward Per-emb 5 and Per-emb 25, their minimum distances are calculated with 5$\sigma$ and 3$\sigma$ contours, respectively.  The solid circles show the sources with detection of \methanol, while the opened circles show the sources without detection of \methanol.}
  \label{fig:Dmin}
\end{figure}

\citet{2018ApJS..236...52H} found a tentative trend where the sources closer to the edge of the cloud or the isolated sources show a higher ratio of \cch\ to \methanol.  Although our ALMA observations cannot fully sample the emission of \cch, we can test if a similar trend exists between the abundance of COMs and the location of the sources in the cloud.  We follow the approach described in \citet{2018ApJS..236...52H} to calculate the minimum distance ($D_\text{min}$) from the source to the edge of the cloud, which is arbitrarily defined as the 10$\sigma$ level in the Planck 217\ghz\ observations.\footnote{Based on observations obtained with Planck (\href{http://www.esa.int/Planck}{http://www.esa.int/Planck}), an ESA science mission with instruments and contributions directly funded by ESA Member States, NASA, and Canada.}  Although the \methanol\ normalized by \tbc\ has a large dynamic range (more than 2 orders of magnitude), the $D_\text{min}$ has no obvious effect on the normalized column density of \methanol\ (Figure\,\ref{fig:Dmin}).  The column density of \methanol\ is normalized with the average continuum brightness temperature so that Figure\,\ref{fig:Dmin} shows no clear threshold for the detection of \methanol\ due to the sensitivity.  Because the emission of COMs appears compact in the PEACHES survey, the non-effect of $D_\text{min}$ indicates that the $D_\text{min}$ only impacts the carbon-chain molecules but not the COMs.  However, two surveys have a factor of $\sim$40 difference in spatial resolution so that the chemical dependence with $D_\text{min}$ need to be further investigated with observations that probe carbon-chain molecules and COMs at similar spatial scale.  The nature of the abundance of \cch\ requires a comprehensive analysis of the \cch\ emission to confirm, which will be presented in future papers.

\subsection{The Abundance Ratios of O-bearing and N-bearing COMs}
\label{sec:ratios}

\begin{figure*}[htbp!]
  \centering
  \includegraphics[width=0.8\textwidth]{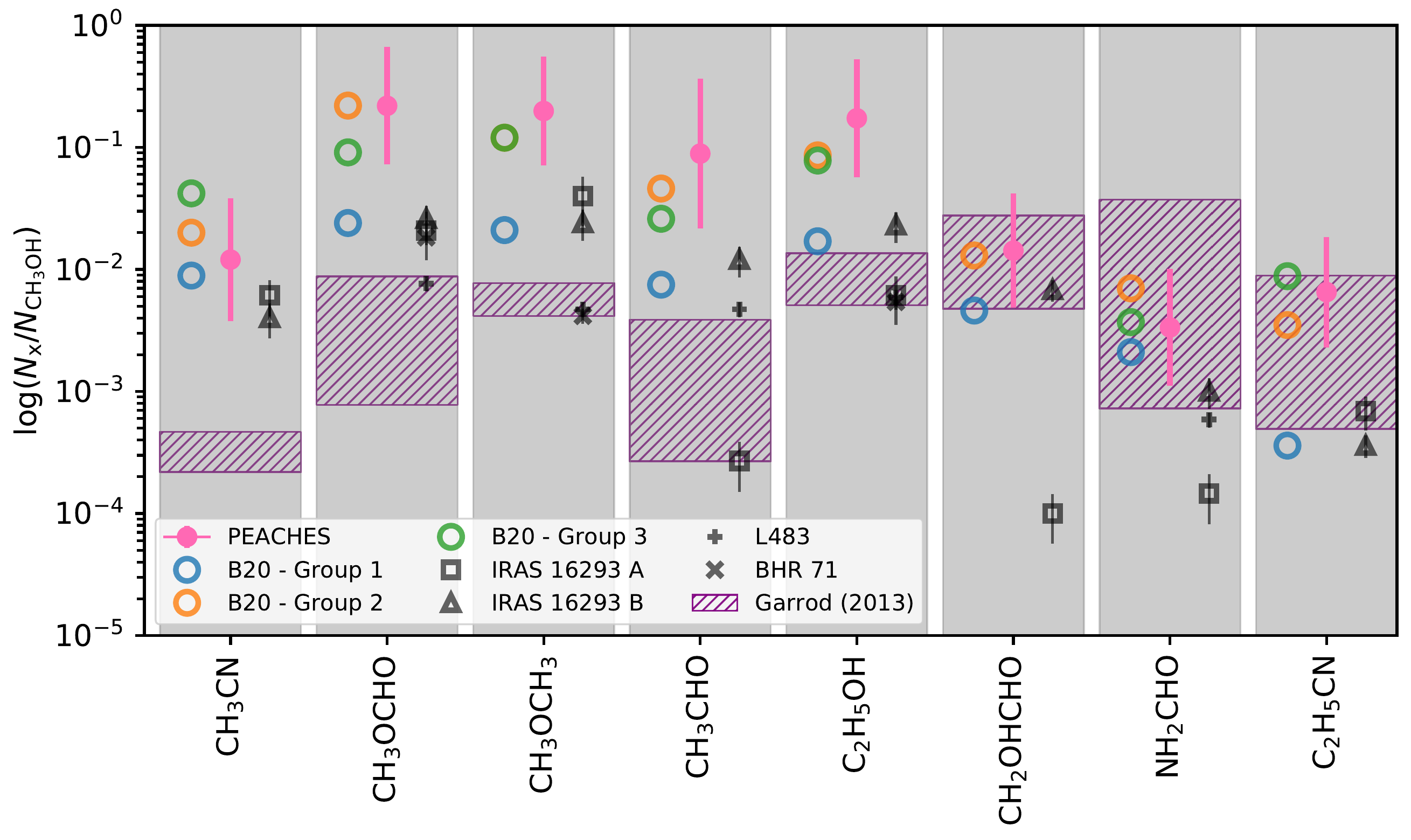}
  \caption{The ratios of the COMs commonly found in the PEACHES survey and the CALYPSO survey \citep{2020AA...635A.198B} as well as a few selected hot corinos.  The CALYPSO sources are categorized into three groups according to their COM abundance to \methanol.  Group 3 has no detection of \glycolaldehyde.  The column densities of COMs toward IRAS 16293$-$2422 B are taken from \citet{2016AA...595A.117J} for \methanol, \citet{2018AA...620A.170J} for \ethanol, \dimethylether, \methylformate, \acetaldehyde, \formamide, and \glycolaldehyde, and \citet{2018AA...616A..90C} for \methylcyanide\ and \ethylcyanide.  The column densities of COMs toward IRAS 16293$-$2422 A are taken from \citet{2020AA...635A..48M} for \methanol, \ethanol, \dimethylether, \methylformate, \acetaldehyde, \formamide, and \glycolaldehyde, \citet{2018AA...616A..90C} for \methylcyanide\ and \ethylcyanide.  The column densities of COMs toward L483 are taken from \citet{2019AA...629A..29J}.  The column densities of COMs toward BHR 71 are taken from \citet{2020ApJ...891...61Y}, assuming a $T_\text{ex}$ of 100\,K for \methanol.  The purple hatched regions indicate the range of the peak abundance ratios to \methanol\ from the warm-up models in \citet{2013ApJ...765...60G} that have three different speeds of the warm-up process.}
  \label{fig:ratios_all}
\end{figure*}

The abundance ratios of COMs reflect the chemistry of COMs.  Figure\,\ref{fig:ratios_all} compares the abundance ratios of COMs toward the PEACHES sample to the ratios from the CALYPSO survey \citep{2020AA...635A.198B}, the observations of individual protostars, and the model predictions from \citet{2013ApJ...765...60G}.  \citet{2020AA...635A.198B} divided up the CALYPSO sample into three groups, where Group 1 has a low abundance of O-bearing COMs to \methanol, Group 2 has a higher abundance ratio of O-bearing COMs to \methanol\ than Group 1, and Group 3 is similar to Group 2, but has a higher abundance of \methylcyanide\ and \ethylcyanide\ relative to \methanol\ and a lower abundance of O-bearing COMs.  The abundance ratios of COMs toward the PEACHES sample generally agree with the ratios from the CALYPSO survey, which is expected becasue the sample overlaps significantly.  Three out of four sources in Group 2 are Perseus protostars.  Compared to one of the archetype hot corinos, IRAS 16293$-$2422 B, the ratios of COMs in the PEACHES sample are 2--18 times higher \citep{2016AA...595A.117J,2018AA...620A.170J,2018AA...616A..90C}.  However, our sample has only a few sources with a comparable column density of \methanol, $\gtrsim 10^{18}$\,cm$^{-2}$, as that in IRAS 16293$-$2422, making the disagreement less robust.  Moreover, our observations may underestimate the column densities of \methanol\ due to the dust optical depth.  Constraining the true difference of the ratios requires future analyses of the effect of the dust opacity as demonstrated by \citet{2020ApJ...896L...3D}.

We also compare the ratios of COMs to \methanol\ with the numerical results of chemo-physical models.  \citet{2013ApJ...765...60G} modeled three different timescales for the warm-up phase in their model, where the formation of COMs becomes efficient \citep{2008ApJ...682..283G}.  The ratios of COMs toward the PEACHES sample differ from the ranges of the peak abundance ratios in the warm-up models by \citet{2013ApJ...765...60G}.  The models underestimate the abundance ratios for most of the O-bearing COMs, while the ratios of \glycolaldehyde\ and N-bearing COMs agree with the ratios derived from the PEACHES sample (Figure\,\ref{fig:ratios_all}).  
Because we may underestimate the column density of \methanol\ due to the optical depth, the abundance ratio to \methanol\ could be overestimated.  If the ratios derived from the PEACHES sample were lower by a factor of 30, they would be consistent with the ratios in \citet{2013ApJ...765...60G} for the five COMs in the left, but the model would then overestimate the abundance ratios of \glycolaldehyde, \formamide, and \ethylcyanide.  This disagreement may point to a dominant gas-phase formation of (some) COMs other than \methanol, as suggested by other studies \citep[e.g., ][]{2015MNRAS.449L..16B,2018ApJ...854..135S,2020MNRAS.499.5547V}.

The well-correlated abundance of N-bearing COMs and O-bearing COMs (Figure\,\ref{fig:corner} and \ref{fig:major_minor}) indicates no ``N-/O-bearing COMs differentiation,'' which has long been discussed for Orion-KL, a massive star-forming region \citep[e.g., ][]{2008ApJ...672..962F,2008Ap&SS.313...45G,2015AA...581A..71F,2018AA...620L...6T}.  The N-bearing COMs and O-bearing COMs segregate toward the hot core and the compact ridge of Orion-KL, respectively, which has been interpreted as a result of different temperatures at the prestellar phase or different chemical evolutionary stage \citep{1993ApJ...408..548C,2011ApJ...728...71L,2011JPCA..115.6472N,2013ApJ...765...60G}.  If the Perseus protostars have different initial conditions at the prestellar stage or are at different chemical stages that would result in a similar differentiation in Orion-KL, we would expect to detect a variety of N-bearing COMs to O-bearing COMs ratios.  However, the well-correlated abundance of N-bearing COMs and O-bearing COMs suggests that the Perseus protostars may form from a similar initial condition or be in a similar chemical stage.  Further studies of region-to-region comparisons await. 

\subsection{Origin of the \methanol/\methylcyanide\ Correlation}
The tight correlation between \methanol\ and \methylcyanide\ is a striking finding of the PEACHES survey, while studies such as \citet{2017ApJ...841..120B} and \citet{2020AA...635A.198B} have shown a similar trend with fewer detections or larger scatter.  \citet{2020AA...635A.198B} argued that this correlation between \methanol\ and \methylcyanide\ may not be due to chemistry because of the somewhat unrelated formation pathways of two molecules.  If the gas-phase chemistry is negligible for the production of \methanol\ and \methylcyanide, the tight correlation between \methanol\ and \methylcyanide\ suggests a similar abundance ratio of two molecules on the icy grains prior to being thermally desorbed, hinting at a common chemistry of COMs in the warm protostellar envelope.  The tight correlation may also reflect a uniform elemental abundance of O and N in the Perseus molecular cloud before the star formation.

The compact emission of \methanol\ has a well-known icy origin, where \methanol\ forms via subsequent CO hydrogenation on the surface of dust grains \citep{1997IAUS..178...45T,2003ApJ...588L.121W,2014AA...572A..70R}.  At a cold temperature, the nonthermal desorption and/or the excessive energy from the formation of \methanol\ that kicks it off from dust grains are thought to be responsible for the presence of gaseous \methanol\ \citep[e.g., ][; but see the discussion in \citet{2020ApJ...897...56P} on the chemical desorption probability]{2007AA...467.1103G,2013ApJ...769...34V,2015ApJ...802...74S}.  \methylcyanide\ can form in the gas phase via radiative association (HCN $+$ CH$_3^+$ $\rightarrow$ CH$_3$CNH$^+$ $+$ $h\nu$) followed by the dissociative recombination of CH$_3$CNH$^+$ or on the grain surface.  The grain-surface reactions to form \methylcyanide\ can occur at cold temperature by the hydrogenation of CCN, which can form via C$+$CN or C$_2+$N.  At a warm temperature, \methylcyanide\ can form directly from reactions between the radicals, CH$_3$ and CN.  For embedded protostars, the grain-surface reactions are thought to dominate the production of \methylcyanide; however, it remains unclear whether the cold-phase or the warm-phase grain-surface pathway regulates the abundance of \methylcyanide.  

Assuming an icy grain origin for both \methanol\ and \methylcyanide, we can further test the efficiency of the cold- and warm-phase pathways by comparing the ratio of \methylcyanide/\methanol\ toward the prestellar and protostellar sources.  In prestellar cores, [\methylcyanide/\methanol] are 0.28, 0.0192$\pm$0.0075, and 0.021$\pm$0.008 for TMC\,1-CP (cyanoployyne peak; \citealt{2004PASJ...56...69K,2016ApJS..225...25G}), L1521 E \citep{2019AA...630A.136N}, and L1544 \citep{2019AA...630A.136N}, respectively.  The ratio of TMC1-CP is substantially higher than that in our survey, 0.012$^{+0.009}_{-0.005}$, while the mean ratios of L1521 E and L1544 seem higher than the ratio in our survey within 1$\sigma$ uncertainty.  Furthermore, we can divide the PEACHES sample into Class 0 and I according to their \tbol\ (Class 0: \tbol\ $<$ 70 K; Class I: \tbol\ $\geq$70 K; \citealt{2014prpl.conf..195D}).  Only one Class I source has emission of both \methanol\ and \methylcyanide, along with another borderline Class I source, which has a \tbol\ of 69 K.  The mean value of \methylcyanide/\methanol\ for Class I sources ($N=2$) is $6.5^{+6.0}_{-3.0}$\ee{-3}, while the mean value for Class 0 sources ($N=12$) is $1.4^{+1.1}_{-1.1}$\ee{-2}.  Only the sources with reliable measurements of \tbol\ were included.  Although the difference is within the uncertainty, the tentative decrease of \methylcyanide/\methanol\ from Class 0 to Class I phase would be consistent with the tentative decrease of \methylcyanide/\methanol\ from prestellar to protostellar phase.  This difference provides constraints on the interplay between the production of \methylcyanide\ in the warm and cold phase.  Testing the chemical pathways of \methylcyanide\ requires more measurements of \methylcyanide\ and \methanol\ in Class I and prestellar sources.

Alternatively, the desorption process may regulate the number of COMs in the gas phase.  Studies find a similar binding energy for \methanol\ and \methylcyanide\ \citep{2004MNRAS.354.1133C,2013AA...550A..36M,2015ApJS..217...20W,2017MolAs...6...22W,2017ApJ...844...71P}.  Recently, \citet{2020ApJ...904...11F} showed a similar range of binding energies for \methanol\ (3770--8618 K) and \methylcyanide\ (4745--7652 K) simulated with amorphous solid water.  Thus, both molecules may simply desorb from grains at a similar rate if the ice mantles have not completely desorbed.  However, the microphysics during the desorption involve the structure of the water ice and how the COMs reside in the ice matrix, requiring further understanding of how the chemical evolution takes place on the grain surface.  Furthermore, the distribution of the binding energy can have a crucial effect on the estimate of the desorbed species \citep{2020ApJ...904...11F,2020AA...643A.155G}.

\section{Conclusions}
\label{sec:conclusions}
This work presents the PEACHES survey, an unbiased chemistry survey of COMs toward 50 embedded protostars at the Perseus molecular cloud using ALMA Band 6 observations.  The main conclusions are listed below.

\begin{enumerate}
  \item Perseus Class 0/I protostars commonly (58\%) show the warm COM emission.  \methanol\ is the most frequently detected species of COMs, while our observations show the emission of 12 O-bearing COMs and 4 N-bearing COMs, including isotopologs.
  \item Complex organic molecules other than \methanol\ are detected for the first time toward Per-emb 35 A and Per-emb 11 A.
  \item The \methanol\ and \methylcyanide\ normalized column density ($N$/\tbc) has a large diversity (more than two orders of magnitude), suggesting the existence of chemical diversity in the sources in the Perseus molecular cloud.  Moreover, the detectability of these COMs is not directly related to the protostellar properties such as \tbol, \lbol, and the gas mass column density.
  \item The column density of \methanol\ is tightly correlated with that of \methylcyanide\ with a Pearson $r$ of 0.87, where two molecules may not have a common chemical origin.  The abundance of the most frequently detected COMs, \methanol, \methylcyanide, \methylformate, and \dimethylether, correlate with other species of COMs, which are detected in fewer sources in the PEACHES sample, hinting at a common chemistry of COMs in the Perseus embedded protostars.  Protostellar properties such as \lbol, \tbol, and the minimum distance to the cloud edge, have little impact on the number of detected COMs and the abundance of COMs.
  \item \methylformate\ and \dimethylether\ are the most abundant COMs other than \methanol, with abundance ratios to \methanol\ of 0.22$^{+0.17}_{-0.10}$ and 0.21$^{+0.15}_{-0.09}$, respectively.  The abundance of COMs to \methanol\ in the PEACHES sample is similar to that in the CALYPSO survey \citep{2020AA...635A.198B}, which includes 14 Perseus sources in 26 samples and 6 of 14 sources exhibit emission of COMs.  The abundance ratios of \formamide\ and \ethylcyanide\ agree with the model predictions by \citet{2013ApJ...765...60G}, which underestimate the abundance of the O-bearing COMs.  However, the abundance ratios of O-bearing COMs and \methylcyanide\ (e.g., \methylcyanide\ to \methylformate\ and \methylformate\ to \dimethylether) seem to be consistent with the abundance ratios predicted in \citet{2013ApJ...765...60G} (Figure\,\ref{fig:ratios_all}).  alternatively, the disagreement may indicate that at least some COMs are the products of gas-phase chemistry \citep{2015MNRAS.449L..16B,2018ApJ...854..135S,2020MNRAS.499.5547V}.
  \item  The \methylcyanide/\methanol\ ratio of Perseus sources seems lower than the same ratio for other prestellar cores, especially L1544.  For the two Class I sources with detections of both molecules, the ratio is $6.5^{+6.0}_{-3.0}$\ee{-3} compared to $1.4^{+1.1}_{-1.1}$\ee{-2} for the Class 0 sources.  This may suggest an evolution of the contributions from the gas-phase chemistry and grain-surface chemistry from the prestellar to protostellar phase.
  \item The ratios of more complex COMs, such as \methylformate\ and \dimethylether, to the less complex COMs, such as \methanol\ and \methylcyanide, increase with the averaged continuum brightness temperature, a proxy of the gas column density, suggesting an enhanced production of more complex COMs toward the sources with more centrally concentrated mass.
\end{enumerate}

\begin{deluxetable*}{ccccccc}
    \tabletypesize{\scriptsize}
    \tablecaption{Column Densities of Molecules \label{tbl:Ncol}}
    \tablewidth{\textwidth}
    \tablehead{ \colhead{Source} & \colhead{CH$_{3}$OH} & \colhead{CH$_{3}$CN} & \colhead{CH$_{3}$OCHO} & \colhead{CH$_{3}$OCH$_{3}$} & \colhead{CH$_{3}$CHO} & \colhead{C$_{2}$H$_{5}$OH}}
    \startdata
    L1448 NW       & 2.1$^{+0.7}_{-0.1}$$\times 10^{15}$ & $<$1.0$\times 10^{14}$ & $<$1.4$\times 10^{15}$ & $<$4.8$\times 10^{15}$ & $<$6.9$\times 10^{14}$ & $<$3.3$\times 10^{15}$ \\
    Per-emb 33 A   & 5.3$^{+1.9}_{-2.8}$$\times 10^{15}$ & 8.5$^{+4.3}_{-3.5}$$\times 10^{13}$ & 1.3$^{+0.4}_{-0.5}$$\times 10^{15}$ & $<$6.1$\times 10^{15}$ & $<$5.2$\times 10^{14}$ & $<$1.4$\times 10^{15}$ \\
    Per-emb 33 B/C & $<$3.9$\times 10^{15}$ & $<$9.8$\times 10^{13}$ & $<$1.2$\times 10^{15}$ & $<$1.2$\times 10^{16}$ & $<$8.5$\times 10^{14}$ & $<$1.3$\times 10^{15}$ \\
    L1448 IRS 3A   & 8.4$^{+4.5}_{-5.3}$$\times 10^{15}$ & 1.7$^{+0.8}_{-0.7}$$\times 10^{14}$ & 2.3$^{+0.1}_{-0.9}$$\times 10^{15}$ & $<$9.9$\times 10^{15}$ & $<$3.3$\times 10^{14}$ & $<$2.5$\times 10^{15}$ \\
    Per-emb 26     & 8.9$^{+3.2}_{-4.4}$$\times 10^{16}$ & 1.5$^{+0.7}_{-0.6}$$\times 10^{15}$ & 1.0$^{+0.5}_{-0.5}$$\times 10^{16}$ & 5.9$^{+0.2}_{-1.0}$$\times 10^{15}$ & $<$2.4$\times 10^{15}$ & $<$2.1$\times 10^{15}$ \\
    Per-emb 42     & 9.0$^{+4.6}_{-5.4}$$\times 10^{15}$ & $<$4.1$\times 10^{13}$ & $<$1.1$\times 10^{15}$ & $<$1.2$\times 10^{16}$ & $<$4.5$\times 10^{14}$ & $<$4.4$\times 10^{15}$ \\
    Per-emb 22 A   & 2.0$^{+1.1}_{-1.2}$$\times 10^{16}$ & 1.8$^{+0.9}_{-0.8}$$\times 10^{14}$ & $<$1.4$\times 10^{15}$ & $<$4.0$\times 10^{15}$ & $<$3.3$\times 10^{14}$ & $<$2.7$\times 10^{15}$ \\
    Per-emb 22 B   & 1.6$^{+0.7}_{-1.0}$$\times 10^{16}$ & 1.6$^{+0.7}_{-0.7}$$\times 10^{14}$ & $<$1.0$\times 10^{15}$ & $<$6.7$\times 10^{15}$ & $<$1.4$\times 10^{13}$ & $<$1.5$\times 10^{15}$ \\
    Per-emb 25     & $<$4.6$\times 10^{15}$ & $<$1.2$\times 10^{14}$ & $<$2.0$\times 10^{15}$ & $<$1.0$\times 10^{16}$ & $<$5.6$\times 10^{14}$ & $<$3.0$\times 10^{15}$ \\
    Per-emb 20     & 1.3$^{+0.8}_{-0.8}$$\times 10^{16}$ & $<$7.4$\times 10^{13}$ & $<$8.7$\times 10^{14}$ & $<$5.1$\times 10^{14}$ & $<$1.5$\times 10^{14}$ & $<$2.8$\times 10^{15}$ \\
    L1455 IRS 2    & $<$4.4$\times 10^{15}$ & $<$1.2$\times 10^{14}$ & $<$1.5$\times 10^{15}$ & $<$5.6$\times 10^{15}$ & $<$4.9$\times 10^{14}$ & $<$2.6$\times 10^{15}$ \\
    Per-emb 44     & 8.1$^{+5.0}_{-3.1}$$\times 10^{17}$ & 5.9$^{+1.8}_{-1.3}$$\times 10^{15}$ & 1.6$^{+0.6}_{-0.4}$$\times 10^{17}$ & 1.1$^{+0.2}_{-0.1}$$\times 10^{17}$ & 2.5$^{+1.7}_{-1.4}$$\times 10^{16}$ & 9.6$^{+1.8}_{-3.1}$$\times 10^{16}$ \\
    SVS 13A2       & 6.9$^{+4.5}_{-4.6}$$\times 10^{15}$ & $<$7.5$\times 10^{13}$ & $<$1.1$\times 10^{15}$ & $<$2.4$\times 10^{15}$ & $<$5.5$\times 10^{14}$ & $<$2.0$\times 10^{15}$ \\
    Per-emb 12 A   & 7.3$^{+6.4}_{-4.9}$$\times 10^{15}$ & $<$5.4$\times 10^{13}$ & 1.3$^{+0.1}_{-0.7}$$\times 10^{14}$ & $<$6.8$\times 10^{15}$ & $<$6.0$\times 10^{14}$ & $<$1.4$\times 10^{15}$ \\
    Per-emb 12 B   & 2.0$^{+1.0}_{-0.1}$$\times 10^{17}$ & 2.9$^{+1.1}_{-1.0}$$\times 10^{15}$ & 9.2$^{+3.6}_{-3.0}$$\times 10^{16}$ & 5.1$^{+0.9}_{-0.7}$$\times 10^{16}$ & 1.7$^{+0.1}_{-0.9}$$\times 10^{16}$ & 5.2$^{+1.2}_{-1.1}$$\times 10^{16}$ \\
    Per-emb 13     & 5.0$^{+3.3}_{-3.2}$$\times 10^{16}$ & 7.3$^{+3.2}_{-2.5}$$\times 10^{14}$ & 3.7$^{+1.6}_{-1.4}$$\times 10^{16}$ & 2.9$^{+0.6}_{-0.4}$$\times 10^{16}$ & 1.1$^{+0.6}_{-0.6}$$\times 10^{16}$ & 1.5$^{+0.4}_{-0.3}$$\times 10^{16}$ \\
    IRAS4B'        & $<$3.5$\times 10^{14}$ & $<$3.9$\times 10^{13}$ & $<$9.6$\times 10^{14}$ & $<$4.8$\times 10^{15}$ & $<$1.5$\times 10^{15}$ & $<$1.7$\times 10^{15}$ \\
    Per-emb 27     & 1.1$^{+0.1}_{-0.5}$$\times 10^{18}$ & 6.5$^{+2.1}_{-1.3}$$\times 10^{15}$ & 8.4$^{+3.6}_{-1.9}$$\times 10^{16}$ & 5.7$^{+0.1}_{-1.0}$$\times 10^{16}$ & 2.9$^{+1.8}_{-1.6}$$\times 10^{16}$ & 4.8$^{+1.3}_{-2.3}$$\times 10^{16}$ \\
    Per-emb 54     & $<$2.3$\times 10^{15}$ & $<$8.4$\times 10^{13}$ & $<$1.1$\times 10^{15}$ & $<$4.9$\times 10^{15}$ & $<$6.1$\times 10^{14}$ & $<$2.7$\times 10^{15}$ \\
    Per-emb 21     & 1.7$^{+0.8}_{-1.0}$$\times 10^{16}$ & $<$8.2$\times 10^{13}$ & $<$1.2$\times 10^{15}$ & $<$4.4$\times 10^{15}$ & $<$3.2$\times 10^{14}$ & $<$3.5$\times 10^{15}$ \\
    Per-emb 14     & $<$3.3$\times 10^{15}$ & $<$3.1$\times 10^{13}$ & $<$9.9$\times 10^{14}$ & $<$4.2$\times 10^{15}$ & $<$7.8$\times 10^{13}$ & $<$1.9$\times 10^{15}$ \\
    Per-emb 35 A   & 5.3$^{+2.2}_{-2.9}$$\times 10^{16}$ & 5.0$^{+2.5}_{-2.1}$$\times 10^{14}$ & 2.0$^{+0.9}_{-0.7}$$\times 10^{15}$ & $<$9.0$\times 10^{15}$ & $<$2.6$\times 10^{14}$ & $<$2.9$\times 10^{15}$ \\
    Per-emb 35 B   & 6.3$^{+3.0}_{-3.5}$$\times 10^{15}$ & 6.9$^{+3.5}_{-2.9}$$\times 10^{13}$ & $<$1.3$\times 10^{15}$ & $<$1.1$\times 10^{15}$ & $<$8.1$\times 10^{14}$ & $<$3.3$\times 10^{15}$ \\
    SVS 13B        & $<$2.4$\times 10^{15}$ & 3.0$^{+1.5}_{-1.1}$$\times 10^{13}$ & 9.4$^{+4.2}_{-3.2}$$\times 10^{14}$ & 5.2$^{+1.7}_{-1.0}$$\times 10^{15}$ & $<$4.1$\times 10^{15}$ & $<$1.6$\times 10^{15}$ \\
    Per-emb 15     & $<$2.8$\times 10^{15}$ & $<$5.5$\times 10^{13}$ & $<$1.1$\times 10^{15}$ & $<$3.9$\times 10^{15}$ & $<$3.6$\times 10^{14}$ & $<$2.0$\times 10^{15}$ \\
    Per-emb 50     & $<$2.2$\times 10^{15}$ & $<$5.9$\times 10^{13}$ & $<$1.5$\times 10^{15}$ & $<$4.0$\times 10^{14}$ & $<$8.4$\times 10^{14}$ & $<$3.7$\times 10^{15}$ \\
    Per-emb 18     & 2.3$^{+1.0}_{-0.1}$$\times 10^{16}$ & 2.3$^{+0.1}_{-1.0}$$\times 10^{14}$ & 2.1$^{+1.1}_{-1.1}$$\times 10^{15}$ & $<$5.8$\times 10^{15}$ & $<$5.3$\times 10^{14}$ & $<$1.2$\times 10^{15}$ \\
    Per-emb 37     & $<$1.3$\times 10^{15}$ & $<$2.3$\times 10^{13}$ & $<$1.2$\times 10^{15}$ & $<$9.0$\times 10^{14}$ & $<$4.7$\times 10^{14}$ & $<$1.2$\times 10^{15}$ \\
    EDJ2009-235    & $<$4.0$\times 10^{15}$ & $<$8.1$\times 10^{13}$ & $<$8.1$\times 10^{14}$ & $<$5.1$\times 10^{15}$ & $<$1.0$\times 10^{15}$ & $<$2.3$\times 10^{15}$ \\
    Per-emb 36     & $<$3.6$\times 10^{15}$ & $<$8.4$\times 10^{13}$ & $<$1.4$\times 10^{15}$ & $<$1.0$\times 10^{15}$ & $<$1.1$\times 10^{13}$ & $<$3.3$\times 10^{15}$ \\
    B1-b S         & 1.5$^{+0.1}_{-0.9}$$\times 10^{16}$ & 3.2$^{+1.6}_{-1.4}$$\times 10^{14}$ & 1.4$^{+0.4}_{-0.2}$$\times 10^{16}$ & 2.0$^{+0.3}_{-0.3}$$\times 10^{16}$ & 2.1$^{+1.5}_{-1.2}$$\times 10^{15}$ & $<$1.5$\times 10^{15}$ \\
    B1-b N         & $<$3.9$\times 10^{15}$ & $<$8.5$\times 10^{13}$ & $<$2.6$\times 10^{14}$ & $<$3.4$\times 10^{15}$ & $<$2.2$\times 10^{12}$ & $<$2.1$\times 10^{15}$ \\
    Per-emb 29     & 9.0$^{+6.3}_{-5.9}$$\times 10^{16}$ & 1.1$^{+0.5}_{-0.4}$$\times 10^{15}$ & 5.9$^{+0.1}_{-0.9}$$\times 10^{16}$ & 2.8$^{+0.9}_{-0.7}$$\times 10^{16}$ & 2.1$^{+1.2}_{-1.3}$$\times 10^{15}$ & 3.6$^{+1.3}_{-1.0}$$\times 10^{16}$ \\
    Per-emb 10     & 2.7$^{+1.3}_{-1.6}$$\times 10^{15}$ & $<$8.8$\times 10^{13}$ & $<$5.2$\times 10^{14}$ & $<$5.9$\times 10^{15}$ & $<$1.4$\times 10^{15}$ & $<$2.9$\times 10^{15}$ \\
    Per-emb 40     & $<$4.8$\times 10^{15}$ & $<$8.6$\times 10^{13}$ & $<$1.1$\times 10^{15}$ & $<$6.0$\times 10^{15}$ & $<$4.5$\times 10^{14}$ & $<$3.5$\times 10^{15}$ \\
    Per-emb 2      & 3.1$^{+0.9}_{-0.1}$$\times 10^{15}$ & $<$5.1$\times 10^{13}$ & $<$9.4$\times 10^{13}$ & $<$8.3$\times 10^{14}$ & $<$2.0$\times 10^{14}$ & $<$1.9$\times 10^{15}$ \\
    Per-emb 5      & 9.9$^{+4.7}_{-6.0}$$\times 10^{15}$ & 1.8$^{+0.8}_{-0.8}$$\times 10^{14}$ & $<$9.1$\times 10^{14}$ & $<$1.1$\times 10^{16}$ & $<$4.7$\times 10^{12}$ & $<$2.6$\times 10^{15}$ \\
    Per-emb 1      & 8.9$^{+3.8}_{-5.4}$$\times 10^{15}$ & 6.2$^{+3.1}_{-2.6}$$\times 10^{13}$ & $<$3.9$\times 10^{14}$ & $<$1.2$\times 10^{15}$ & $<$4.8$\times 10^{15}$ & $<$3.4$\times 10^{15}$ \\
    Per-emb 11 A   & 9.1$^{+5.3}_{-5.6}$$\times 10^{15}$ & 1.4$^{+0.7}_{-0.6}$$\times 10^{14}$ & 7.3$^{+1.7}_{-1.3}$$\times 10^{15}$ & 3.4$^{+0.4}_{-0.3}$$\times 10^{15}$ & 8.4$^{+3.7}_{-3.8}$$\times 10^{15}$ & $<$2.3$\times 10^{15}$ \\
    Per-emb 11 B   & $<$2.6$\times 10^{15}$ & $<$1.9$\times 10^{13}$ & $<$2.7$\times 10^{14}$ & $<$3.7$\times 10^{15}$ & $<$4.2$\times 10^{12}$ & $<$9.3$\times 10^{14}$ \\
    Per-emb 11 C   & 6.4$^{+3.9}_{-4.0}$$\times 10^{15}$ & $<$4.6$\times 10^{13}$ & 1.1$^{+0.2}_{-0.3}$$\times 10^{15}$ & $<$4.0$\times 10^{14}$ & $<$5.1$\times 10^{14}$ & $<$2.4$\times 10^{15}$ \\
    Per-emb 8      & $<$3.4$\times 10^{15}$ & $<$8.0$\times 10^{13}$ & $<$1.5$\times 10^{14}$ & $<$5.2$\times 10^{14}$ & $<$5.2$\times 10^{13}$ & $<$4.2$\times 10^{15}$ \\
    Per-emb 55     & $<$3.9$\times 10^{15}$ & $<$7.9$\times 10^{13}$ & $<$1.1$\times 10^{15}$ & $<$4.9$\times 10^{15}$ & $<$5.7$\times 10^{14}$ & $<$3.0$\times 10^{15}$ \\
    Per-emb 16     & $<$2.0$\times 10^{15}$ & $<$6.1$\times 10^{13}$ & $<$1.1$\times 10^{15}$ & $<$3.7$\times 10^{15}$ & $<$5.6$\times 10^{12}$ & $<$3.3$\times 10^{15}$ \\
    Per-emb 28     & $<$1.7$\times 10^{16}$ & $<$4.0$\times 10^{14}$ & $<$3.5$\times 10^{14}$ & $<$1.7$\times 10^{16}$ & $<$1.8$\times 10^{13}$ & $<$7.1$\times 10^{15}$ \\
    Per-emb 53     & 5.1$^{+2.9}_{-3.2}$$\times 10^{15}$ & $<$7.7$\times 10^{13}$ & $<$1.0$\times 10^{14}$ & $<$3.8$\times 10^{15}$ & $<$4.5$\times 10^{12}$ & $<$1.5$\times 10^{15}$ \\
    \enddata
    \tablenotetext{a}{The column densities of molecules are unconstrained due to the exclusion of the spectral window contaminated by the SiO emission.}
    \tablecomments{The column density is fitted with an aperture of 0\farcs{5}.}
\end{deluxetable*}
\renewcommand{\thetable}{\arabic{table} (Cont.)}
\addtocounter{table}{-1}
\begin{deluxetable*}{ccccccc}
    \tabletypesize{\scriptsize}
    \tablecaption{Column Densities of Molecules}
    \tablewidth{\textwidth}
    \tablehead{ \colhead{Source} & \colhead{CH$_{3}$OCHO $v=1$} & \colhead{CH$_{3}$COCH$_{3}$} & \colhead{$t$-HCOOH} & \colhead{C$_{2}$H$_{5}$CN} & \colhead{NH$_{2}$CHO} & \colhead{CH$_{2}$DCN}}
    \startdata
    L1448 NW       & $<$1.7$\times 10^{16}$ & $<$5.0$\times 10^{14}$ & $<$1.2$\times 10^{15}$ & $<$2.5$\times 10^{14}$ & $<$2.1$\times 10^{14}$ & $<$7.5$\times 10^{13}$ \\
    Per-emb 33 A   & $<$7.6$\times 10^{15}$ & $<$7.6$\times 10^{14}$ & $<$4.3$\times 10^{14}$ & $<$2.3$\times 10^{14}$ & $<$2.0$\times 10^{14}$ & $<$1.2$\times 10^{14}$ \\
    Per-emb 33 B/C & $<$4.7$\times 10^{15}$ & $<$1.0$\times 10^{15}$ & $<$1.2$\times 10^{15}$ & $<$2.3$\times 10^{14}$ & $<$5.1$\times 10^{13}$ & $<$1.5$\times 10^{14}$ \\
    L1448 IRS 3A   & $<$9.0$\times 10^{15}$ & $<$5.2$\times 10^{14}$ & $<$9.8$\times 10^{14}$ & $<$2.0$\times 10^{14}$ & $<$9.4$\times 10^{13}$ & $<$7.3$\times 10^{13}$ \\
    Per-emb 26     & $<$1.2$\times 10^{16}$ & $<$9.8$\times 10^{14}$ & 1.0$^{+0.6}_{-0.5}$$\times 10^{15}$ & 2.6$^{+0.8}_{-0.5}$$\times 10^{14}$ & 1.4$^{+0.5}_{-0.5}$$\times 10^{14}$ & 1.5$^{+0.7}_{-0.6}$$\times 10^{14}$ \\
    Per-emb 42     & $<$5.3$\times 10^{15}$ & $<$1.3$\times 10^{15}$ & $<$1.2$\times 10^{15}$ & $<$2.7$\times 10^{14}$ & $<$2.3$\times 10^{14}$ & $<$1.1$\times 10^{14}$ \\
    Per-emb 22 A   & $<$1.4$\times 10^{16}$ & $<$3.9$\times 10^{13}$ & $<$7.8$\times 10^{14}$ & $<$1.4$\times 10^{14}$ & $<$1.8$\times 10^{14}$ & $<$1.3$\times 10^{14}$ \\
    Per-emb 22 B   & $<$9.4$\times 10^{15}$ & $<$9.9$\times 10^{14}$ & $<$1.0$\times 10^{15}$ & $<$2.3$\times 10^{14}$ & $<$1.9$\times 10^{14}$ & $<$6.7$\times 10^{13}$ \\
    Per-emb 25     & $<$4.3$\times 10^{15}$ & $<$1.2$\times 10^{15}$ & $<$9.4$\times 10^{14}$ & $<$1.9$\times 10^{14}$ & $<$8.9$\times 10^{13}$ & $<$1.4$\times 10^{14}$ \\
    Per-emb 20     & $<$8.0$\times 10^{15}$ & $<$4.9$\times 10^{14}$ & $<$3.8$\times 10^{14}$ & $<$1.8$\times 10^{14}$ & 9.6$^{+3.1}_{-3.6}$$\times 10^{13}$ & $<$8.6$\times 10^{13}$ \\
    L1455 IRS 2    & $<$1.1$\times 10^{16}$ & $<$5.6$\times 10^{14}$ & $<$1.4$\times 10^{15}$ & $<$2.7$\times 10^{14}$ & $<$7.7$\times 10^{13}$ & $<$7.9$\times 10^{13}$ \\
    Per-emb 44     & 1.7$^{+0.6}_{-0.2}$$\times 10^{17}$ & 6.3$^{+3.7}_{-4.4}$$\times 10^{16}$ & 1.5$^{+0.6}_{-0.7}$$\times 10^{16}$ & 2.8$^{+0.9}_{-0.6}$$\times 10^{15}$ & 1.9$^{+0.6}_{-0.7}$$\times 10^{15}$ & 1.9$^{+1.0}_{-0.8}$$\times 10^{15}$ \\
    SVS 13A2       & $<$3.3$\times 10^{15}$ & $<$7.9$\times 10^{14}$ & $<$6.3$\times 10^{14}$ & $<$2.1$\times 10^{14}$ & $<$1.9$\times 10^{14}$ & $<$1.2$\times 10^{14}$ \\
    Per-emb 12 A   & $<$6.6$\times 10^{15}$ & $<$6.5$\times 10^{14}$ & $<$7.5$\times 10^{14}$ & $<$1.5$\times 10^{14}$ & $<$1.4$\times 10^{14}$ & $<$7.8$\times 10^{13}$ \\
    Per-emb 12 B   & 1.1$^{+0.3}_{-0.2}$$\times 10^{17}$ & 3.9$^{+5.5}_{-3.0}$$\times 10^{16}$ & 9.1$^{+4.8}_{-4.2}$$\times 10^{15}$ & 2.3$^{+0.6}_{-0.5}$$\times 10^{15}$ & 1.1$^{+0.4}_{-0.4}$$\times 10^{15}$ & 1.3$^{+0.6}_{-0.5}$$\times 10^{15}$ \\
    Per-emb 13     & 3.8$^{+0.1}_{-0.6}$$\times 10^{16}$ & 9.6$^{+1.4}_{-7.8}$$\times 10^{15}$ & 1.4$^{+0.8}_{-0.7}$$\times 10^{15}$ & 5.7$^{+1.5}_{-1.2}$$\times 10^{14}$ & 2.3$^{+0.8}_{-0.9}$$\times 10^{14}$ & 2.3$^{+0.1}_{-0.9}$$\times 10^{14}$ \\
    IRAS4B'        & $<$8.2$\times 10^{15}$ & $<$5.5$\times 10^{14}$ & $<$3.8$\times 10^{14}$ & $<$1.9$\times 10^{14}$ & $<$1.5$\times 10^{14}$ & \nodata \tablenotemark{a} \\
    Per-emb 27     & 6.9$^{+3.7}_{-1.2}$$\times 10^{16}$ & 3.0$^{+4.3}_{-2.4}$$\times 10^{16}$ & 1.6$^{+0.8}_{-0.7}$$\times 10^{16}$ & 3.1$^{+0.1}_{-0.8}$$\times 10^{15}$ & 2.1$^{+0.7}_{-0.8}$$\times 10^{15}$ & 1.4$^{+0.7}_{-0.5}$$\times 10^{15}$ \\
    Per-emb 54     & $<$1.6$\times 10^{16}$ & $<$1.0$\times 10^{15}$ & $<$4.0$\times 10^{14}$ & $<$2.4$\times 10^{14}$ & $<$2.4$\times 10^{14}$ & $<$1.0$\times 10^{14}$ \\
    Per-emb 21     & $<$1.4$\times 10^{16}$ & $<$9.0$\times 10^{14}$ & $<$3.5$\times 10^{14}$ & $<$2.2$\times 10^{14}$ & $<$2.1$\times 10^{14}$ & $<$4.3$\times 10^{13}$ \\
    Per-emb 14     & $<$1.5$\times 10^{16}$ & $<$7.4$\times 10^{14}$ & $<$9.8$\times 10^{14}$ & $<$1.2$\times 10^{14}$ & $<$1.2$\times 10^{14}$ & $<$8.8$\times 10^{13}$ \\
    Per-emb 35 A   & $<$8.8$\times 10^{15}$ & $<$8.0$\times 10^{14}$ & 9.1$^{+5.0}_{-4.5}$$\times 10^{14}$ & $<$1.5$\times 10^{14}$ & 1.3$^{+0.4}_{-0.5}$$\times 10^{14}$ & $<$1.0$\times 10^{14}$ \\
    Per-emb 35 B   & $<$8.6$\times 10^{15}$ & $<$9.1$\times 10^{14}$ & $<$8.5$\times 10^{14}$ & $<$1.9$\times 10^{14}$ & $<$2.0$\times 10^{14}$ & $<$6.8$\times 10^{13}$ \\
    SVS 13B        & $<$1.9$\times 10^{15}$ & $<$7.6$\times 10^{14}$ & $<$2.4$\times 10^{14}$ & $<$1.7$\times 10^{14}$ & $<$5.3$\times 10^{13}$ & \nodata \tablenotemark{a} \\
    Per-emb 15     & $<$9.5$\times 10^{15}$ & $<$7.9$\times 10^{14}$ & $<$7.6$\times 10^{14}$ & $<$6.2$\times 10^{13}$ & $<$1.7$\times 10^{14}$ & $<$1.2$\times 10^{14}$ \\
    Per-emb 50     & $<$8.1$\times 10^{15}$ & $<$3.6$\times 10^{14}$ & $<$8.6$\times 10^{14}$ & $<$2.3$\times 10^{14}$ & $<$6.9$\times 10^{13}$ & $<$6.2$\times 10^{13}$ \\
    Per-emb 18     & $<$3.2$\times 10^{15}$ & $<$6.9$\times 10^{14}$ & $<$2.9$\times 10^{14}$ & $<$1.6$\times 10^{14}$ & $<$1.6$\times 10^{14}$ & $<$8.2$\times 10^{13}$ \\
    Per-emb 37     & $<$1.3$\times 10^{16}$ & $<$2.7$\times 10^{13}$ & $<$9.2$\times 10^{14}$ & $<$1.8$\times 10^{14}$ & $<$1.8$\times 10^{14}$ & $<$5.6$\times 10^{13}$ \\
    EDJ2009-235    & $<$1.5$\times 10^{16}$ & $<$1.1$\times 10^{15}$ & $<$1.1$\times 10^{15}$ & $<$2.5$\times 10^{14}$ & $<$1.6$\times 10^{14}$ & $<$8.1$\times 10^{13}$ \\
    Per-emb 36     & $<$4.6$\times 10^{15}$ & $<$6.8$\times 10^{14}$ & $<$5.8$\times 10^{14}$ & $<$1.5$\times 10^{14}$ & $<$1.1$\times 10^{14}$ & $<$8.2$\times 10^{13}$ \\
    B1-b S         & 2.1$^{+0.8}_{-0.3}$$\times 10^{16}$ & 6.4$^{+8.8}_{-5.0}$$\times 10^{15}$ & $<$7.8$\times 10^{14}$ & 2.9$^{+0.7}_{-0.7}$$\times 10^{14}$ & $<$1.6$\times 10^{14}$ & 2.2$^{+0.1}_{-0.9}$$\times 10^{14}$ \\
    B1-b N         & $<$8.5$\times 10^{15}$ & $<$9.5$\times 10^{14}$ & $<$3.7$\times 10^{14}$ & $<$1.7$\times 10^{14}$ & $<$2.0$\times 10^{14}$ & $<$1.1$\times 10^{14}$ \\
    Per-emb 29     & 7.0$^{+2.8}_{-1.1}$$\times 10^{16}$ & 2.5$^{+3.7}_{-2.0}$$\times 10^{16}$ & 1.5$^{+0.8}_{-0.7}$$\times 10^{15}$ & 1.5$^{+0.4}_{-0.2}$$\times 10^{15}$ & 3.5$^{+1.2}_{-1.3}$$\times 10^{14}$ & 5.3$^{+2.6}_{-2.1}$$\times 10^{14}$ \\
    Per-emb 10     & $<$1.9$\times 10^{16}$ & $<$1.3$\times 10^{15}$ & $<$1.4$\times 10^{15}$ & $<$2.7$\times 10^{14}$ & $<$2.2$\times 10^{14}$ & $<$2.2$\times 10^{14}$ \\
    Per-emb 40     & $<$2.0$\times 10^{16}$ & $<$1.3$\times 10^{15}$ & $<$1.4$\times 10^{15}$ & $<$2.9$\times 10^{14}$ & $<$2.4$\times 10^{14}$ & $<$1.0$\times 10^{14}$ \\
    Per-emb 2      & $<$8.6$\times 10^{15}$ & $<$6.0$\times 10^{14}$ & $<$2.3$\times 10^{14}$ & $<$1.3$\times 10^{14}$ & $<$1.2$\times 10^{14}$ & $<$8.7$\times 10^{13}$ \\
    Per-emb 5      & $<$1.5$\times 10^{16}$ & $<$1.0$\times 10^{15}$ & $<$1.1$\times 10^{15}$ & $<$1.8$\times 10^{14}$ & $<$1.8$\times 10^{14}$ & $<$1.6$\times 10^{14}$ \\
    Per-emb 1      & $<$7.1$\times 10^{15}$ & $<$8.3$\times 10^{14}$ & $<$3.3$\times 10^{14}$ & $<$2.2$\times 10^{14}$ & $<$1.7$\times 10^{14}$ & \nodata \tablenotemark{a} \\
    Per-emb 11 A   & 8.7$^{+3.8}_{-1.6}$$\times 10^{15}$ & 2.0$^{+2.9}_{-1.6}$$\times 10^{15}$ & $<$8.5$\times 10^{14}$ & $<$1.3$\times 10^{14}$ & $<$1.7$\times 10^{14}$ & \nodata \tablenotemark{a} \\
    Per-emb 11 B   & $<$1.1$\times 10^{16}$ & $<$6.8$\times 10^{14}$ & $<$2.6$\times 10^{14}$ & $<$6.0$\times 10^{13}$ & $<$1.3$\times 10^{14}$ & $<$4.8$\times 10^{13}$ \\
    Per-emb 11 C   & $<$5.5$\times 10^{15}$ & $<$3.8$\times 10^{14}$ & $<$6.3$\times 10^{14}$ & $<$1.5$\times 10^{14}$ & $<$1.1$\times 10^{14}$ & $<$3.5$\times 10^{13}$ \\
    Per-emb 8      & $<$2.0$\times 10^{16}$ & $<$5.6$\times 10^{14}$ & $<$1.1$\times 10^{15}$ & $<$3.0$\times 10^{14}$ & $<$1.9$\times 10^{14}$ & $<$9.6$\times 10^{13}$ \\
    Per-emb 55     & $<$1.4$\times 10^{16}$ & $<$1.3$\times 10^{15}$ & $<$1.1$\times 10^{15}$ & $<$1.9$\times 10^{14}$ & $<$6.1$\times 10^{13}$ & $<$1.8$\times 10^{14}$ \\
    Per-emb 16     & $<$1.3$\times 10^{16}$ & $<$5.2$\times 10^{14}$ & $<$7.8$\times 10^{14}$ & $<$1.3$\times 10^{14}$ & $<$1.4$\times 10^{14}$ & $<$8.1$\times 10^{13}$ \\
    Per-emb 28     & $<$3.4$\times 10^{16}$ & $<$2.0$\times 10^{15}$ & $<$4.5$\times 10^{15}$ & $<$1.3$\times 10^{15}$ & $<$8.5$\times 10^{14}$ & $<$2.9$\times 10^{14}$ \\
    Per-emb 53     & $<$8.1$\times 10^{15}$ & $<$1.6$\times 10^{14}$ & $<$6.7$\times 10^{14}$ & $<$1.2$\times 10^{14}$ & $<$1.5$\times 10^{14}$ & $<$5.4$\times 10^{13}$ \\
    \enddata
    \tablenotetext{a}{The column densities of molecules are unconstrained because the spectral window contaminated by the SiO emission is excluded.}
    \tablecomments{The column density is fitted with an aperture of 0\farcs{5}.}
\end{deluxetable*}
\renewcommand{\thetable}{\arabic{table}}

\acknowledgements
Y.-L. Yang acknowledges the support from the JSPS Postdoctoral Fellowship from the Japan Society for the Promotion of Science and the Virginia Initiative of Cosmic Origins Postdoctoral Fellowship.  Y.-L. Yang also acknowledges the visitor support from the Institute of Astronomy and Astrophysics, Academia Sinica.  Y.-L. Yang thanks L. I. Cleeves for fruitful discussion.  This work was supported by JSPS KAKENHI grants 16H03964, 18H05222, and 20H05845.  This paper makes use of the following ALMA data: ADS/JAO.ALMA\#2016.1.01501.S and \#2017.1.01462.S. ALMA is a partnership of ESO (representing its member states), NSF (USA), and NINS (Japan), together with NRC (Canada), MOST and ASIAA (Taiwan), and KASI (Republic of Korea), in cooperation with the Republic of Chile. The Joint ALMA Observatory is operated by ESO, AUI/NRAO, and NAOJ.  The National Radio Astronomy Observatory is a facility of the National Science Foundation operated under cooperative agreement by Associated Universities, Inc.

\facilities{ALMA}

\software{astropy \citep{2018AJ....156..123A}, CASA \citep{2007ASPC..376..127M}, RADEX \citep{2007AA...468..627V}, spectral-cube \citep{2016ascl.soft09017R}, and XCLASS \citep{2017AA...598A...7M}}

\appendix
\setcounter{table}{0}
\renewcommand{\thetable}{A\arabic{table}}

\section{Intensity maps of \methanol, \methylcyanide, \methylformate, and \dimethylether}
\begin{figure*}[htbp!]
  \centering
  \includegraphics[width=\textwidth]{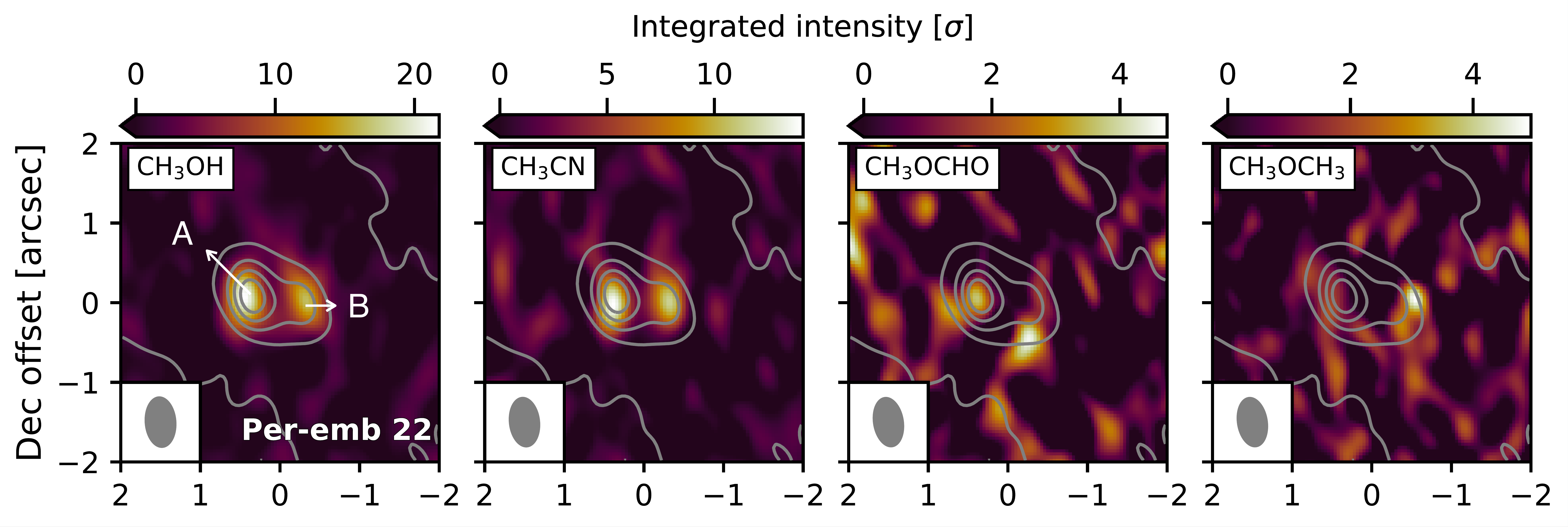}
  \includegraphics[width=\textwidth]{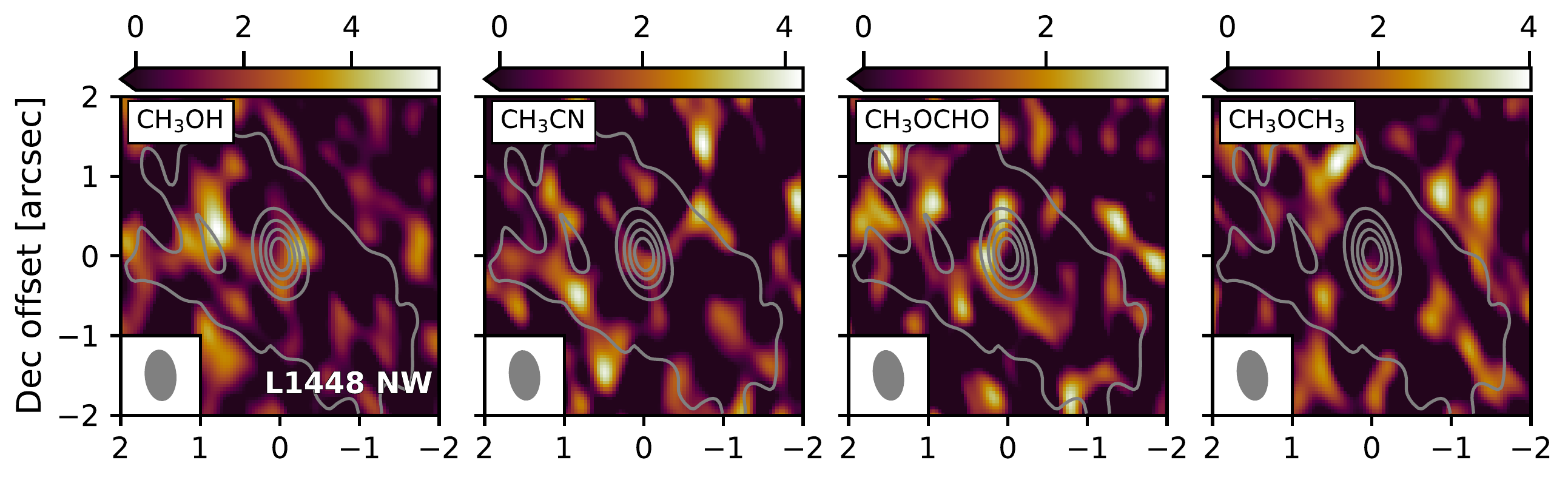}
  \includegraphics[width=\textwidth]{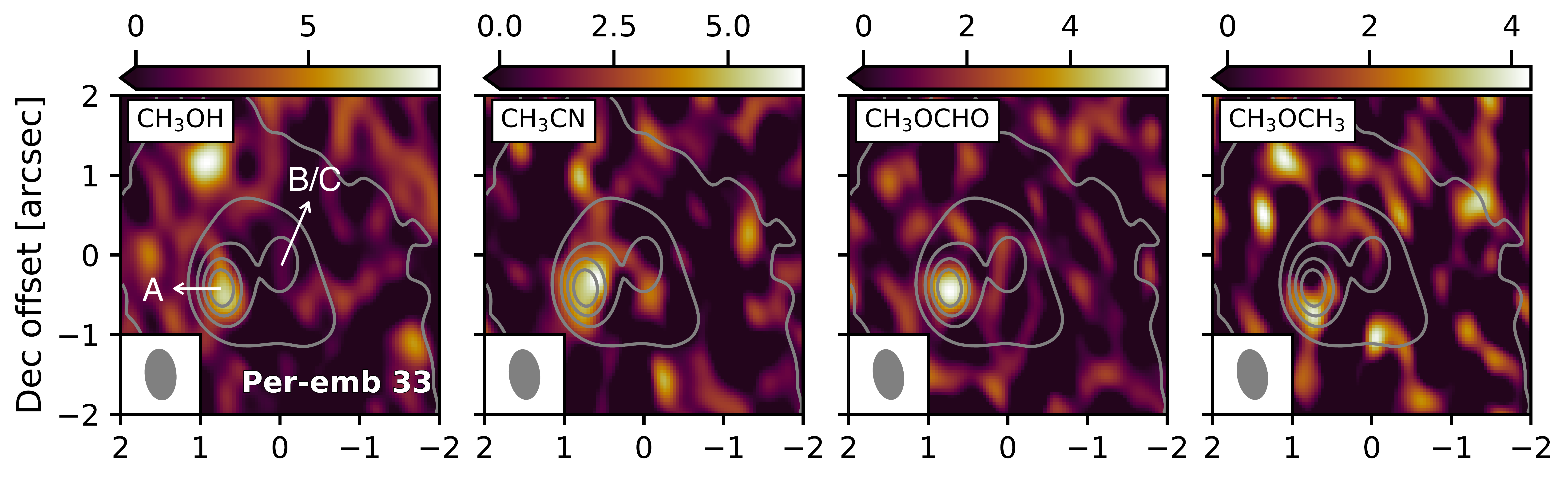}
  \includegraphics[width=\textwidth]{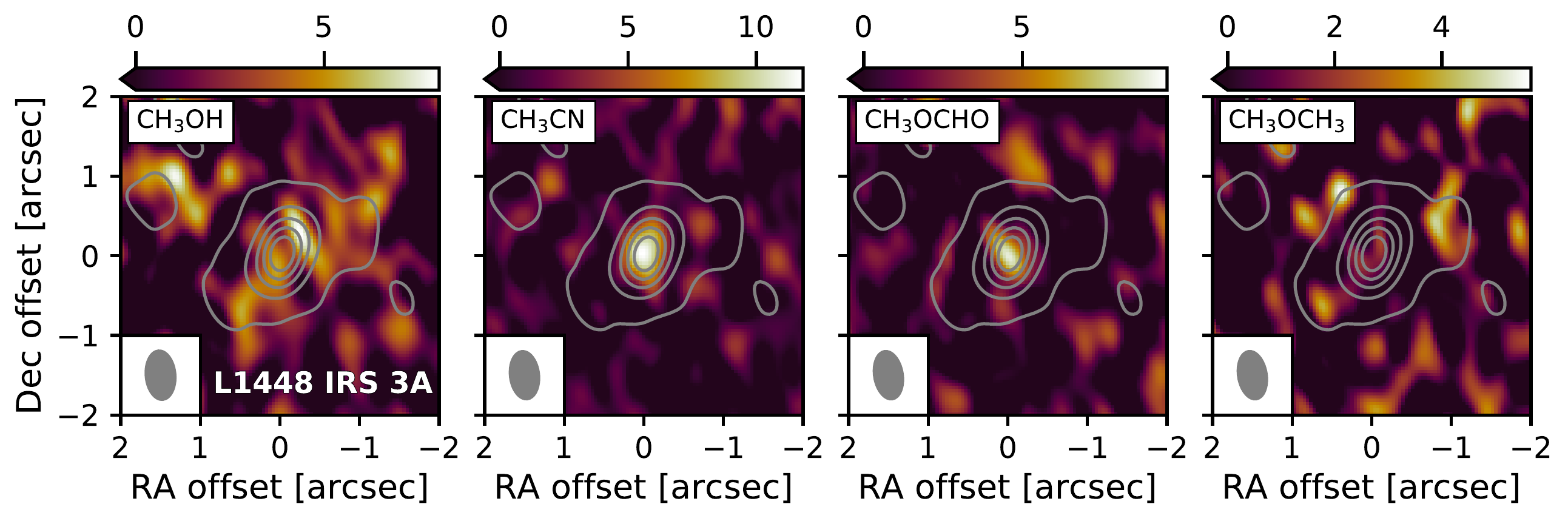}
  \caption{The intensity maps of most detected COMs, \methanol, \methylcyanide, \methylformate, and \dimethylether\ (from left to right).  
           The figures and legends are similar as the ones in Figure\,\ref{fig:coms_map_sample}.}
  \label{fig:coms_map_all}
\end{figure*}

Figure\,\ref{fig:coms_map_all} shows the intensity maps of \methanol, \methylcyanide, \methylformate, and \dimethylether\ for the PEACHES sources not shown in Figure\,\ref{fig:coms_map_sample}.  
\label{sec:coms_maps}
\renewcommand{\thefigure}{\arabic{figure} (Cont.)}
\addtocounter{figure}{-1}
\begin{figure*}[htbp!]
  \centering
  \includegraphics[width=\textwidth]{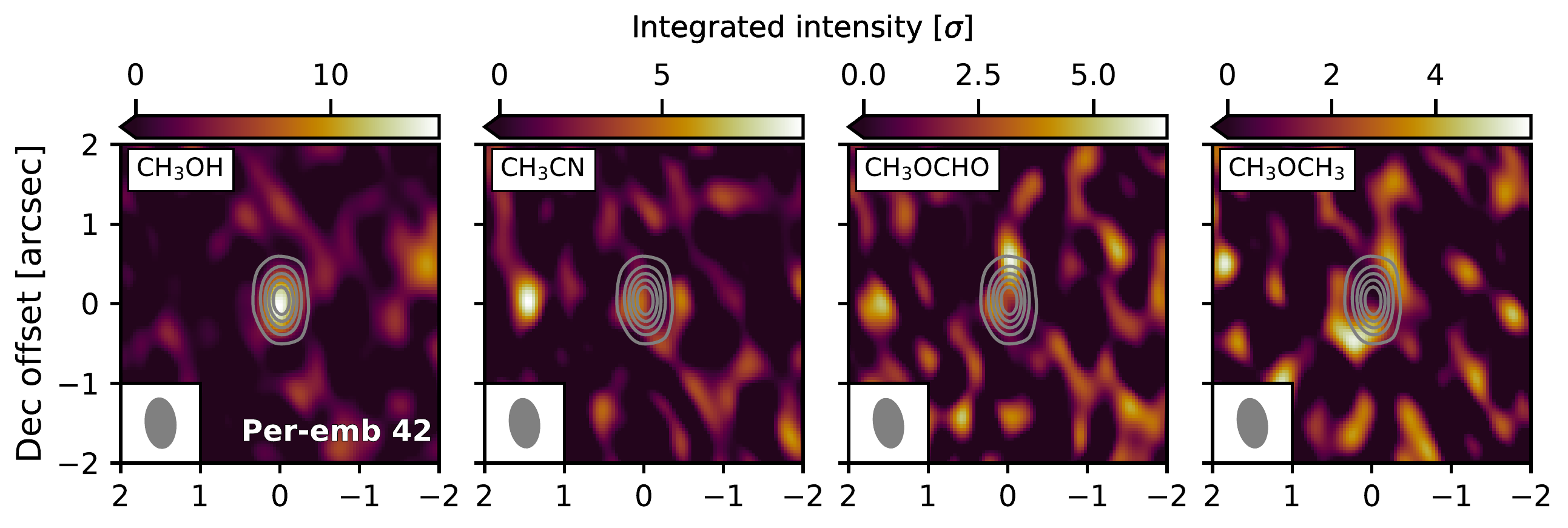}
  \includegraphics[width=\textwidth]{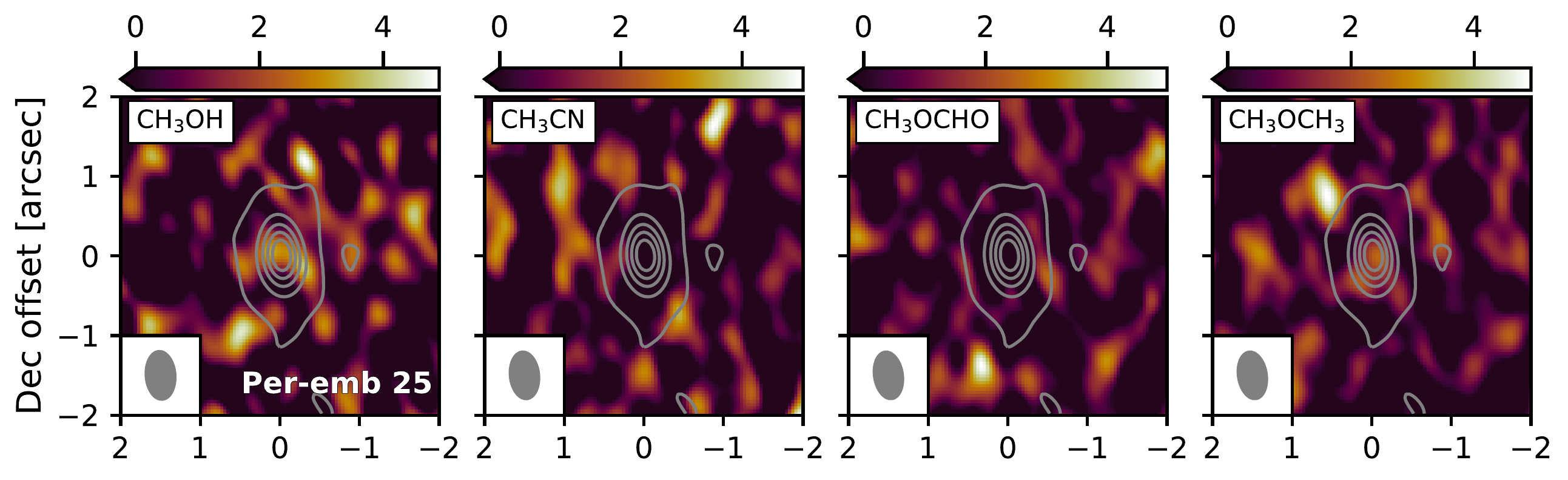}
  \includegraphics[width=\textwidth]{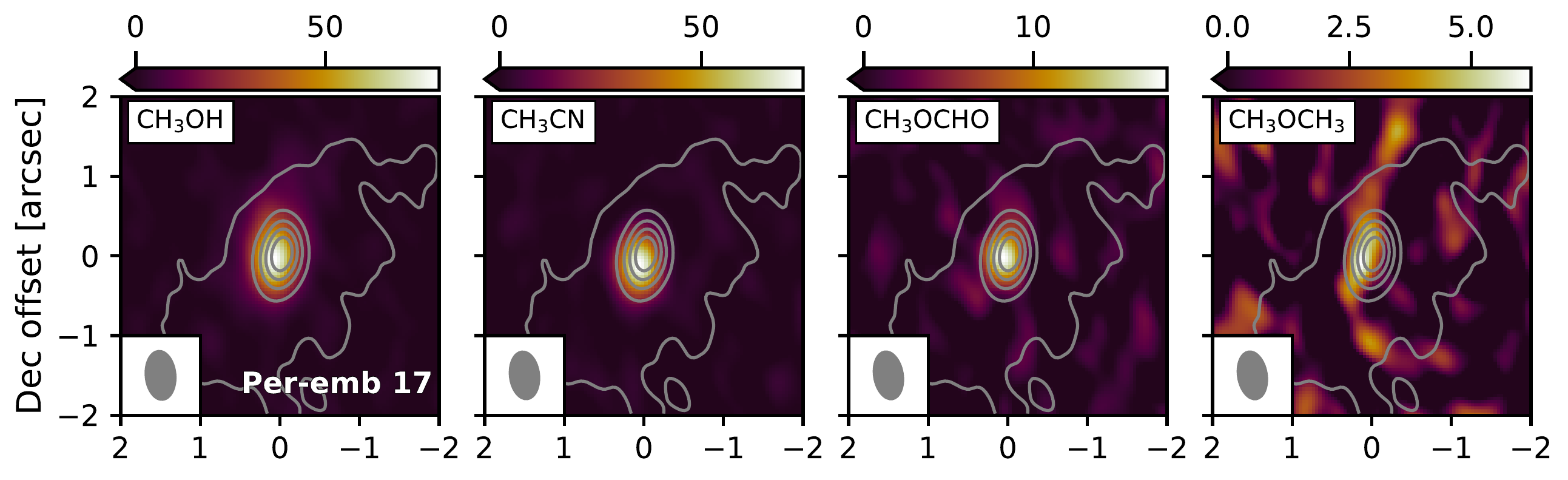}
  \includegraphics[width=\textwidth]{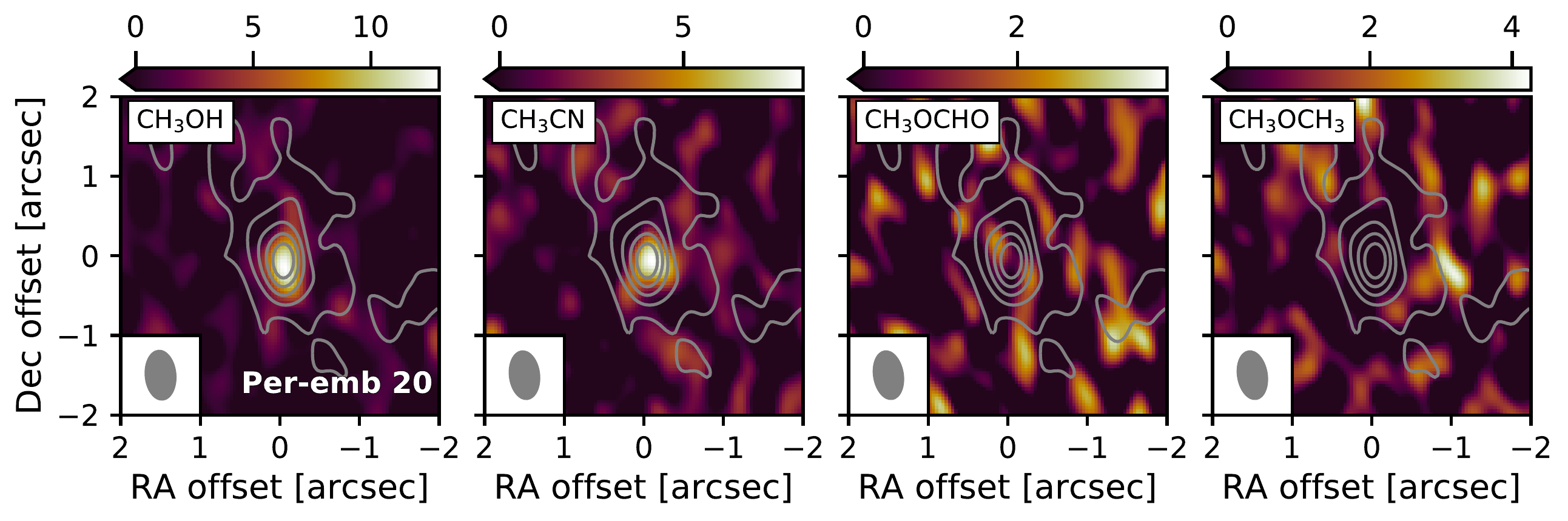}
  \caption{}
\end{figure*}
\addtocounter{figure}{-1}
\begin{figure*}[htbp!]
  \centering
  \includegraphics[width=\textwidth]{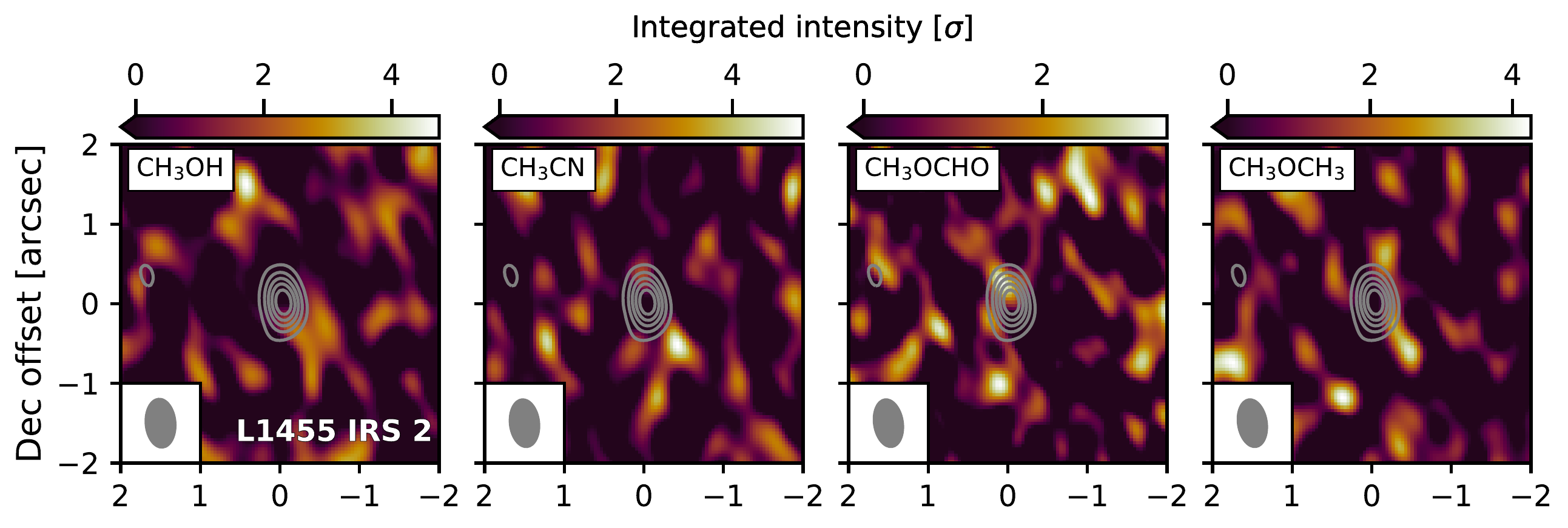}
  \includegraphics[width=\textwidth]{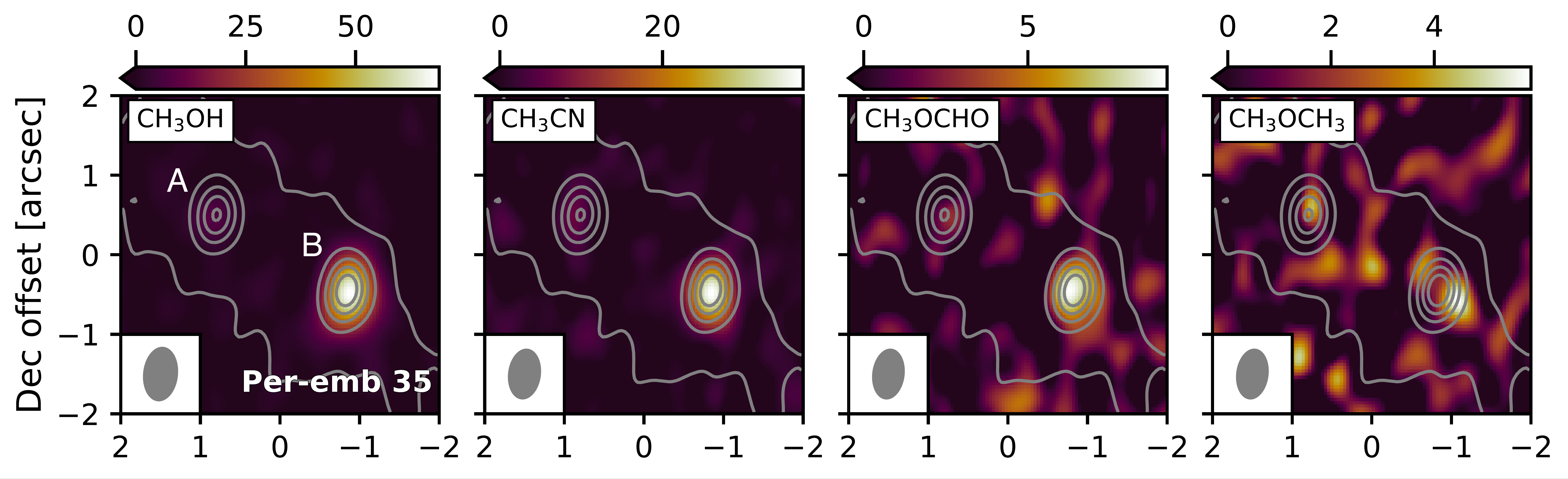}
  \includegraphics[width=\textwidth]{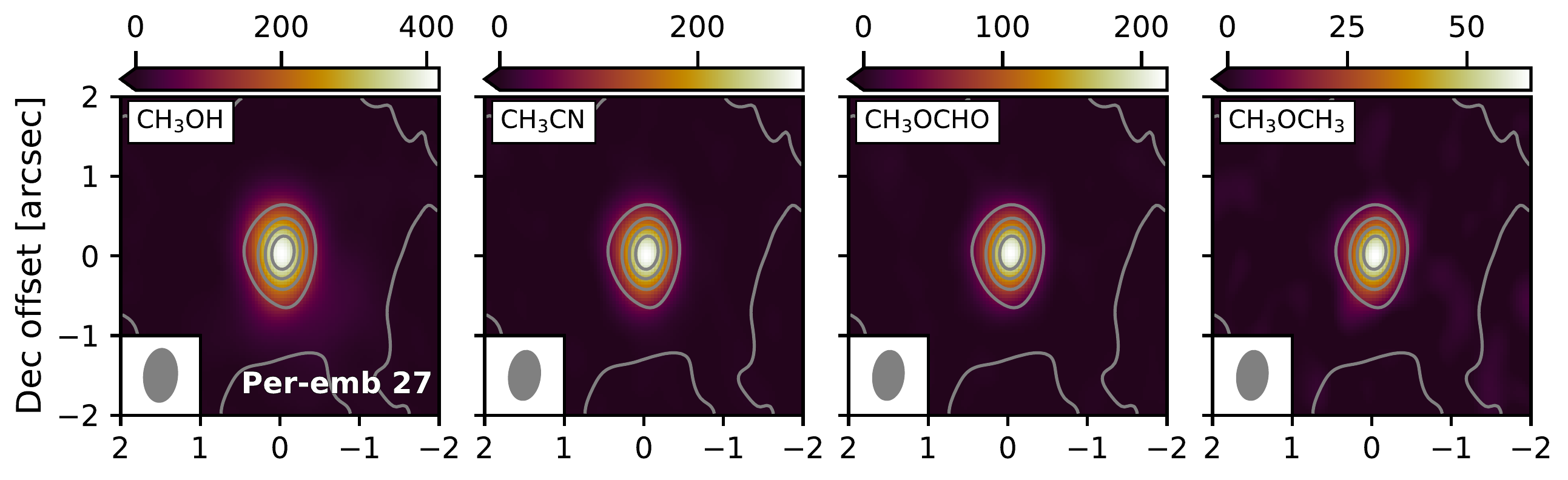}
  \includegraphics[width=\textwidth]{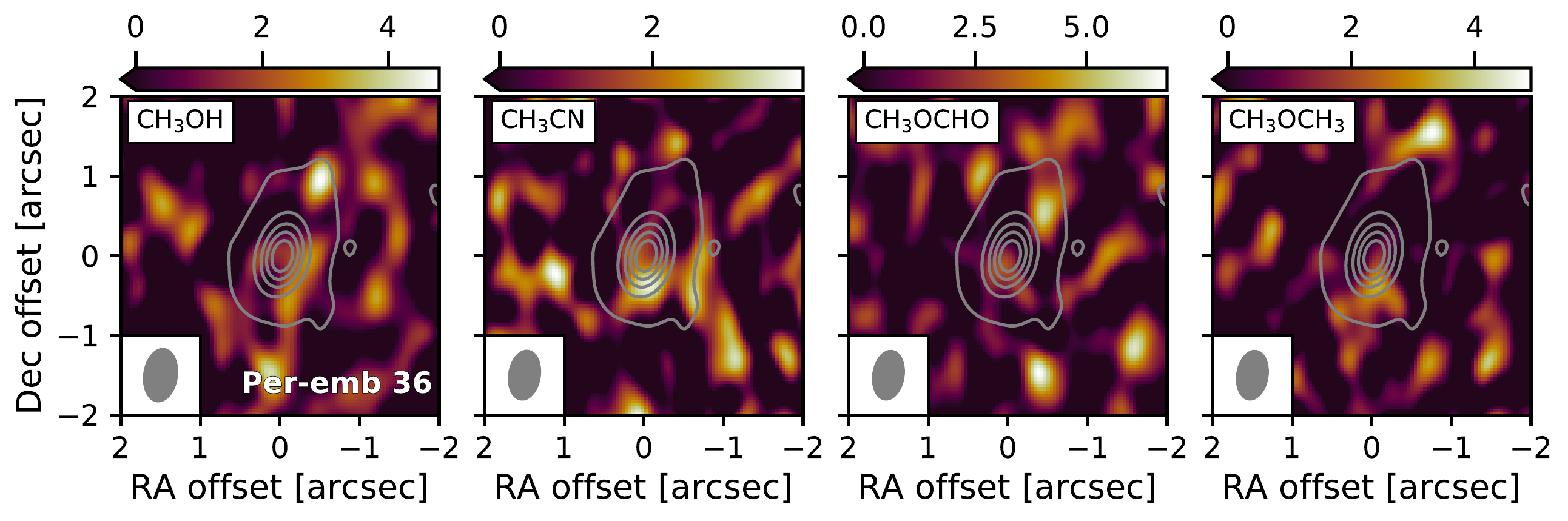}
  \caption{}
\end{figure*}
\addtocounter{figure}{-1}
\begin{figure*}[htbp!]
  \centering
  \includegraphics[width=\textwidth]{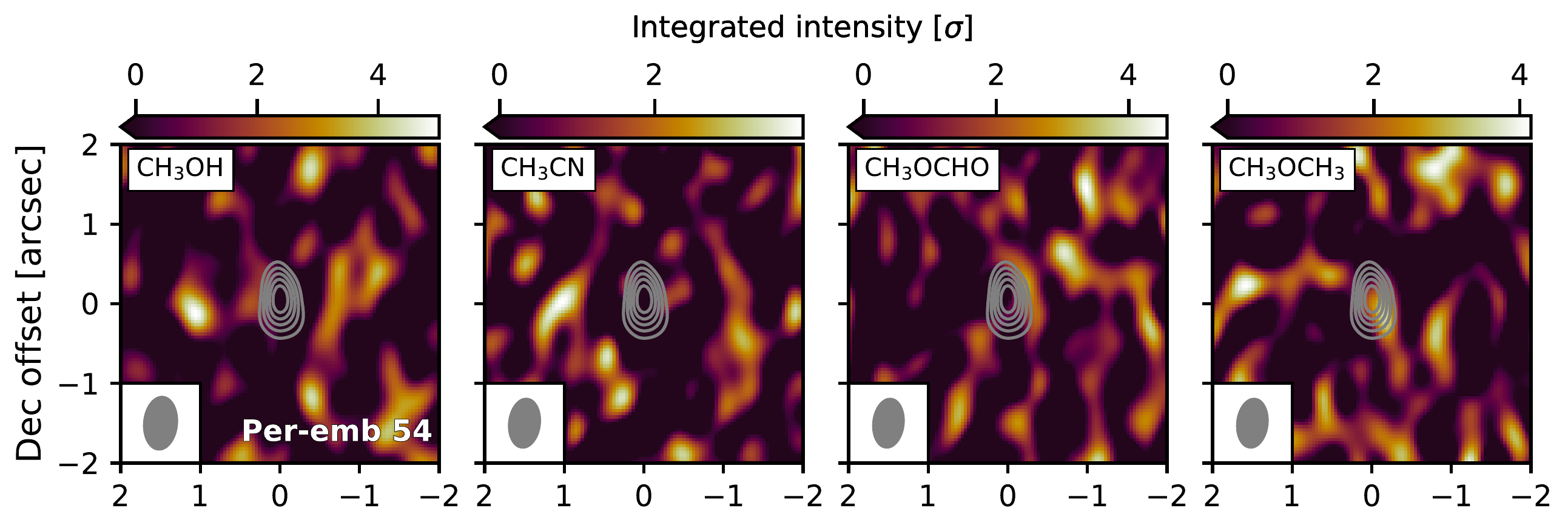}
  \includegraphics[width=\textwidth]{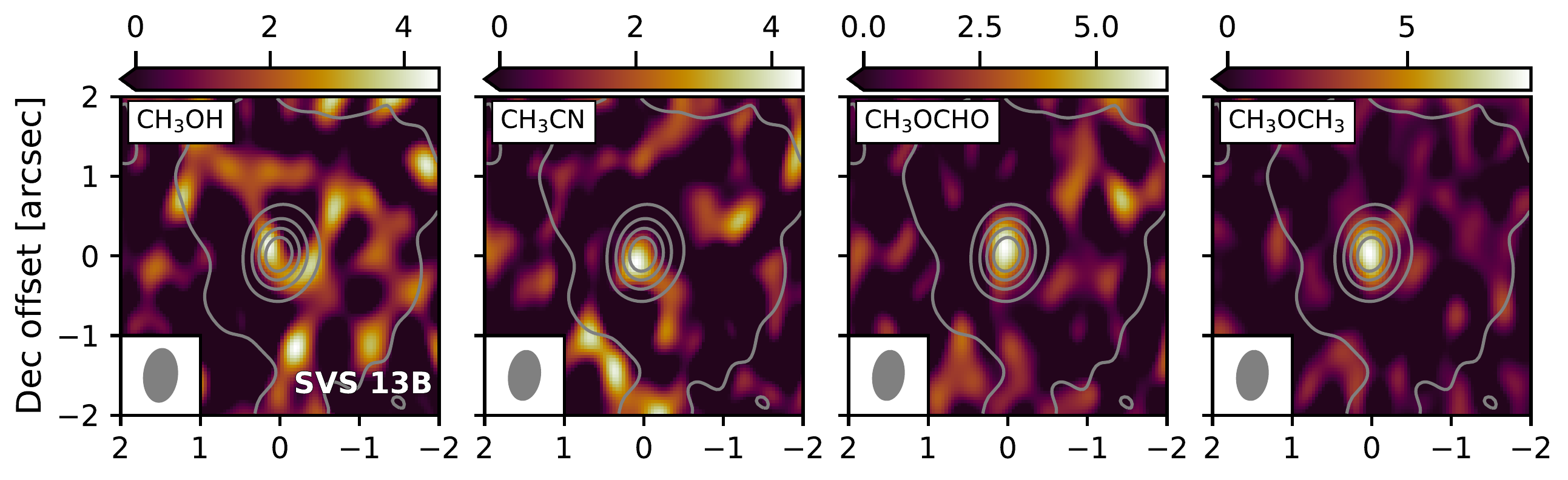}
  \includegraphics[width=\textwidth]{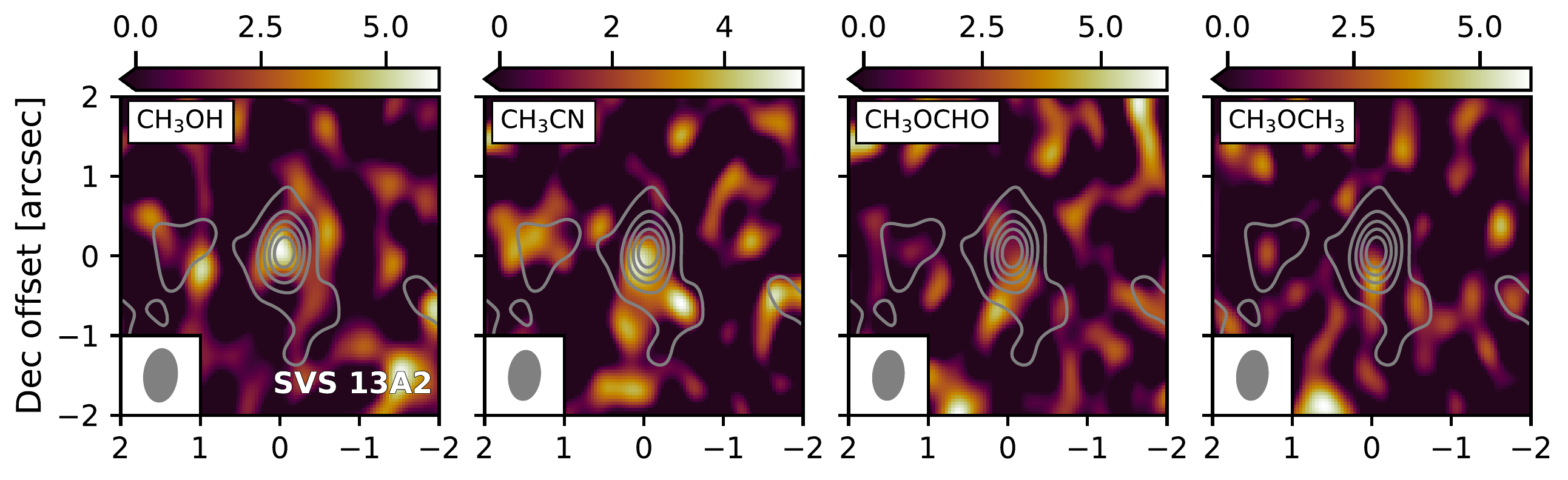}
  \includegraphics[width=\textwidth]{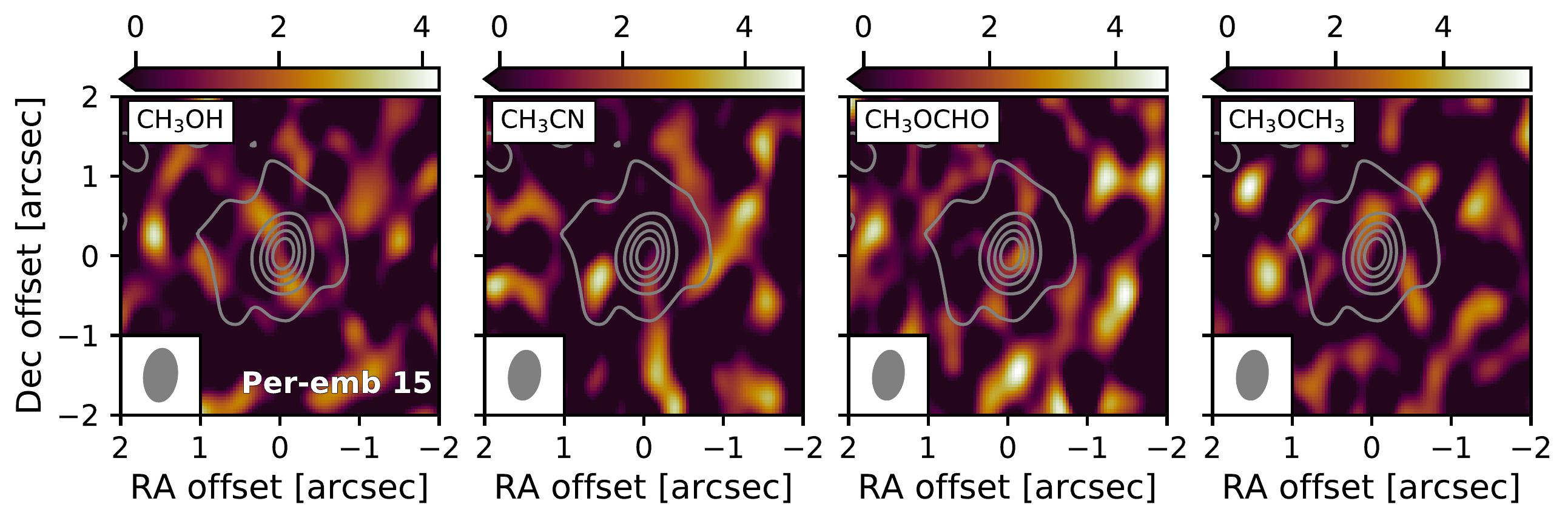}
  \caption{}
\end{figure*}
\addtocounter{figure}{-1}
\begin{figure*}[htbp!]
  \centering
  \includegraphics[width=\textwidth]{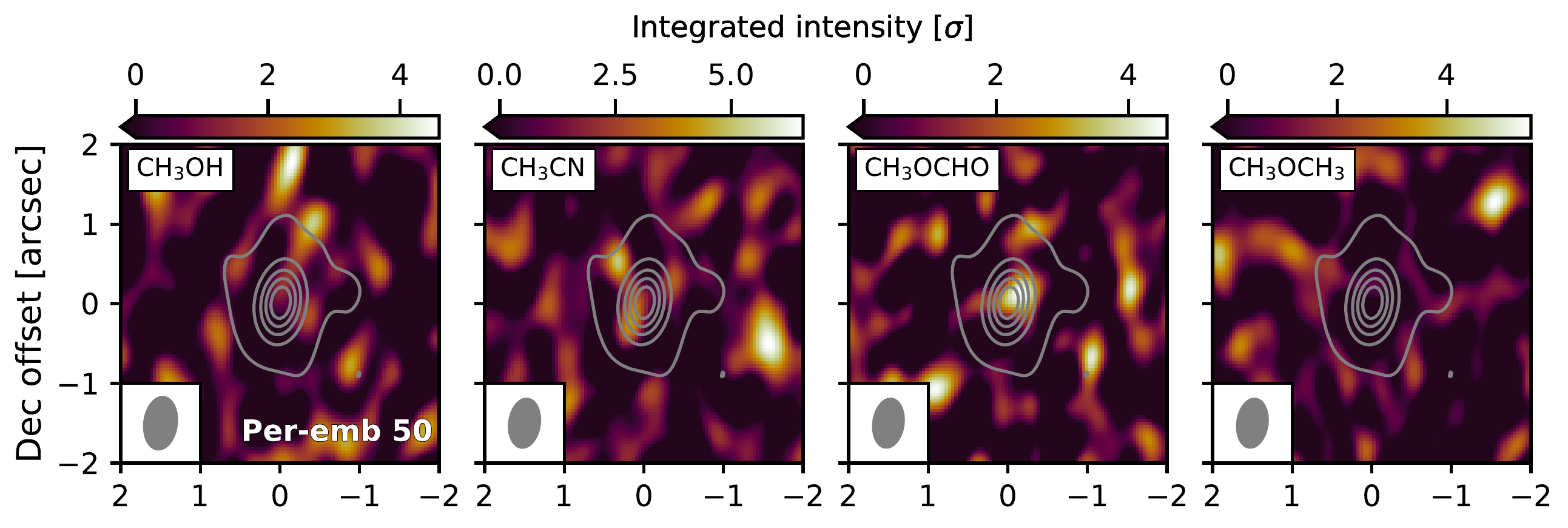}
  \includegraphics[width=\textwidth]{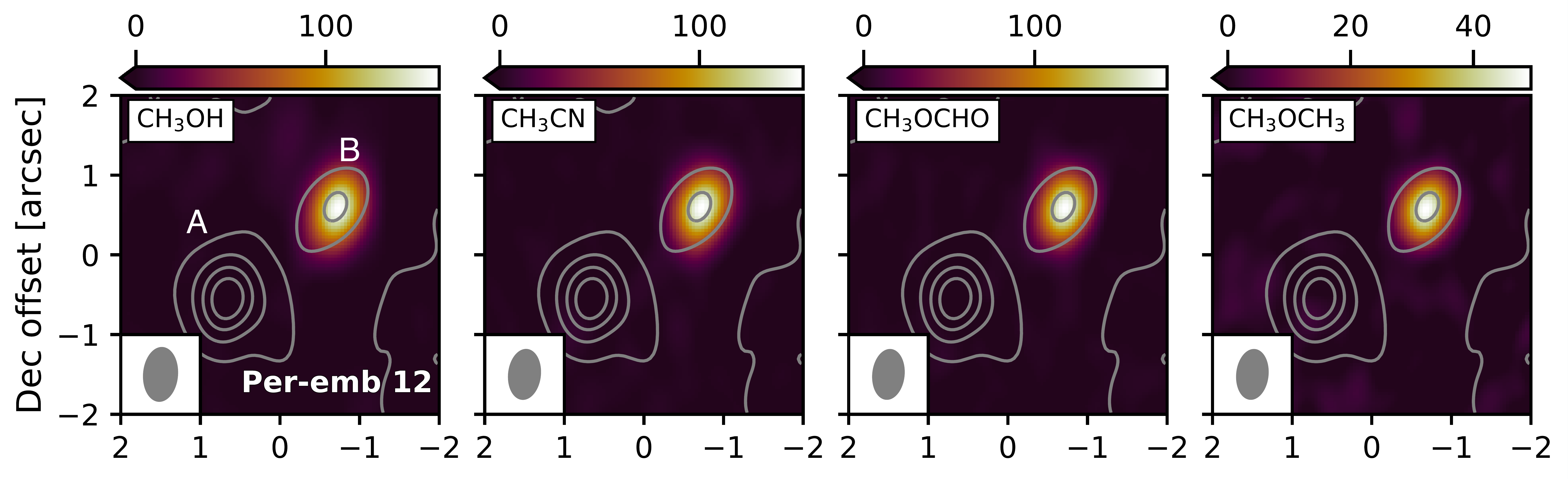}
  \includegraphics[width=\textwidth]{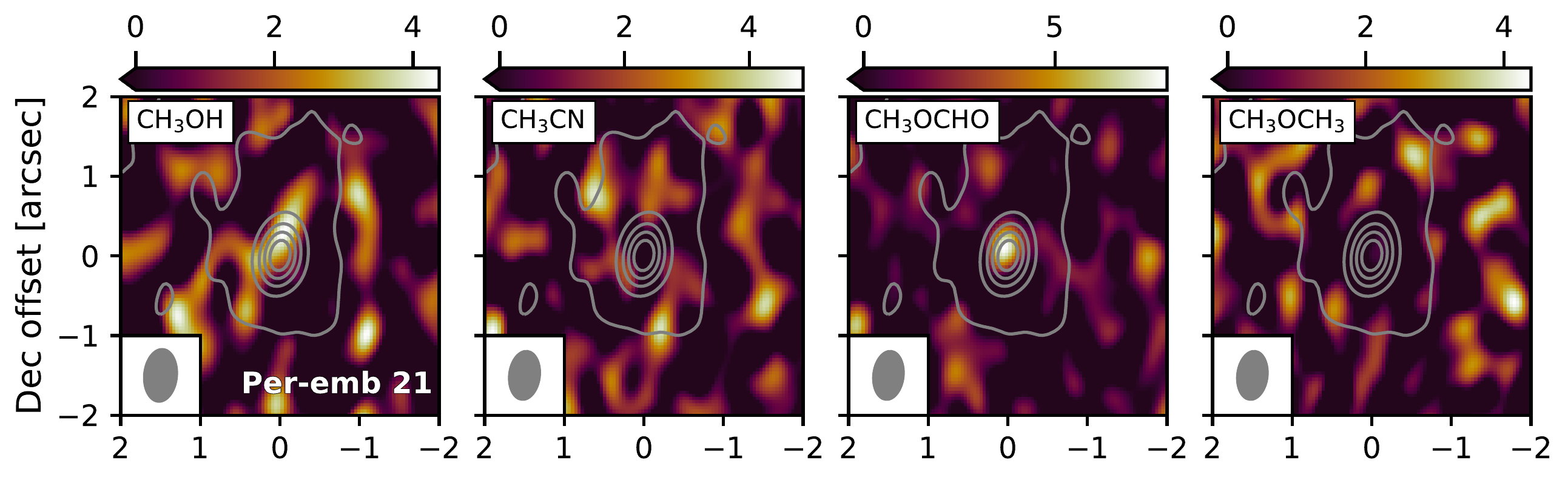}
  \includegraphics[width=\textwidth]{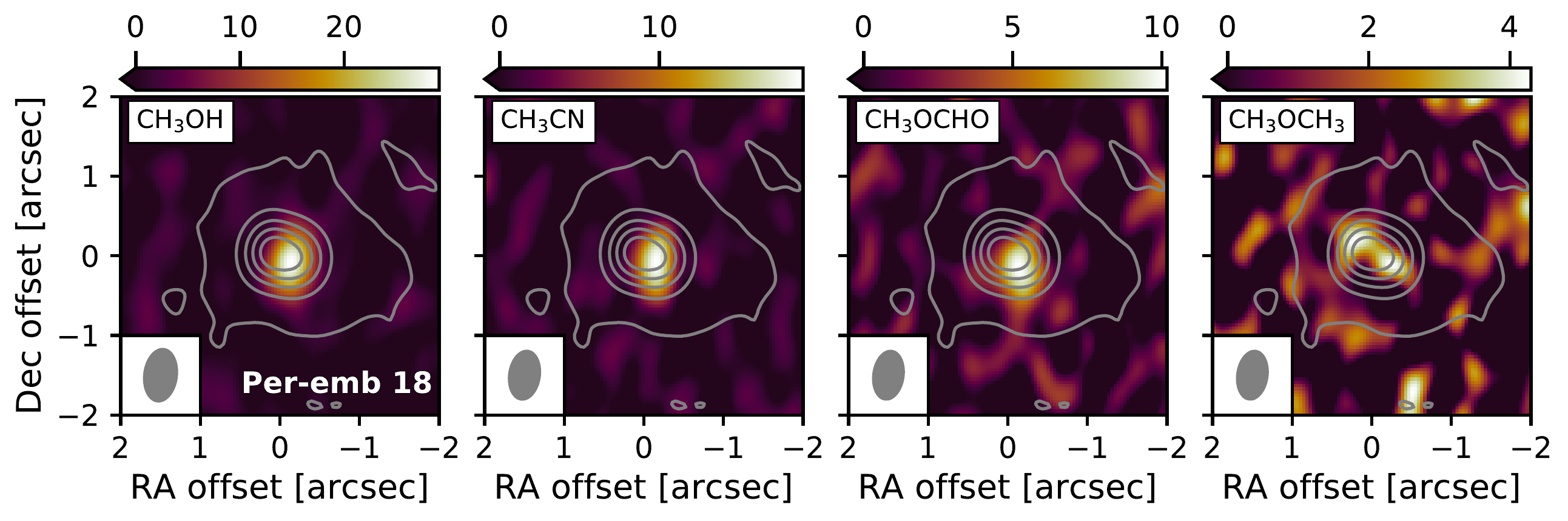}
  \caption{}
\end{figure*}
\addtocounter{figure}{-1}
\begin{figure*}[htbp!]
  \centering
  \includegraphics[width=\textwidth]{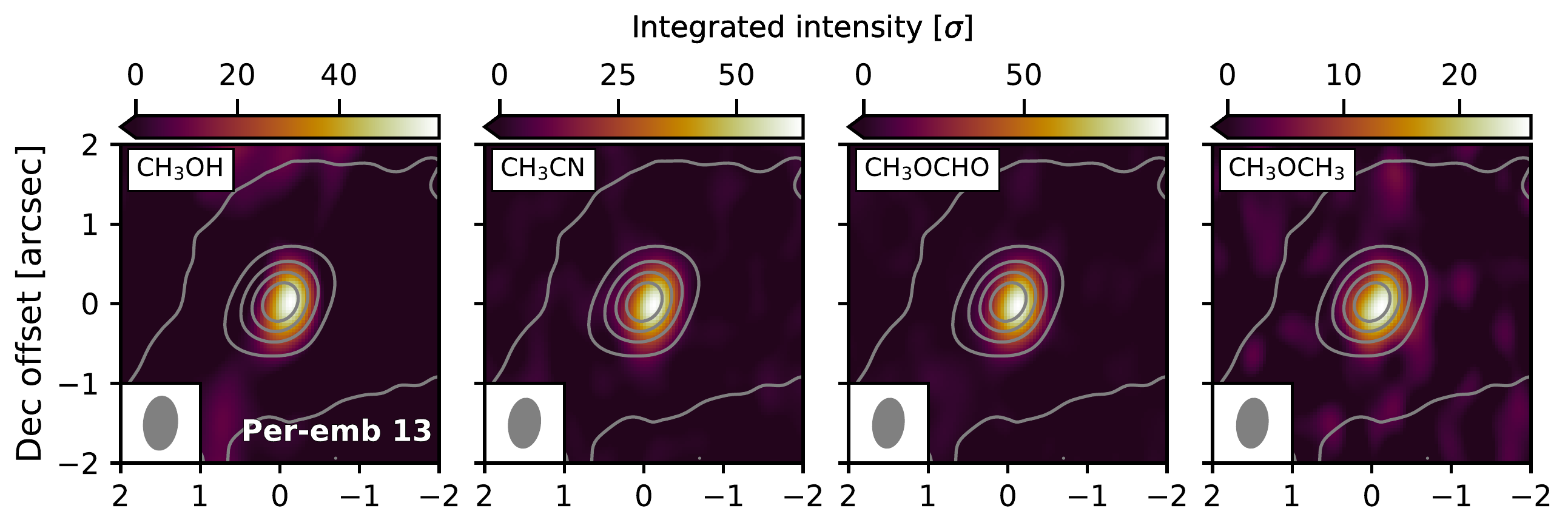}
  \includegraphics[width=\textwidth]{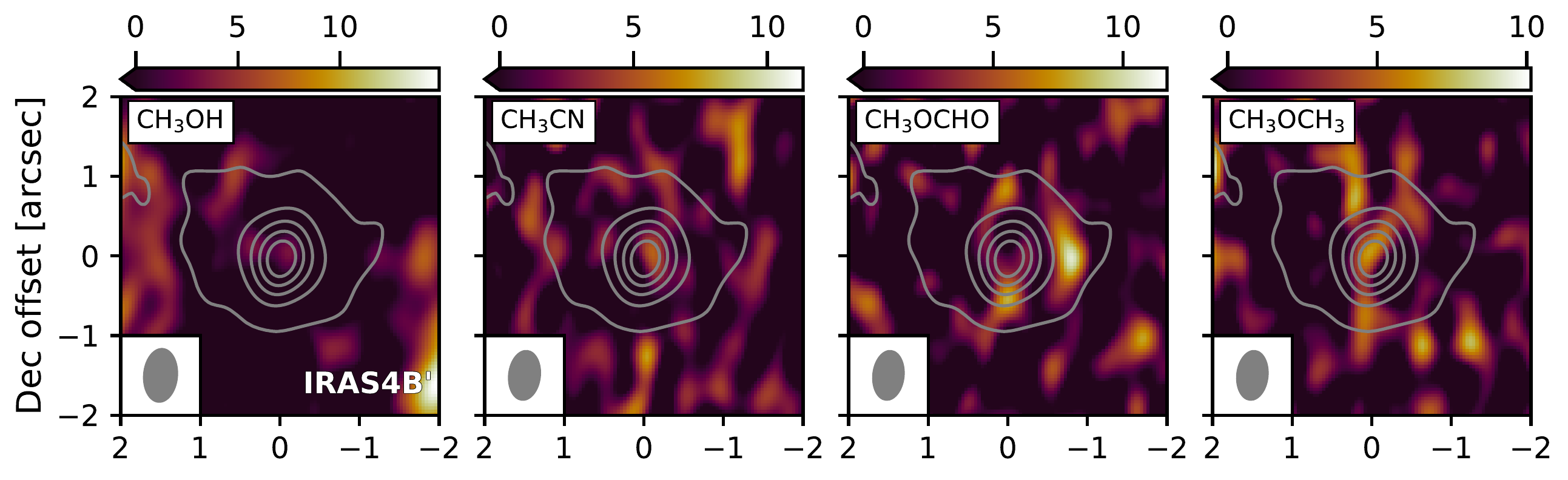}
  \includegraphics[width=\textwidth]{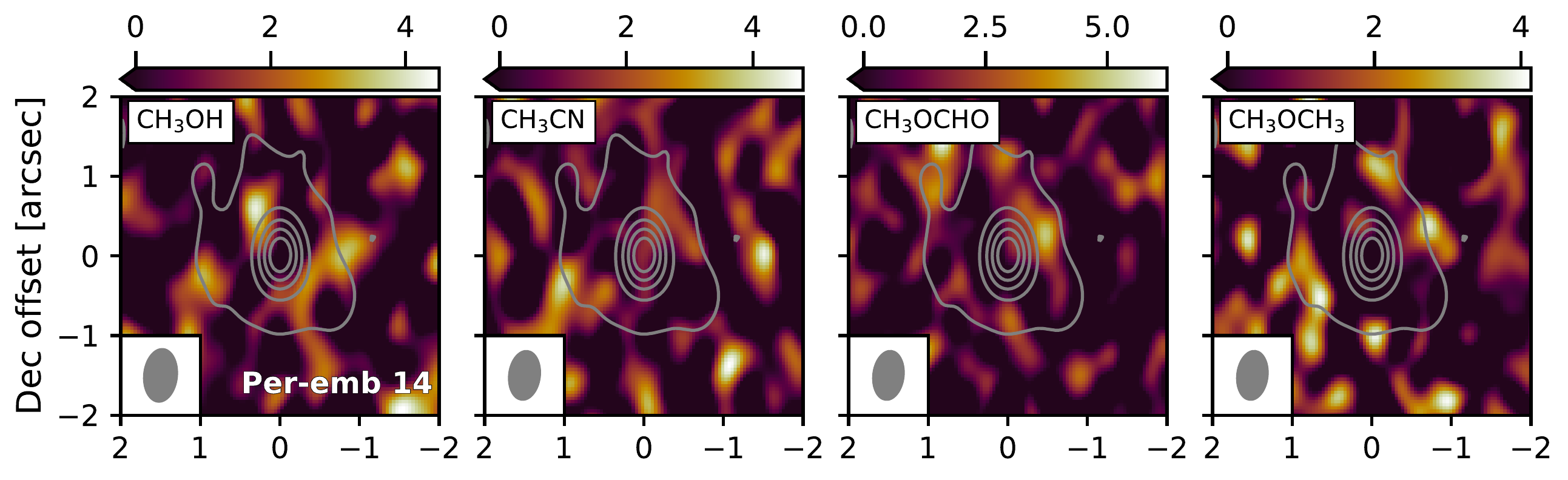}
  \includegraphics[width=\textwidth]{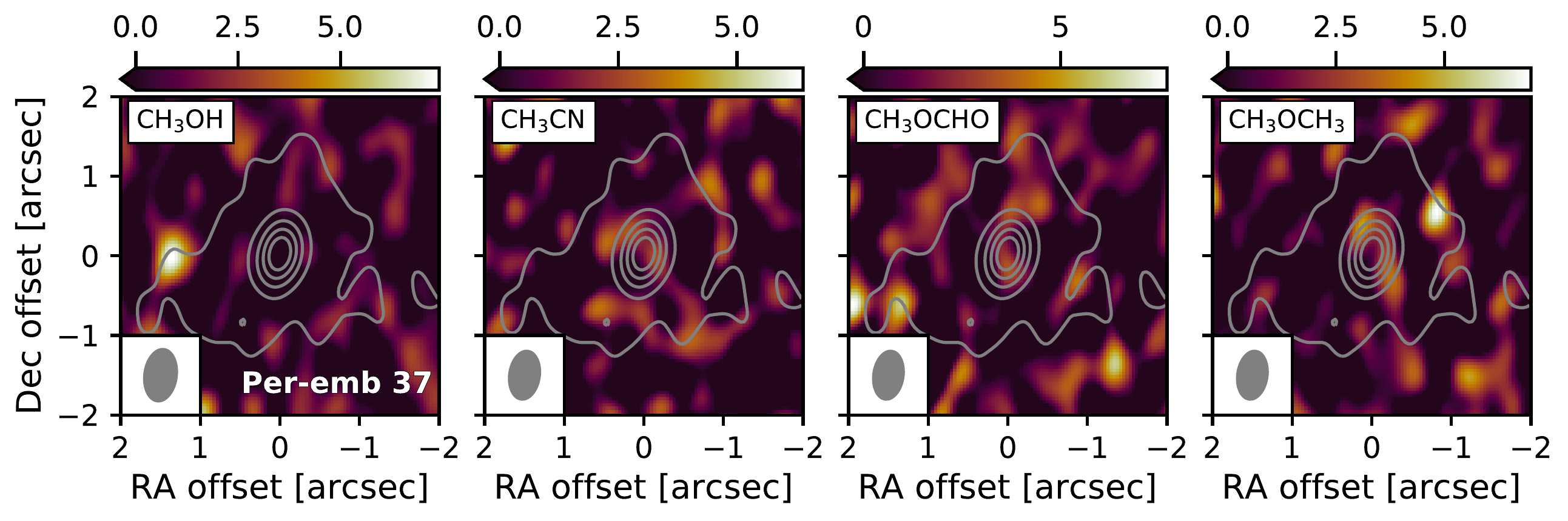}
  \caption{}
\end{figure*}
\addtocounter{figure}{-1}
\begin{figure*}[htbp!]
  \centering
  \includegraphics[width=\textwidth]{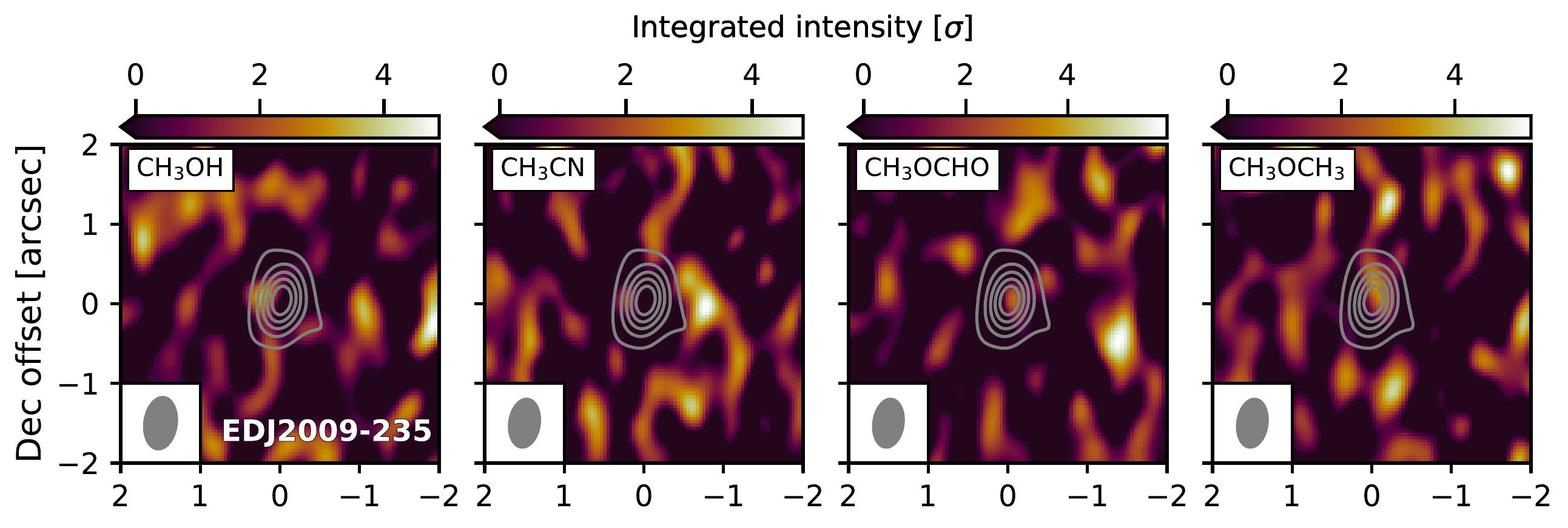}
  \includegraphics[width=\textwidth]{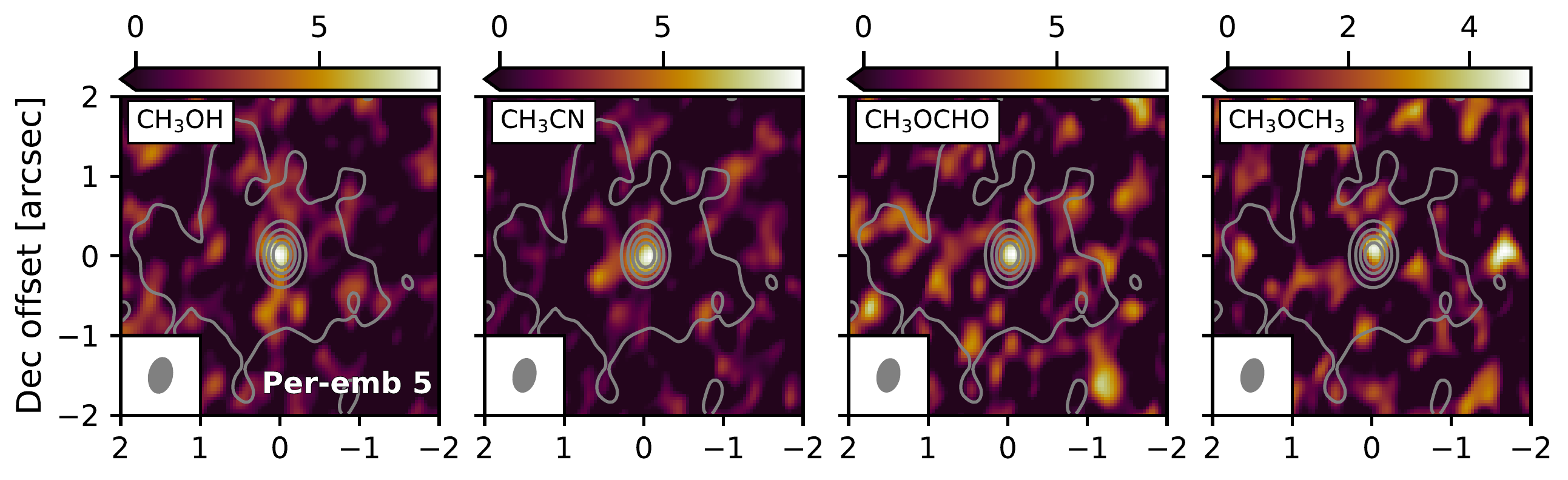}
  \includegraphics[width=\textwidth]{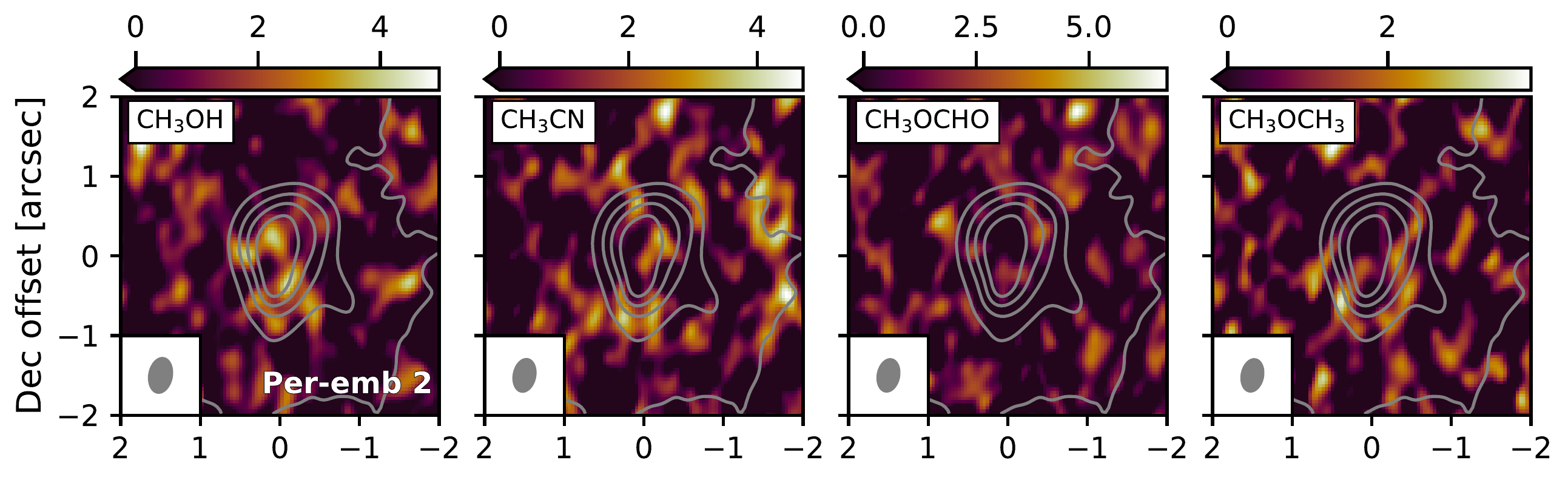}
  \includegraphics[width=\textwidth]{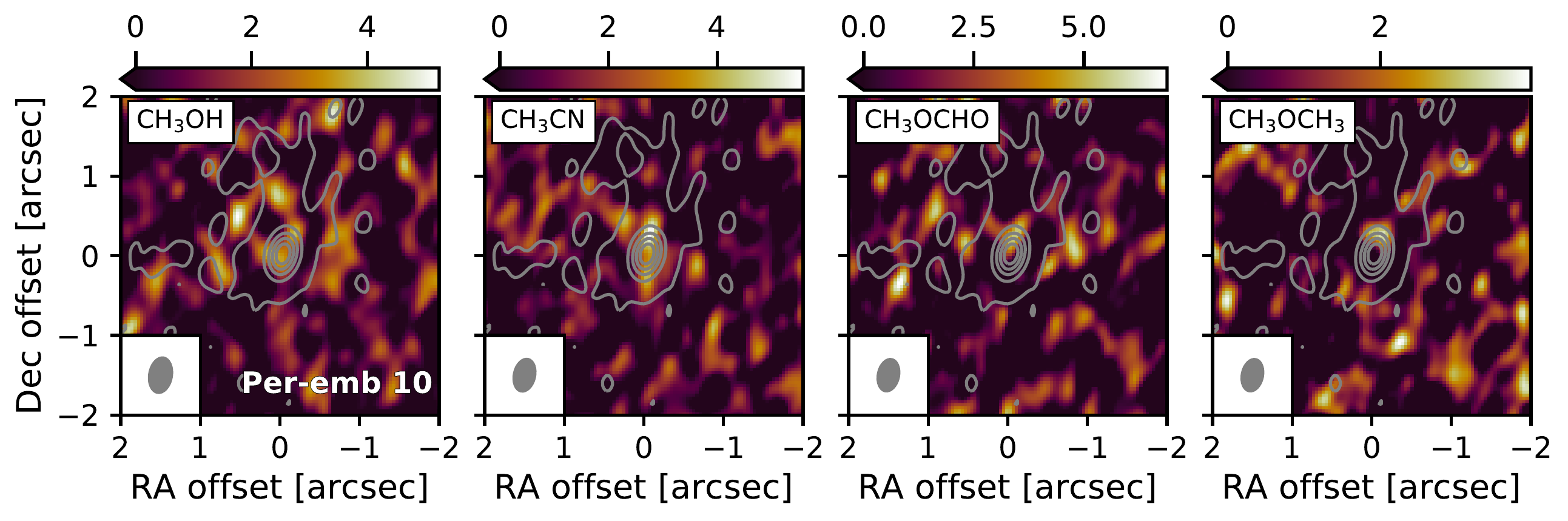}
  \caption{}
\end{figure*}
\addtocounter{figure}{-1}
\begin{figure*}[htbp!]
  \centering
  \includegraphics[width=\textwidth]{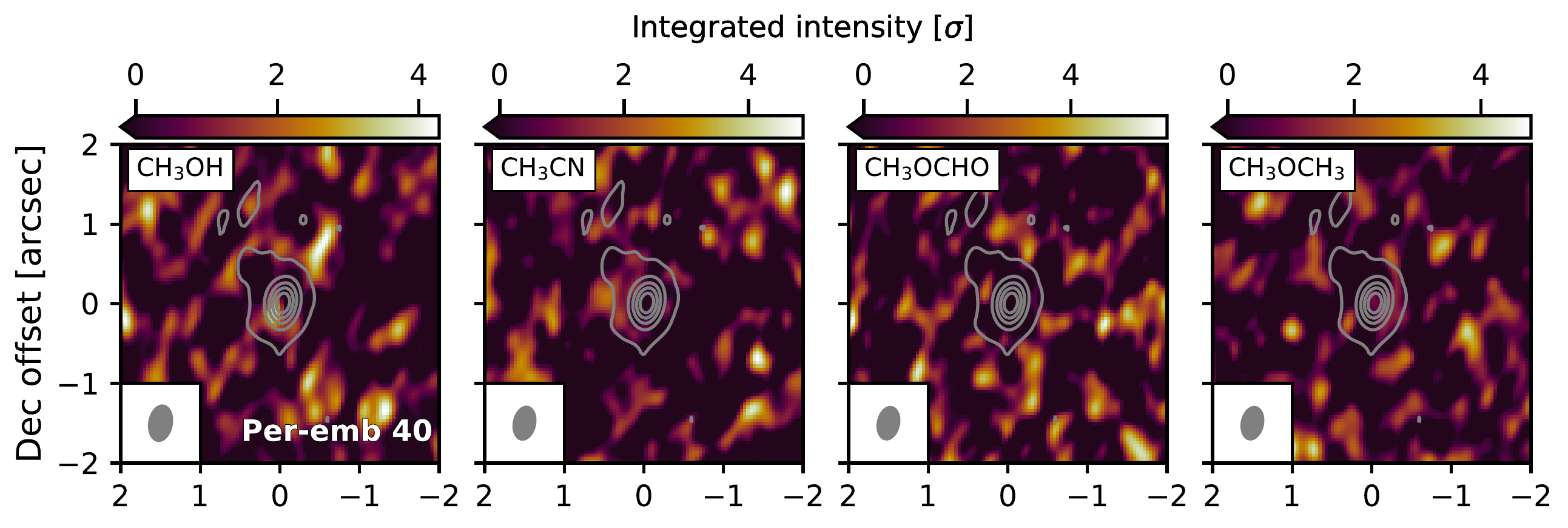}
  \includegraphics[width=\textwidth]{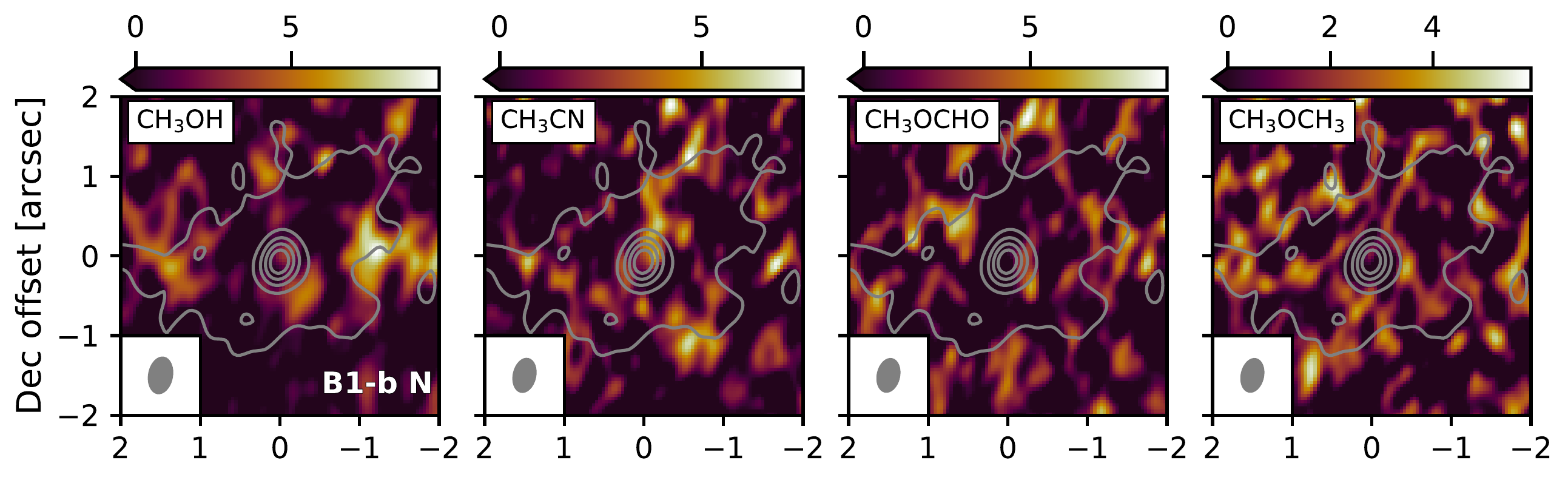}
  \includegraphics[width=\textwidth]{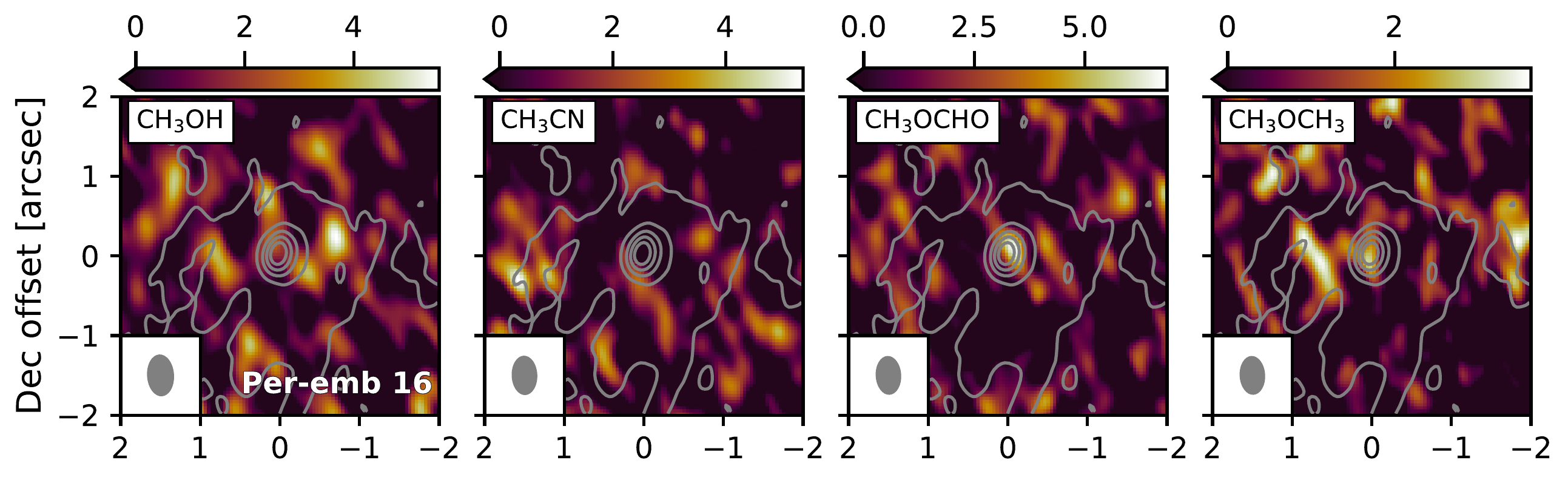}
  \includegraphics[width=\textwidth]{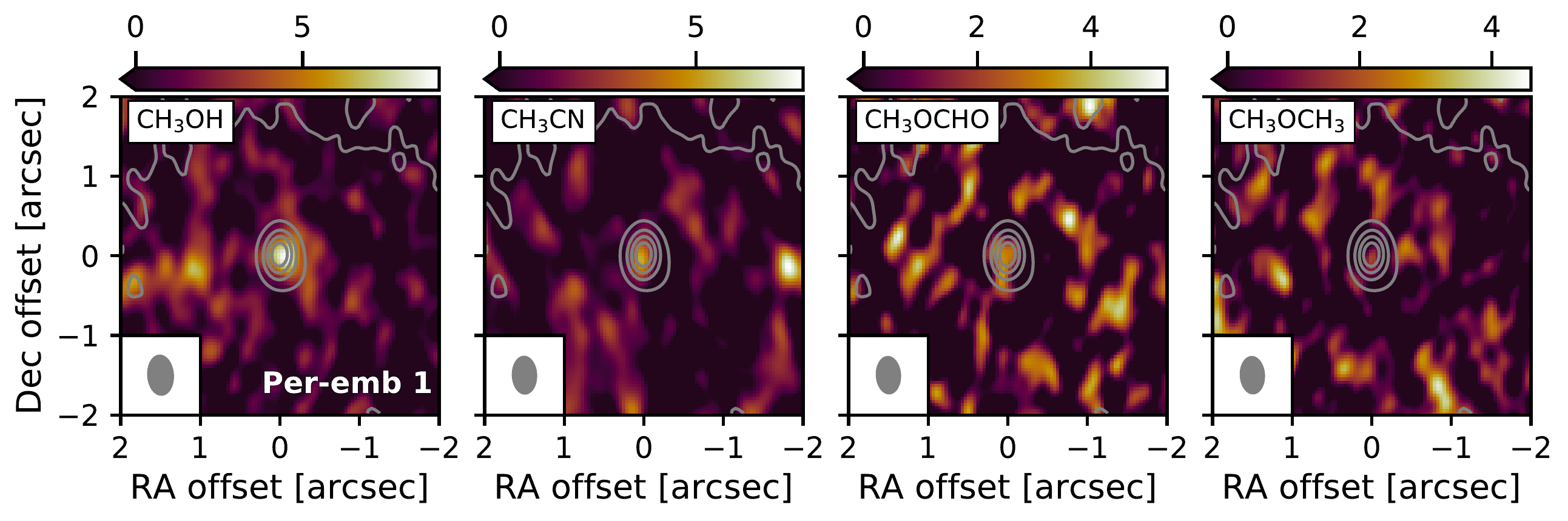}
  \caption{}
\end{figure*}
\addtocounter{figure}{-1}
\begin{figure*}[htbp!]
  \centering
  \includegraphics[width=\textwidth]{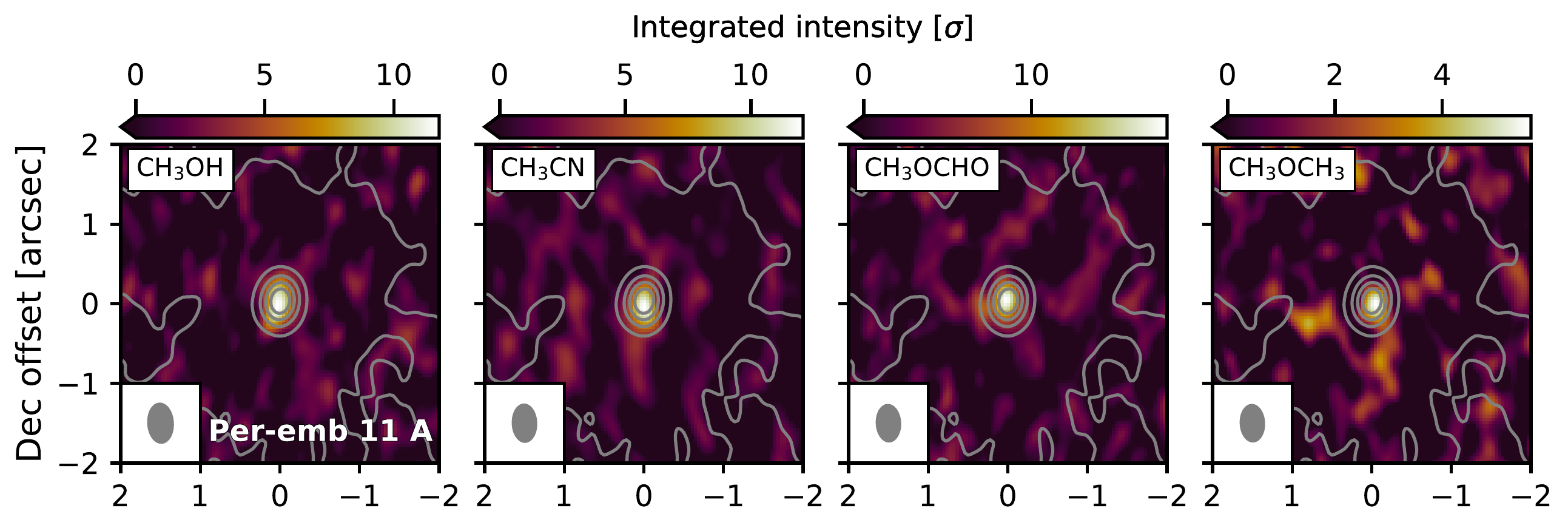}
  \includegraphics[width=\textwidth]{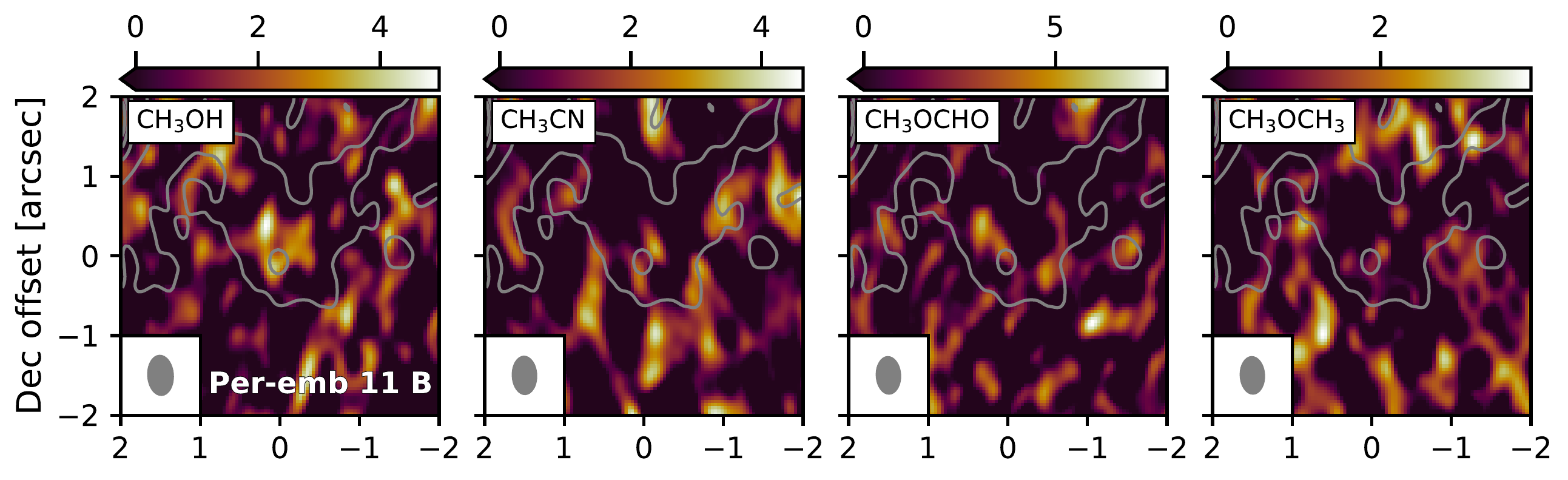}
  \includegraphics[width=\textwidth]{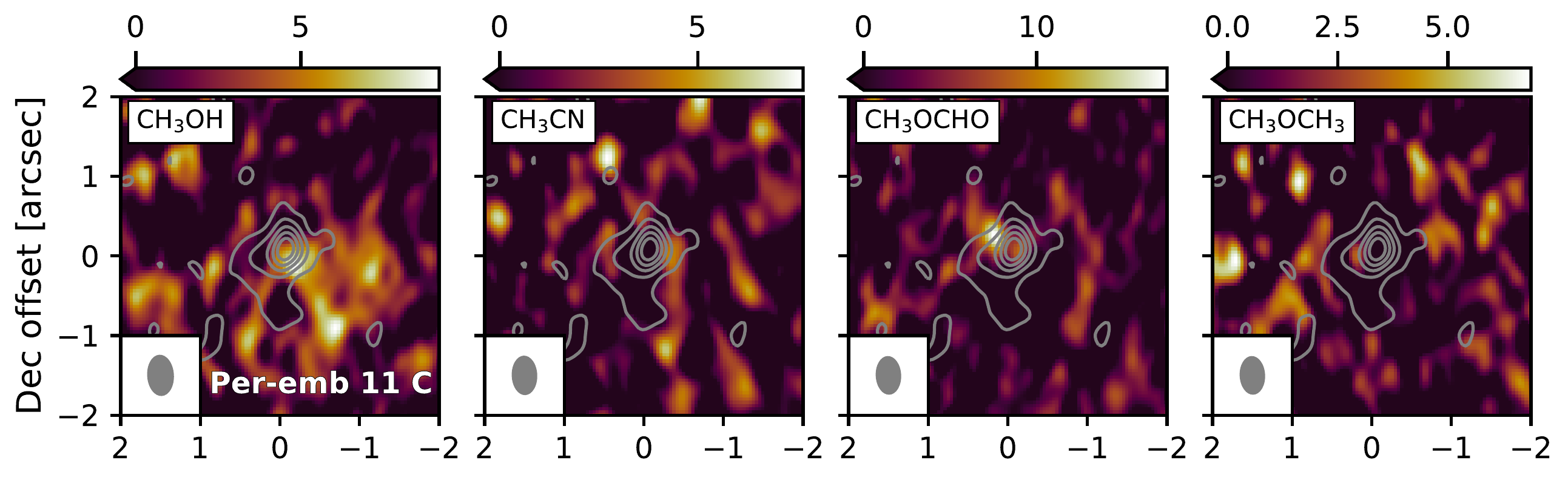}
  \includegraphics[width=\textwidth]{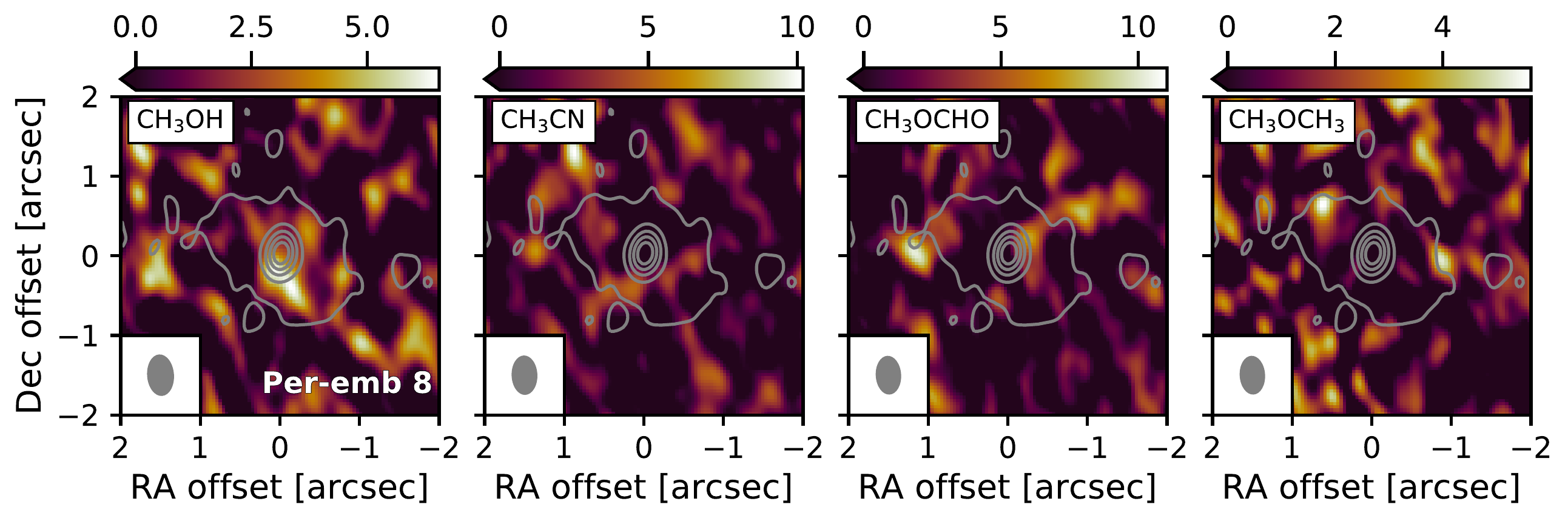}
  \caption{}
\end{figure*}
\addtocounter{figure}{-1}
\begin{figure*}[htbp!]
  \centering
  \includegraphics[width=\textwidth]{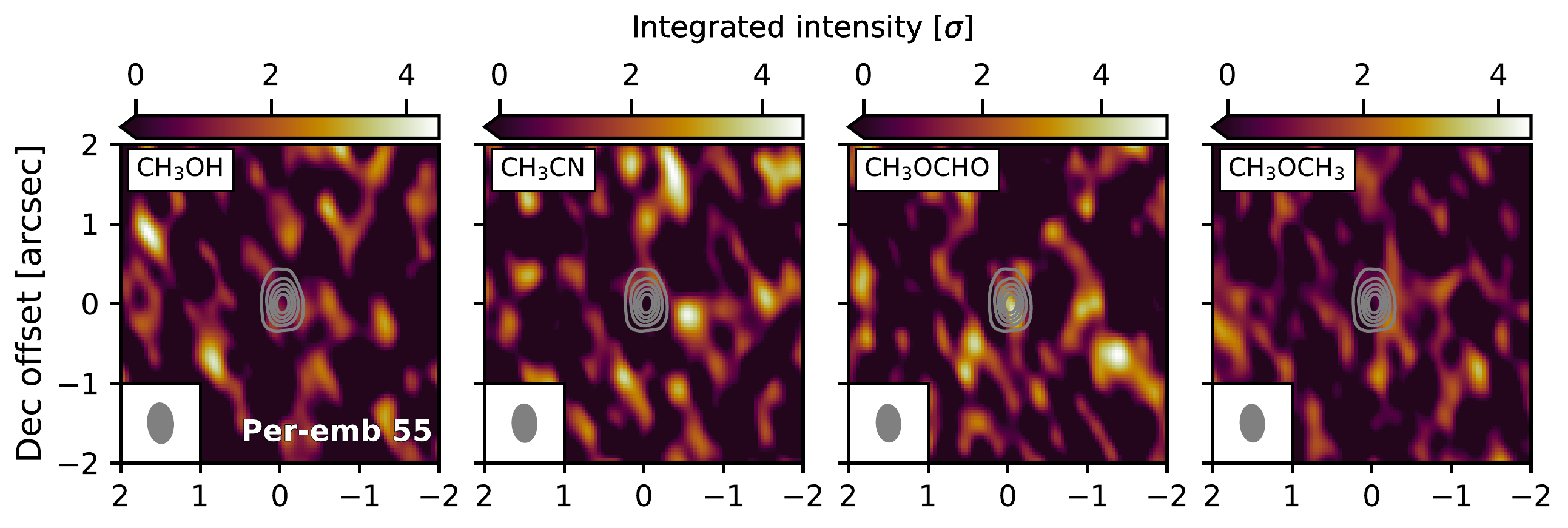}
  \includegraphics[width=\textwidth]{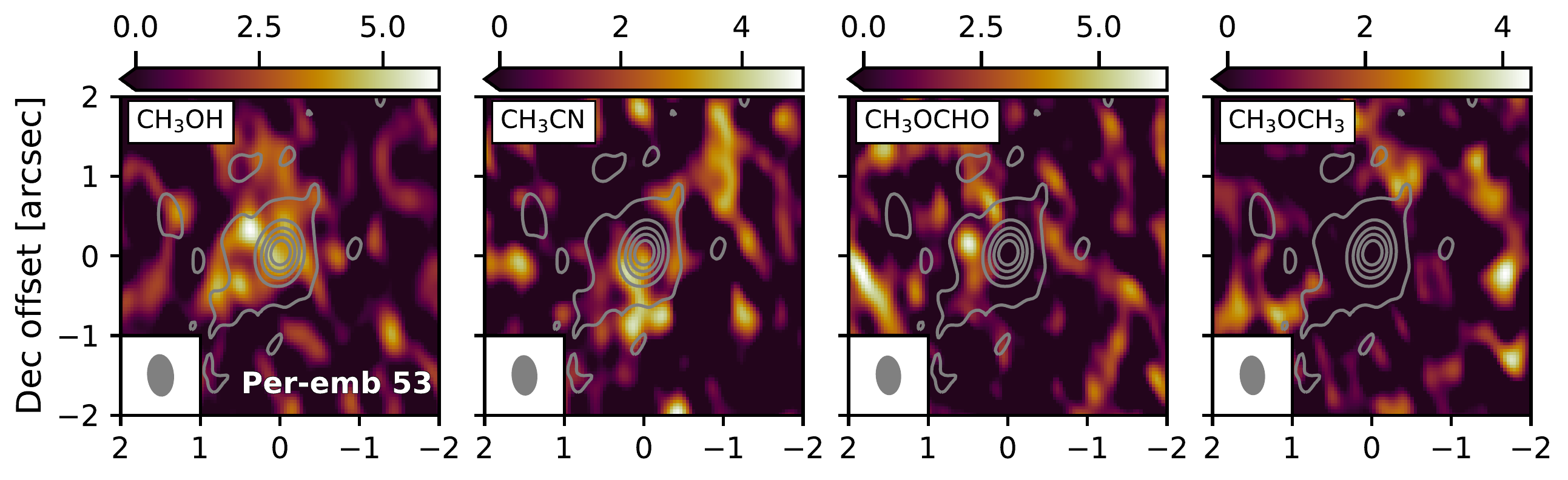}
  \caption{}
\end{figure*}
\renewcommand{\thefigure}{\arabic{figure}}

\section{Catalogs for Molecular Data}
\renewcommand{\thetable}{B\arabic{table}}
\label{sec:catalogs}
The spectroscopic data are taken from the Cologne Database of Molecular Spectroscopy (CDMS; \citealt{2001AA...370L..49M,2005JMoSt.742..215M,2016JMoSp.327...95E}) and the Jet Propulsion Laboratory (JPL; \citealt{1998JQSRT..60..883P}).  Table\,\ref{tbl:molcat} lists only the references that cover the frequencies relevant to this study.  

\begin{table*}[htbp!]
  \caption{Molecular Catalogs}
  \label{tbl:molcat}
  \centering
  \begin{tabular}{cp{2in}cp{2in}}
  Molecule & References & Molecule & References \\
  \toprule
  \cch & \citet{1981ApJ...251L.119S,2000AA...357L..65M} &
  c-C$_3$H$_2$ & \citet{1986CPL...125..383B,1987ApJ...314..716V} \\
  SO$_2$ & \citet{1979JChPh..70.2740P,1985JMoSp.111...66H,1985JPCRD..14..395L,1996JMoSp.176..316A} & 
  \tmethanol & \citet{1984ApJS...55..633P,1999ApJ...521..255O,2007JMoSp.246..158C,2008JMoSp.251..293M,2009JMoSp.255...32I} \\
  \dmethanol & \citet{2012JMoSp.280..119P} &
  \methanol & \citet{2008JMoSp.251..305X} \\
  CH$_3^{18}$OH & \citet{2007JMoSp.245....7F} &
  \sosigma & \citet{1964JChPh..41.1413P,1974JMoSp..53..346A,1976JMoSp..60..332C,1992ApJ...399..325L,1994JMoSp.167..468C} \\
  $t$-HCOOH & \citet{1957JChPh..26..680L,1958JChPh..28..736T,1962JChPh..37.2748K,1971JMoSt...9...49B,1982JMoSp..93..248K} &
  \htcn & \citet{2000JMoSp.202..166M,2004ZNatA..59..861F,2005JMoSp.233..280C} \\
  CS & \citet{2005JMoSt.742..215M} &
  HDCO & \citet{1999JMoSp.195..345B} \\
  \methylformate\ ($v=0, 1$) & \citet{2009JMoSp.255...32I} &
  \dimethylether & \citet{2009AA...504..635E} \\
  \acetone & \citet{2002ApJS..142..145G} & 
  \methylcyanide & \citet{2006JMoSp.240..153C,2009AA...506.1487M} \\
  CH$_2$DCN & \citet{2013AA...553A..84N} & 
  \ethylcyanide & \citet{1994ApJS...93..589P,2009ApJS..184..133B} \\
  \acetaldehyde & \citet{1996JPCRD..25.1113K} &
  \ethanol & \citet{1996JMoSp.175..246P,2008JMoSp.251..394P} \\
  \glycolaldehyde & \citet{2001ApJS..134..319B} &
  \formamide & \cite{blanco2006microsolvation,2009JMoSp.254...28K} \\

  \bottomrule
  \end{tabular}
\end{table*}

\newpage
\section{Identified Species and Transitions}
\renewcommand{\thetable}{C\arabic{table}}
\label{sec:line_id}
Table\,\ref{tbl:line_id} lists the species and their transitions identified from the PEACHES spectra.
\newpage
\startlongtable
\begin{deluxetable*}{cccccc}
    \tabletypesize{\scriptsize}
    \tablecaption{Line Identification \label{tbl:line_id}}
    \tablewidth{\textwidth}
    \tablehead{\colhead{Frequency (MHz)} & \colhead{Transition\tablenotemark{a}} & 
               \colhead{log(Einstein-$A$)} & \colhead{$E_\text{u}$ (K)} & \colhead{$g_\text{u}$} & \colhead{Reference}}
    \startdata
    \multicolumn{6}{c}{Ethynyl (CCH)} \\
    \hline
    262065.00 (0.05) & [3, 5/2, 3]\rt[2, 3/2, 2]\tablenotemark{b}   & $-$4.31 & 25.16  & 7  & CDMS \\
    262067.47 (0.05) & [3, 5/2, 2]\rt[2, 3/2, 1]\tablenotemark{b}   & $-$4.35 & 25.16  & 5  & CDMS \\
    262078.93 (0.02) & [3, 5/2, 2]\rt[2, 3/2, 2]\tablenotemark{b}   & $-$5.22 & 25.16  & 5  & CDMS \\
    \hline
    \multicolumn{6}{c}{Cyclopropenylidene (\cctht)} \\
    \hline
    244222.15 (0.01) & [3, 2, 1]\rt[2, 1, 2]                        & $-$4.23 & 18.17  & 21 & CDMS \\
    246557.77 (0.02) & [16, 10, 7]\rt[16, 9, 8]                     & $-$3.36 & 397.83 & 99 & CDMS \\
    260479.75 (0.02) & [5, 3, 2]\rt[4, 4, 1]                        & $-$3.79 & 44.72  & 33 & CDMS \\
    \hline
    \multicolumn{6}{c}{Methanol (\methanol\ $v_\text{t}=0$)} \\
    \hline
    243915.79 (0.01) & [5, 1, 4]\rt[4, 1, 3] A                      & $-$4.22 & 49.66  & 44 & CDMS \\
    246074.61 (0.02) & [20, 3, 17]\rt[20, 2, 18] A                  & $-$4.08 & 537.03 & 164& CDMS \\
    246873.30 (0.02) & [19, 3, 16]\rt[19, 2, 17] A                  & $-$4.08 & 490.65 & 156& CDMS \\
    261805.68 (0.01) & [2, 1, 1]\rt[1, 0, 1] E                      & $-$4.25 & 28.01  & 20 & CDMS \\
    \hline
    \multicolumn{6}{c}{Methanol (\tmethanol\ $v_\text{t}=0$)} \\
    \hline
    246426.12 (0.22) & [23, 4, 19]\rt[22, 5, 18]                    & $-$4.58 & 721.02 & 47 & CDMS \\
    247086.3 (0.5)   & [23, 3, 20]\rt[23, 2, 21] A$-$\rt\ A$+$      & $-$4.07 & 674.86 & 47 & CDMS \\
    259036.49 (0.17) & [17, 3, 15]\rt[17, 2, 16] A$+$\rt\ A$-$      & $-$4.04 & 396.48 & 35 & CDMS \\
    \hline
    \multicolumn{6}{c}{Methanol (\dmethanol\ $v_\text{t}=0$)} \\
    \hline
    243514.31 (0.01) & [9, 2, 8]\rt[10, 1, 10] o$_\text{1}$         & $-$5.17 & 131.85 & 19 & JPL  \\
    246973.11 (0.01) & [4, 1, 4]\rt[4, 1, 3] e$_\text{1}$           & $-$4.67 & 37.69  & 9  & JPL  \\
    260543.63 (0.01) & [3, 2, 1]\rt[3, 1, 2] o$_\text{1}$           & $-$4.65 & 48.34  & 7  & JPL  \\
    \hline
    \multicolumn{6}{c}{Methanol (\etmethanol\ $v_\text{t}=0$)} \\
    \hline
    246256.60 (0.04) & [11. 2. 10]\rt[10, 3, 7] A                   & $-$4.64 & 184.27 & 92 & CDMS \\
    \hline
    \multicolumn{6}{c}{Sulfur monoxide (\sosigma)} \\
    \hline
    258255.83 (0.01) & [\N, \J]$=$[6, 6]\rt[5, 5]                   & $-$3.67 & 56.50  & 13 & CDMS \\
    261843.72 (0.03) & [\N, \J]$=$[7, 6]\rt[6, 5]                   & $-$3.64 & 47.55  & 15 & CDMS \\
    \hline
    \multicolumn{6}{c}{Sulfur monoxide (\tfso)} \\
    \hline
    246663.47 (0.1)  & [\N, \J]$=$[5, 6]\rt[4, 5]                   & $-$3.74 & 49.89  & 11 & CDMS \\
    \hline
    \multicolumn{6}{c}{Sulfur dioxide (\sotwo)} \\
    \hline
    244254.22 (0.01) & [14, 0, 14]\rt[13, 1, 13]                    & $-$3.79 & 93.90  & 29 & CDMS \\
    \hline
    \multicolumn{6}{c}{Hydrogen cyanide (\htcn)} \\
    \hline
    259010.26 (0.01) & [\J, \F]$=$[3, 3]\rt[2, 3]                   & $-$4.07 & 24.86  & 7  & CDMS \\
    259011.55 (0.01) & [\J, \F]$=$[3, 2]\rt[2, 1]                   & $-$3.19 & 24.86  & 5  & CDMS \\
    259011.80 (0.01) & [\J, \F]$=$[3, 3]\rt[2, 2]                   & $-$3.16 & 24.86  & 7  & CDMS \\
    259011.86 (0.01) & [\J, \F]$=$[3, 4]\rt[2, 3]                   & $-$3.11 & 24.86  & 9  & CDMS \\
    259012.34 (0.01) & [\J, \F]$=$[3, 2]\rt[2, 3]                   & $-$5.46 & 24.86  & 5  & CDMS \\
    259013.89 (0.01) & [\J, \F]$=$[3, 2]\rt[2, 2]                   & $-$3.92 & 24.86  & 5  & CDMS \\
    \hline
    \multicolumn{6}{c}{Carbon monosulfide (CS)} \\
    \hline
    244935.56 (0.01) & [\J]$=$[5]\rt[4]                             & $-$3.53 & 35.27  & 11 & CDMS \\
    \hline
    \multicolumn{6}{c}{Formaldehyde (HDCO)} \\
    \hline
    246924.6 (0.1)   & [4, 1, 4]\rt[3, 1, 3]                        & $-$3.40 & 37.60  & 9  & CDMS \\
    259034.9 (0.1)   & [4, 2, 2]\rt[3, 2, 1]                        & $-$3.44 & 62.86  & 9  & CDMS \\
    \hline
    \multicolumn{6}{c}{Methyl formate (\methylformate)} \\
    \hline
    245883.2 (0.1)   & [20, 13, 7]\rt[19, 13, 6] E                  & $-$3.89 & 235.98 & 82 & JPL  \\
    245885.2 (0.1)   & [20, 13, 7]\rt[19, 13, 6] A                  & $-$3.89 & 235.98 & 82 & JPL  \\
    245885.2 (0.1)   & [20, 13, 8]\rt[19, 13, 7] A                  & $-$3.89 & 235.98 & 82 & JPL  \\
    245903.7 (0.1)   & [20, 13, 8]\rt[19, 13, 7] E                  & $-$3.89 & 235.97 & 82 & JPL  \\
    246027.5 (0.1)   & [21, 2, 19]\rt[20, 3, 18] E                  & $-$4.63 & 139.85 & 86 & JPL  \\
    246038.9 (0.1)   & [21, 2, 19]\rt[20, 3, 18] A                  & $-$4.63 & 139.85 & 86 & JPL  \\
    246054.8 (0.1)   & [20, 12, 8]\rt[19, 12, 7] E                  & $-$3.84 & 219.43 & 82 & JPL  \\
    246060.8 (0.1)   & [20, 12, 8/9]\rt[19, 12, 7/8] A              & $-$3.84 & 219.43 & 82 & JPL  \\
    246076.9 (0.1)   & [20, 12, 9]\rt[19, 12, 8] E                  & $-$3.84 & 219.41 & 82 & JPL  \\
    246285.4 (0.1)   & [20, 11, 9]\rt[19, 11, 8] E                  & $-$3.80 & 204.21 & 82 & JPL  \\
    246295.1 (0.1)   & [20, 11, 10]\rt[19, 11, 9] A                 & $-$3.80 & 204.21 & 82 & JPL  \\
    246295.1 (0.1)   & [20, 11, 9]\rt[19, 11, 8] A                  & $-$3.80 & 204.21 & 82 & JPL  \\
    246308.3 (0.1)   & [20, 11, 10]\rt[19, 11, 9] E                 & $-$3.80 & 204.20 & 82 & JPL  \\
    246456.1 (0.1)   & [10, 5, 6]\rt[9, 4, 5] E                     & $-$5.52 & 49.09  & 42 & JPL  \\
    246600.0 (0.1)   & [20, 10, 10]\rt[19, 10, 9] E                 & $-$3.77 & 190.34 & 82 & JPL  \\
    246613.4 (0.1)   & [20, 10, 11]\rt[19, 10, 10] A                & $-$3.77 & 190.34 & 82 & JPL  \\
    246613.4 (0.1)   & [20, 10, 10]\rt[19, 10, 9] A                 & $-$3.77 & 190.34 & 82 & JPL  \\
    246623.2 (0.1)   & [20, 10, 11]\rt[19, 10, 10] E                & $-$3.77 & 190.34 & 82 & JPL  \\
    % 246630.0 (0.1)   & [35, 6, 30]\rt[35, 5, 31] A                  & $-$4.77 & 397.98 &142 & JPL  \\
    246660.5 (0.1)   & [10, 5, 6]\rt[9, 4, 5] A                     & $-$4.74 & 49.08  & 42 & JPL  \\
    246675.4 (0.1)   & [15, 4, 12]\rt[14, 3, 11] E                  & $-$4.93 & 81.85  & 62 & JPL  \\
    246683.5 (0.1)   & [15, 4, 12]\rt[14, 3, 11] A                  & $-$4.93 & 81.84  & 62 & JPL  \\
    246752.9 (0.1)   & [10, 5, 5]\rt[9, 4, 5] E                     & $-$4.90 & 49.10  & 42 & JPL  \\
    246891.6 (0.1)   & [19, 4, 15]\rt[18, 4, 14] E                  & $-$3.66 & 126.22 & 78 & JPL  \\
    246914.7 (0.1)   & [19, 4, 15]\rt[18, 4, 14] A                  & $-$3.66 & 126.22 & 78 & JPL  \\
    246945.7 (0.1)   & [10, 5, 6]\rt[9, 4, 6] E                     & $-$4.90 & 49.09  & 42 & JPL  \\
    247040.7 (0.1)   & [20, 9, 11]\rt[19, 9, 10] E                  & $-$3.74 & 177.83 & 82 & JPL  \\
    247044.1 (0.1)   & [21, 3, 19]\rt[20, 3, 18] E                  & $-$3.66 & 139.90 & 86 & JPL  \\
    247053.5 (0.1)   & [21, 3, 19]\rt[20, 3, 18] A                  & $-$3.66 & 139.89 & 86 & JPL  \\
    247057.3 (0.1)   & [20, 9, 12]\rt[19, 9, 11] A                  & $-$3.74 & 177.83 & 82 & JPL  \\
    247057.7 (0.1)   & [20, 9, 11]\rt[19, 9, 10] A                  & $-$3.74 & 177.83 & 82 & JPL  \\
    247063.7 (0.1)   & [20, 9, 12]\rt[19, 9, 11] E                  & $-$3.74 & 177.83 & 82 & JPL  \\
    247124.3 (0.1)   & [10, 5, 5]\rt[9, 4, 6] E                     & $-$4.74 & 49.08  & 42 & JPL  \\
    258275.0 (0.1)   & [21, 13, 8]\rt[20, 13, 7] E                  & $-$3.79 & 248.37 & 86 & JPL  \\
    258277.4 (0.1)   & [21, 13, 8]\rt[20, 13, 7] A                  & $-$3.79 & 248.37 & 86 & JPL  \\
    258277.4 (0.1)   & [21, 13, 9]\rt[20, 13, 8] A                  & $-$3.79 & 248.37 & 86 & JPL  \\
    259341.9 (0.1)   & [24, 0, 24]\rt[23, 1, 23] E                  & $-$4.37 & 158.23 & 98 & JPL  \\
    259342.0 (0.1)   & [24, 1, 24]\rt[23, 1, 23] E                  & $-$3.58 & 158.23 & 98 & JPL  \\
    259342.1 (0.1)   & [24, 0, 24]\rt[23, 0, 23] E                  & $-$3.58 & 158.23 & 98 & JPL  \\
    259342.3 (0.1)   & [24, 1, 24]\rt[23, 0, 23] E                  & $-$4.37 & 158.23 & 98 & JPL  \\
    259342.7 (0.1)   & [24, 0, 24]\rt[23, 1, 23] A                  & $-$4.37 & 158.22 & 98 & JPL  \\
    259342.9 (0.1)   & [24, 1, 24]\rt[23, 1, 23] A                  & $-$3.58 & 158.22 & 98 & JPL  \\
    259343.0 (0.1)   & [24, 0, 24]\rt[23, 0, 23] A                  & $-$3.58 & 158.22 & 98 & JPL  \\
    259343.2 (0.1)   & [24, 1, 24]\rt[23, 0, 23] A                  & $-$4.37 & 158.22 & 98 & JPL  \\
    261822.3 (0.1)   & [17, 10, 7]\rt[17, 9, 8] A                   & $-$4.73 & 156.63 & 70 & JPL  \\
    262088.2 (0.1)   & [16, 10, 6]\rt[16, 9, 7] A                   & $-$4.76 & 146.59 & 66 & JPL  \\
    262088.2 (0.1)   & [16, 10, 7]\rt[16, 9, 8] A                   & $-$4.76 & 146.59 & 66 & JPL  \\
    \hline
    \multicolumn{6}{c}{Methyl formate (\methylformatev)} \\
    \hline
    243511.5 (0.1)   & [20, 12, 8]\rt[19, 12, 7] E                  & $-$3.85 & 407.25 & 82 & JPL  \\
    245846.9 (0.1)   & [21, 3, 19]\rt[20, 3, 18] E                  & $-$3.66 & 326.30 & 86 & JPL  \\
    246106.8 (0.1)   & [20, 7, 14]\rt[19, 7, 13] A                  & $-$3.70 & 343.77 & 82 & JPL  \\
    246184.2 (0.1)   & [20, 8, 13]\rt[19, 8, 12] E                  & $-$3.72 & 353.27 & 82 & JPL  \\
    246187.0 (0.1)   & [21, 2, 19]\rt[20, 2, 18] A                  & $-$3.66 & 326.62 & 86 & JPL  \\
    246233.6 (0.1)   & [20, 7, 13]\rt[19, 7, 12] A                  & $-$3.70 & 343.79 & 82 & JPL  \\
    246274.9 (0.1)   & [20, 7, 13]\rt[19, 7, 12] E                  & $-$3.70 & 343.86 & 82 & JPL  \\
    246410.95 (0.01) & [10, 5, 5]\rt[9, 4, 6] A                     & $-$4.73 & 236.70 & 42 & JPL  \\
    246422.7 (0.1)   & [22, 1, 21]\rt[21, 2, 20] A                  & $-$4.51 & 330.43 & 90 & JPL  \\
    246461.2 (0.1)   & [22, 2, 21]\rt[21, 2, 20] A                  & $-$3.65 & 330.43 & 90 & JPL  \\
    246488.4 (0.1)   & [22, 1, 21]\rt[21, 1, 20] A                  & $-$3.65 & 330.43 & 90 & JPL  \\
    246562.9 (0.1)   & [21, 2, 19]\rt[20, 2, 18] E                  & $-$3.66 & 326.24 & 86 & JPL  \\
    246706.5 (0.1)   & [22, 2, 21]\rt[21, 2, 20] E                  & $-$3.65 & 329.89 & 90 & JPL  \\
    246731.7 (0.1)   & [22, 1, 21]\rt[21, 1, 20] E                  & $-$3.65 & 329.89 & 90 & JPL  \\
    246985.2 (0.1)   & [20, 6, 15]\rt[19, 6, 14] A                  & $-$3.68 & 335.37 & 82 & JPL  \\
    259003.9 (0.1)   & [21, 7, 14]\rt[20, 7, 13] A                  & $-$3.63 & 356.22 & 86 & JPL  \\
    259025.8 (0.1)   & [21, 7, 14]\rt[20, 7, 13] E                  & $-$3.63 & 356.29 & 86 & JPL  \\
    260479.6 (0.1)   & [44, 9, 36]\rt[44, 8, 37] A                  & $-$4.59 & 828.74 & 178& JPL  \\
    \hline
    \multicolumn{6}{c}{Dimethyl ether (\dimethylether)} \\
    \hline
    246499.29 (0.01) & [37, 6, 31]\rt[37, 5, 12] AA                 & $-$4.01 & 693.72 & 750& CDMS \\
    246505.09 (0.01) & [37, 6, 31]\rt[37, 5, 12] AE                 & $-$4.01 & 693.72 & 450& CDMS \\
    246505.09 (0.01) & [37, 6, 31]\rt[37, 5, 12] EA                 & $-$4.01 & 693.72 & 300& CDMS \\
    246697.43 (0.01) & [27, 4, 23]\rt[26, 5, 21] AA                 & $-$4.70 & 367.61 & 330& CDMS \\
    246697.87 (0.01) & [27, 4, 23]\rt[26, 5, 21] EE                 & $-$4.70 & 367.61 & 880& CDMS \\
    246698.31 (0.01) & [27, 4, 23]\rt[26, 5, 21] AE                 & $-$4.70 & 367.61 & 110& CDMS \\
    246698.31 (0.01) & [27, 4, 23]\rt[26, 5, 21] EA                 & $-$4.70 & 367.61 & 220& CDMS \\
    259305.22 (0.01) & [33, 3, 31]\rt[34, 6, 28] AA                 & $-$6.61 & 563.02 & 670& CDMS \\
    259308.39 (0.01) & [33, 3, 31]\rt[34, 6, 28] AE                 & $-$6.61 & 563.02 & 402& CDMS \\
    259308.39 (0.01) & [33, 3, 31]\rt[34, 6, 28] EA                 & $-$6.61 & 563.02 & 268& CDMS \\
    259309.47 (0.01) & [17, 5, 12]\rt[17, 4, 13] AE                 & $-$4.06 & 174.54 & 210& CDMS \\
    259309.76 (0.01) & [17, 5, 12]\rt[17, 4, 13] EA                 & $-$4.06 & 174.54 & 140& CDMS \\
    259311.95 (0.01) & [17, 5, 12]\rt[17, 4, 13] EE                 & $-$4.06 & 174.54 & 560& CDMS \\
    259314.28 (0.01) & [17, 5, 12]\rt[17, 4, 13] AA                 & $-$4.06 & 174.54 & 350& CDMS \\
    \hline
    \multicolumn{6}{c}{Acetone (\acetone)} \\
    \hline
    244218.91 (0.01) & [20, 5, 15]\rt[19, 6, 14] AE                 & $-$3.32 & 139.69 & 82 & JPL \\
    244218.91 (0.01) & [20, 6, 15]\rt[19, 5, 14] AE                 & $-$3.32 & 139.69 & 250& JPL \\
    244218.92 (0.01) & [20, 5, 15]\rt[19, 6, 14] EA                 & $-$3.32 & 139.69 & 160& JPL \\
    244218.92 (0.01) & [20, 6, 15]\rt[19, 5, 14] EA                 & $-$3.32 & 139.69 & 160& JPL \\
    245831.34 (0.09) & [13, 10, 3]\rt[12, 9, 4] EE                  & $-$3.80 & 77.84  & 432& JPL \\ 
    246400.99 (0.05) & [34, 7, 28]\rt[34, 5, 29] EE                 & $-$4.17 & 364.98 &1100& JPL \\
    246400.99 (0.05) & [34, 6, 28]\rt[34, 5, 29] EE                 & $-$4.03 & 364.98 &1100& JPL \\
    246400.99 (0.05) & [34, 7, 28]\rt[34, 6, 29] EE                 & $-$4.03 & 364.98 &1100& JPL \\
    246400.99 (0.05) & [34, 6, 28]\rt[34, 6, 29] EE                 & $-$4.17 & 364.98 &1100& JPL \\
    246404.27 (0.01) & [22, 3, 19]\rt[21, 4, 18] AE                 & $-$3.23 & 149.62 & 90 & JPL \\
    246404.27 (0.01) & [22, 4, 19]\rt[21, 3, 18] AE                 & $-$3.23 & 149.62 & 270& JPL \\
    246404.29 (0.01) & [22, 3, 19]\rt[21, 4, 18] EA                 & $-$3.23 & 149.62 & 180& JPL \\
    246404.29 (0.01) & [22, 4, 19]\rt[21, 3, 18] EA                 & $-$3.23 & 149.62 & 180& JPL \\
    246450.40 (0.01) & [22, 4, 19]\rt[21, 3, 18] EE                 & $-$3.23 & 149.57 & 720& JPL \\
    246450.40 (0.01) & [22, 3, 19]\rt[21, 3, 18] EE                 & $-$5.09 & 149.57 & 720& JPL \\
    246450.40 (0.01) & [22, 3, 19]\rt[21, 4, 18] EE                 & $-$3.24 & 149.57 & 720& JPL \\
    246450.40 (0.01) & [22, 4, 19]\rt[21, 4, 18] EE                 & $-$4.92 & 149.57 & 720& JPL \\
    246496.17 (0.46) & [25, 14, 12]\rt[24, 15, 9] AE                & $-$5.01 & 257.11 & 100& JPL \\
    246496.47 (0.02) & [22, 3, 19]\rt[21, 4, 18] AA                 & $-$3.23 & 149.51 & 270& JPL \\
    246496.47 (0.02) & [22, 4, 19]\rt[21, 3, 18] AA                 & $-$3.23 & 149.51 & 450& JPL \\
    246714.12 (0.05) & [9, 8, 1]\rt[8, 5, 4] EA                     & $-$5.84 & 40.59  & 76 & JPL \\
    246714.94 (0.05) & [32, 4, 28]\rt[32, 4, 29] EA                 & $-$3.97 & 305.61 & 260& JPL \\
    246714.94 (0.05) & [32, 5, 28]\rt[32, 3, 29] EA                 & $-$3.97 & 305.61 & 260& JPL \\
    246715.04 (0.05) & [32, 5, 28]\rt[32, 4, 29] AE                 & $-$3.97 & 305.61 & 390& JPL \\
    246715.04 (0.05) & [32, 4, 28]\rt[32, 3, 29] EA                 & $-$3.97 & 305.61 & 130& JPL \\
    246719.92 (0.04) & [33, 6, 28]\rt[33, 4, 29] EE                 & $-$5.62 & 344.85 &1100& JPL \\
    246719.92 (0.04) & [33, 5, 28]\rt[33, 4, 29] EE                 & $-$3.87 & 344.85 &1100& JPL \\
    246719.92 (0.04) & [33, 6, 28]\rt[33, 5, 29] EE                 & $-$3.87 & 344.85 &1100& JPL \\
    246719.92 (0.04) & [33, 5, 28]\rt[33, 5, 29] EE                 & $-$5.61 & 344.85 &1100& JPL \\
    261818.11 (0.01) & [20, 7, 13]\rt[19, 8, 12] EA                 & $-$3.31 & 151.17 & 160& JPL \\
    261818.17 (0.01) & [20, 7, 13]\rt[19, 8, 12] AE                 & $-$3.31 & 151.17 & 82 & JPL \\
    261819.09 (0.01) & [20, 8, 13]\rt[19, 7, 12] EA                 & $-$3.31 & 151.17 & 160& JPL \\
    261819.17 (0.01) & [20, 8, 13]\rt[19, 7, 12] AE                 & $-$3.31 & 151.17 & 250& JPL \\
    \hline
    \multicolumn{6}{c}{Methyl cyanide (\methylcyanide)} \\
    \hline
    257507.56 (0.01) & [\N, \K]$=$[14, 2]\rt[13, 2]                 & $-$3.00 & 121.28 & 58 & JPL \\
    257522.43 (0.01) & [\N, \K]$=$[14, 1]\rt[13, 1]                 & $-$2.99 & 99.84  & 58 & JPL \\
    257527.38 (0.01) & [\N, \K]$=$[14, 0]\rt[13, 0]                 & $-$2.99 & 92.70  & 58 & JPL \\
    \hline
    \multicolumn{6}{c}{Acetaldehyde (\acetaldehyde\ $v_\text{t}=0$)} \\
    \hline
    246330.73 (0.01) & [15, 3, 13]\rt[15, 2, 14] A                  & $-$4.29 & 131.49 & 62 & JPL \\
    260530.40 (0.01) & [14, 1, 14]\rt[13, 1, 13] E                  & $-$3.20 & 96.39  & 58 & JPL \\
    260544.02 (0.01) & [14, 1, 14]\rt[13, 1, 13] A                  & $-$3.20 & 96.32  & 58 & JPL \\
    260547.46 (2.07) & [9, 4, 5]\rt[9, 3, 7] E, $v_\text{t}=2$      & $-$6.06 & 456.38 & 38 & JPL \\
    \hline
    \multicolumn{6}{c}{gauche-Ethanol ($g$-\ethanol)} \\
    \hline
    246414.76 (0.05) & [14, 3, 11]\rt[13, 3, 10] $v_\text{t}=0$\rt0 & $-$3.89 & 155.72 & 29 & JPL \\
    246524.28 (0.01) & [13, 2, 12]\rt[12, 1, 12] $v_\text{t}=0$\rt1 & $-$4.50 & 136.95 & 27 & JPL \\
    246658.18 (0.01) & [32, 5, 28]\rt[32, 4, 29] $v_\text{t}=0$\rt0 & $-$6.33 & 527.94 & 65 & JPL \\
    246662.98 (0.01) & [4, 2, 3]\rt[3, 1, 3] $v_\text{t}=1$\rt0     & $-$4.36 & 74.77  & 9  & JPL \\
    259322.64 (0.01) & [14, 3, 11]\rt[13, 2, 11] $v_\text{t}=0$\rt1 & $-$4.39 & 155.72 & 29 & JPL \\
    260457.73 (0.01) & [15. 4. 12]\rt[14, 4, 11] $v_\text{t}=1$\rt1 & $-$3.83 & 181.10 & 31 & JPL \\
    \hline
    \multicolumn{6}{c}{trans-Ethanol (\ethanol)} \\
    \hline
    246663.62 (0.05) & [24, 1, 23]\rt[24, 0, 24]                    & $-$3.73 & 252.35 & 49 & JPL \\
    261815.99 (0.05) & [28, 3, 26]\rt[28, 2, 27]                    & $-$3.96 & 350.98 & 57 & JPL \\
    \hline
    \multicolumn{6}{c}{Glycolaldehyde ($cis$-\glycolaldehyde)} \\
    \hline
    246773.09 (0.02) & [30, 2, 28]\rt[30, 1, 29]                    & $-$4.04 & 252.68 & 61 & CDMS \\
    246778.28 (0.02) & [30, 3, 28]\rt[30, 2, 29]                    & $-$4.04 & 252.68 & 61 & CDMS \\
    262056.78 (0.01) & [25, 2, 24]\rt[24, 1, 23]                    & $-$3.34 & 158.25 & 51 & CDMS \\
    261795.48 (0.01) & [25, 11, 14]\rt[25, 10, 15]                  & $-$3.57 & 254.23 & 51 & CDMS \\
    261798.96 (0.01) & [25, 11, 15]\rt[25, 10, 16]                  & $-$3.57 & 254.23 & 51 & CDMS \\
    \hline
    \multicolumn{6}{c}{Methyl cyanide (\dmethylcyanide)} \\
    \hline
    259315.51 (0.01) & [15, 1, 15]\rt[14, 1, 14]                    & $-$2.82 & 104.97 & 31 & CDMS \\
    260523.05 (0.01) & [15, 2, 13]\rt[14, 2, 12]                    & $-$2.82 & 121.60 & 31 & CDMS \\
    \hline
    \multicolumn{6}{c}{Ethyl cyanide (\ethylcyanide)} \\
    \hline
    246268.74 (0.01) & [27, 2, 25]\rt[26, 2, 24]                    & $-$2.90 & 169.80 & 55 & CDMS \\
    246421.92 (0.01) & [28, 2, 27]\rt[27, 2, 26]                    & $-$2.90 & 177.26 & 57 & CDMS \\
    246548.70 (0.01) & [27, 3, 24]\rt[26, 3, 23]                    & $-$2.90 & 174.06 & 55 & CDMS \\
    260535.69 (0.05) & [29, 5, 25]\rt[28, 5, 24]                    & $-$2.84 & 215.06 & 59 & CDMS \\
    \hline
    \multicolumn{6}{c}{Formamide (\formamide)} \\
    \hline
    243521.04 (0.01) & [12, 1, 12]\rt[11, 1, 11]                    & $-$2.98 & 79.19  & 25 & CDMS \\
    \hline
    \multicolumn{6}{c}{Formic acid (\thcooh)} \\
    \hline
    262103.48 (0.01) & [12, 0, 12]\rt[11, 0, 11]                    & $-$3.69 & 82.77  & 25 & CDMS \\
    \enddata
    \tablenotetext{a}{The typical quantum numbers are listed as [\J, \Ka, \Kc] unless specified.}
    \tablenotetext{b}{The quantum numbers are [\N, \J, \F]}
\end{deluxetable*}

\section{Estimation of the Column and Volume Gas Density}
\begin{figure*}[htbp!]
  \centering
  \includegraphics[width=\textwidth]{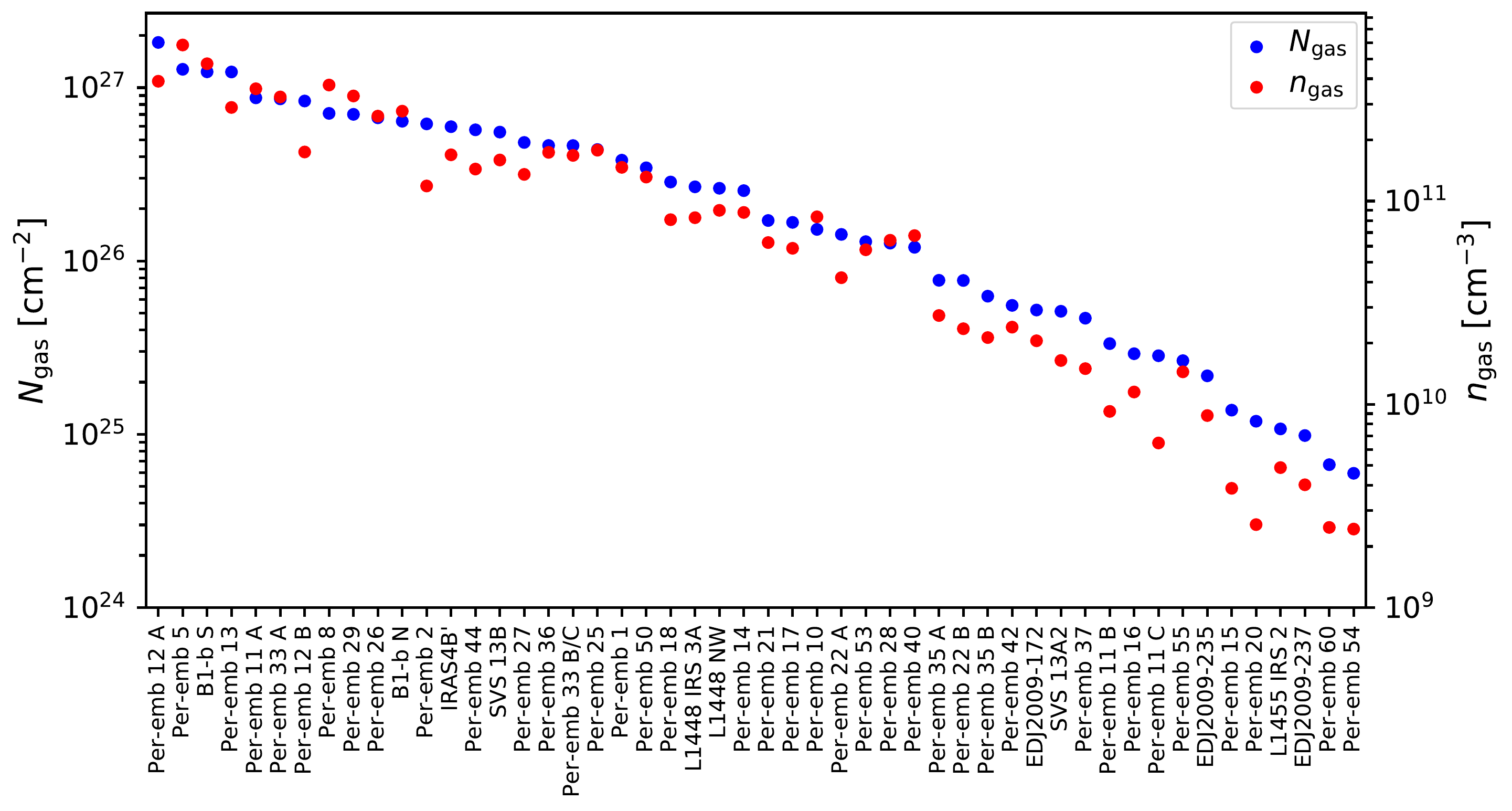}
  \caption{The derived gas column and volume densities for the PEACHES sample sorted by their gas column density.}
  \label{fig:Ngas_ngas}
\end{figure*}

\renewcommand{\thetable}{D\arabic{table}}
\label{sec:n_gas}
The continuum brightness temperature indirectly measures the gas column density assuming an optically thin emission.  Within the extraction region for the 1D spectra, we can estimate the gas column density from the averaged continuum brightness temperature using
\begin{align}
  N_{\rm gas} \text{(g cm$^{-2}$)} & = \frac{R_{\rm g2d}<I_{\nu}>}{\kappa_{\nu}B_{\nu}(T_{\rm d})} \nonumber \\
  N_{\rm gas} \text{(cm$^{-2}$)} & = \frac{R_{\rm g2d}<I_{\nu}>}{\kappa_{\nu}B_{\nu}(T_{\rm d})\mu m_{\rm H}},
\end{align}
where $R_{\rm g2d}$ is the gas-to-dust mass ratio of 100, $<I_{\nu}>$ is the averaged intensity, $\kappa_{\nu}$ is the dust opacity, $B_{\nu}(T_{\rm d})$ is the Planck function at $T_{\rm d}$, which is assumed to be 30\,K, and $\mu$ is the mean molecular weight of 2.37.  We assume a frequency of 250\ghz.  Furthermore, we can estimate that the volume density assuming the emitting gas has the same spatial extent along the line of sight as that along the plane of sky.  Figure\,\ref{fig:Ngas_ngas} shows the estimated column and volume gas densities of the PEACHES sample.

\section{Excitation Temperatures}
\renewcommand{\thetable}{E\arabic{table}}
\label{sec:Tex}
\subsection{\methanol}
The excitation temperature is a key parameter for modeling the COM spectra.  For most of the identified COMs, our observations only cover one transition or a few blended transitions, making their excitation temperature unconstrained.  Thus we assume excitation temperatures ranging from 100 to 300 K in our spectral modeling (Section\,\ref{sec:modeling}).  Fortunately, a few molecules have several transitions detected in our observations, such as \methanol\ and \methylformate, allowing us to verify the assumption of the excitation temperature.

\begin{figure}[htbp!]
  \centering
  \includegraphics[width=0.48\textwidth]{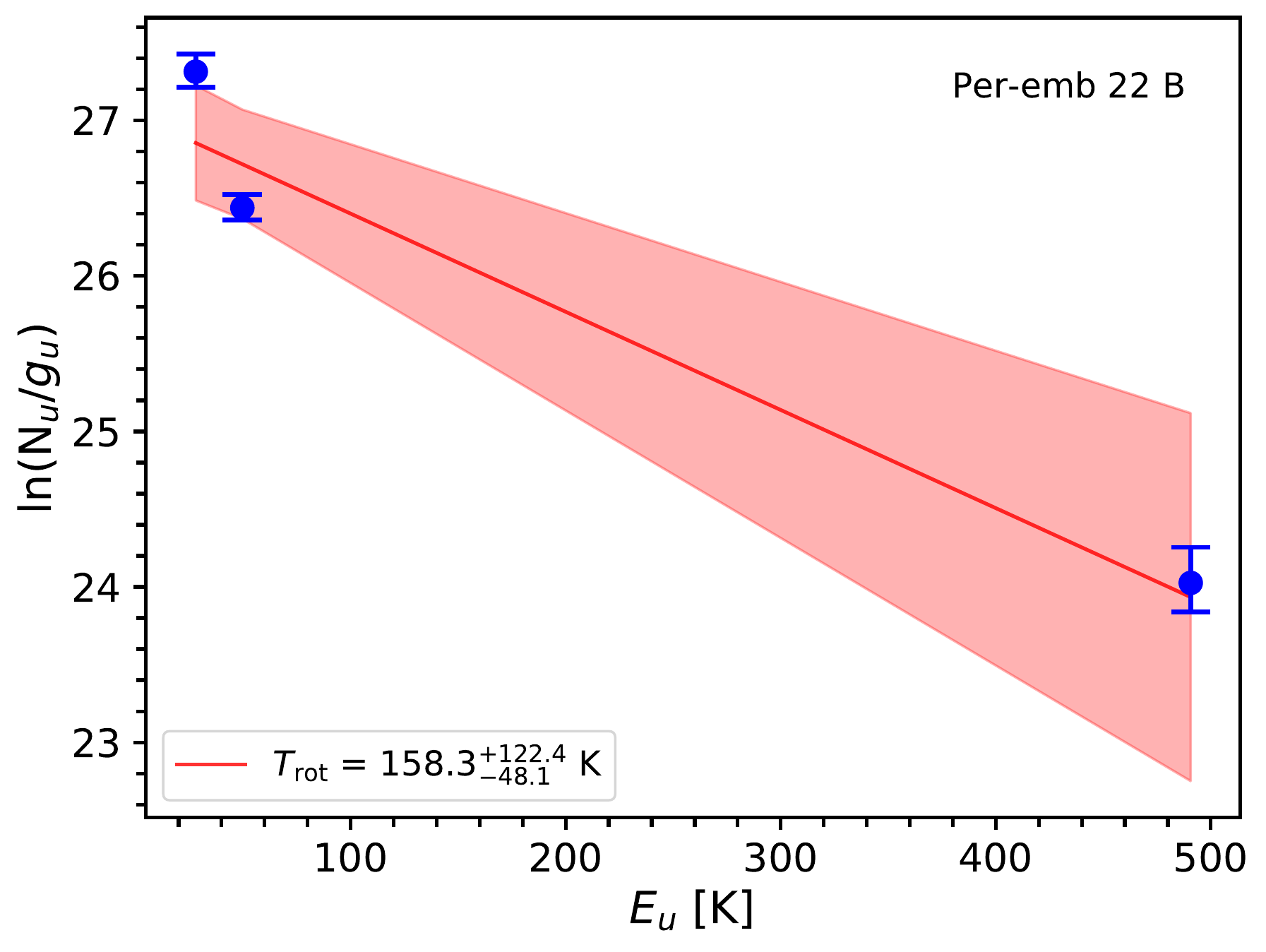}
  \caption{The \methanol\ rotational diagram for Per-emb 22B and the fitted rotational temperature.  The shaded red region indicates the uncertainty of the fitting.}
  \label{fig:rot_dia_example}
\end{figure}

Although the difference in the spectral coverage of the continuum spectral window results in different combinations of transitions, our spectra cover three \methanol\ transitions whose have their upper energy ranges from $\sim$50\,K to $\sim$500\,K, allowing us to estimate the rotational temperature ($T_\text{rot}$) of \methanol\ as a proxy of the excitation temperature ($T_\text{ex}$).  To construct the \methanol\ rotational diagram, we fit the \methanol\ emission with a Gaussian profile; then, we estimated the $T_\text{rot}$ by fitting a linear line to the rotational diagram \citep[e.g., ][]{1999ApJ...517..209G}.  We further assumed that the E- and A-species of \methanol\ have the same rotational temperature so that a single temperature component can describe the rotational diagram, where the statistical weights due to the nuclear spins were considered.  This derivation assumes no optical depth correction because we only have three transitions to fit a straight line that has two parameters.  The \methanol\ emission at 243915\mhz\ may be affected by optical depth, especially in Per-emb 44, Per-emb 12 B, Per-emb 18, and Per-emb 29.  Therefore we excluded the emission between 243914\mhz\ and 342918\mhz\ in the Gaussian fitting to minimize the impact of the optical depth.  Figure\,\ref{fig:rot_dia_example} shows the rotational diagram of Per-emb 22 B along with the fitted rotational temperature.  The derived rotational temperature of \methanol\ ranges from 140 to 350\,K, consistent with our assumption of $T_\text{ex}$ for the spectral modeling (Table\,\ref{tbl:methanol_rot_temps}).

\begin{deluxetable}{cc}
  \tabletypesize{\scriptsize}
  \tablecaption{Rotational Temperatures of \methanol\ \label{tbl:methanol_rot_temps}}
  \tablewidth{0.5\textwidth}
  \tablehead{\colhead{Source} & \colhead{$T_\text{rot}$ (K)}}
  \startdata
  Per-emb 26   & 120\unc{70}{30} \\
  Per-emb 22 A & 190\unc{60}{40} \\
  Per-emb 22 B & 160\unc{120}{50} \\
  Per-emb 20   & 170\unc{70}{40} \\
  Per-emb 44   & 200\unc{10}{10} \\
  SVS 13 A2    & 190\unc{70}{40} \\
  Per-emb 12 B & 170\unc{20}{10} \\
  Per-emb 13   & 210\unc{110}{50} \\
  Per-emb 27   & 190\unc{100}{50} \\
  Per-emb 21   & 150\unc{10}{10} \\
  Per-emb 35 A & 140\unc{110}{40} \\
  Per-emb 18   & 360\unc{140}{80} \\
  B1-bS        & 250\unc{290}{90} \\
  Per-emb 11 A & 240\unc{50}{40} \\
  Per-emb 11 C & 310\unc{340}{100} \\
  Per-emb 29   & 190\unc{10}{10} \\
  \enddata
\end{deluxetable}

\subsection{\methylformate}
\begin{figure}[htbp!]
  \centering
  \includegraphics[width=0.48\textwidth]{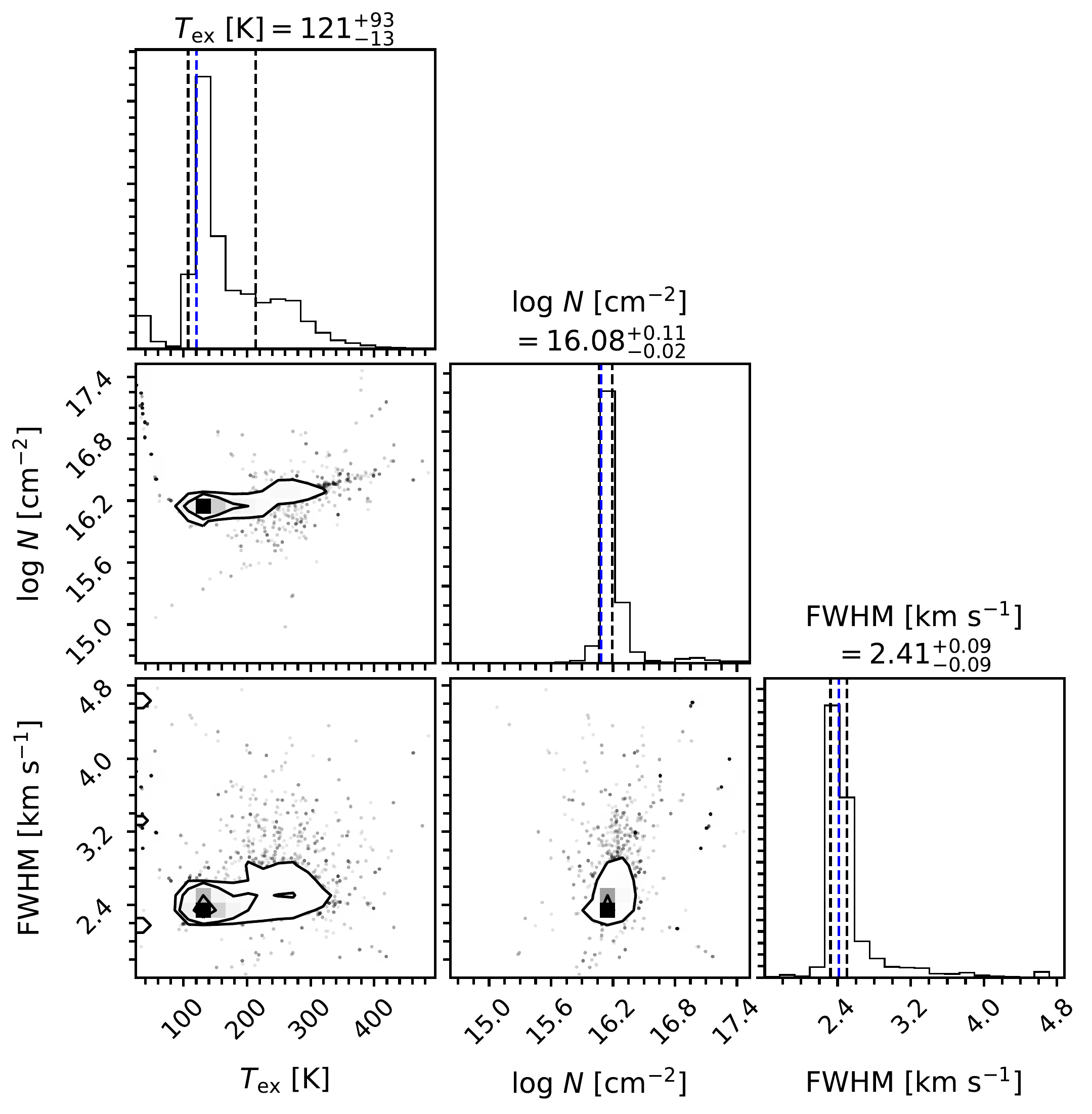}
  \caption{The posterior distributions of the MCMC fitting of the \methylformate\ emission toward B1-b S.}
  \label{fig:corner_mf}
\end{figure}

Of the identified COMs, \methylformate\ has more than 10 detected transitions.  Thus, we can use the spectra of \methylformate\ to test our assumption of $T_\text{ex}$.  With many transitions detected, we can employ a more complex model to constrain the column density and the $T_\text{ex}$ of \methylformate.  We used the \textsc{xclass} model described in Section\,\ref{sec:modeling}, which includes the effect of optical depth.  The model was optimized with the MCMC method (see description in Section\,\ref{sec:modeling}), instead of a combination of genetic and Levenberg--Marquardt $\chi^{2}$ minimization to characterize the uncertainties of the fitted properties.  Table\,\ref{tbl:mf_temp} lists the sources that have sufficient detections of \methylformate\ and the derived properties of \methylformate.  The best-fitting temperatures range from 100-300\,K, consistent with our assumption of the excitation temperatures.  Figure\,\ref{fig:corner_mf} shows an example of the posterior distributions of the fitting parameters toward B1-b S.  The posterior distribution of the excitation temperature has a longer tail toward high temperature, an upper averaged uncertainty of 71.6 K compared to a lower averaged uncertainty of 29.3 K.

\begin{deluxetable*}{ccccc}
    % \tabletypesize{\scriptsize}
    \tablecaption{MCMC Fitting of Methyl Formate \label{tbl:mf_temp}}
    \tablewidth{\textwidth}
    \tablehead{\colhead{Source} & \colhead{Temperature} & 
               \colhead{log($\mathcal{N}$)\tablenotemark{a}} & \colhead{Line Width\tablenotemark{b}} & \colhead{Line Count} \\
               \colhead{} & \colhead{(K)} & \colhead{(cm$^{-2}$)} & \colhead{(km s$^{-1}$)} & \colhead{} }
    \startdata
    Per-emb 26 & 80$^{+50}_{-10}$      & 15.9 & 4.8$^{+0.2}_{-0.6}$ & 17 \\
    Per-emb 44 & 210$^{+10}_{-20}$    & 17.2 & 3.8$\pm$0.1 & 17 \\
    Per-emb 12 B & 200$^{+50}_{-10}$   & 16.9 & 2.4$\pm$0.1 & 17 \\
    Per-emb 13 & 220$^{+40}_{-10}$    & 16.6 & 2.1$\pm$0.1 & 16 \\
    Per-emb 27 & 250$^{+40}_{-20}$    & 17.0 & 4.2$^{+0.1}_{-0.2}$ & 15 \\
    B1-b S & 120$^{+100}_{-10}$        & 16.1 & 2.4$\pm$0.1 & 12 \\
    Per-emb 29 & 190$^{+60}_{-10}$    & 16.7 & 3.5$\pm$0.1 & 14 \\
    Per-emb 11 A & 330$^{+60}_{-160}$ & 16.0 & 2.4$^{+0.2}_{-0.1}$ & 12 \\
    \enddata
    \tablenotetext{a}{The fitting uncertainty is smaller than the calibration uncertainty of 10\%.  
                      Thus, a 10\%\ uncertainty should be adopted for the fitted column densities.}
    \tablenotetext{b}{When the fitting uncertainty is smaller than the channel width, 
                      an uncertainty of the channel width, $\sim$0.1\kms, is adopted.}
\end{deluxetable*}

\section{The Spectra of \cch}
\begin{figure*}[htbp!]
  \centering
  \includegraphics[width=\textwidth]{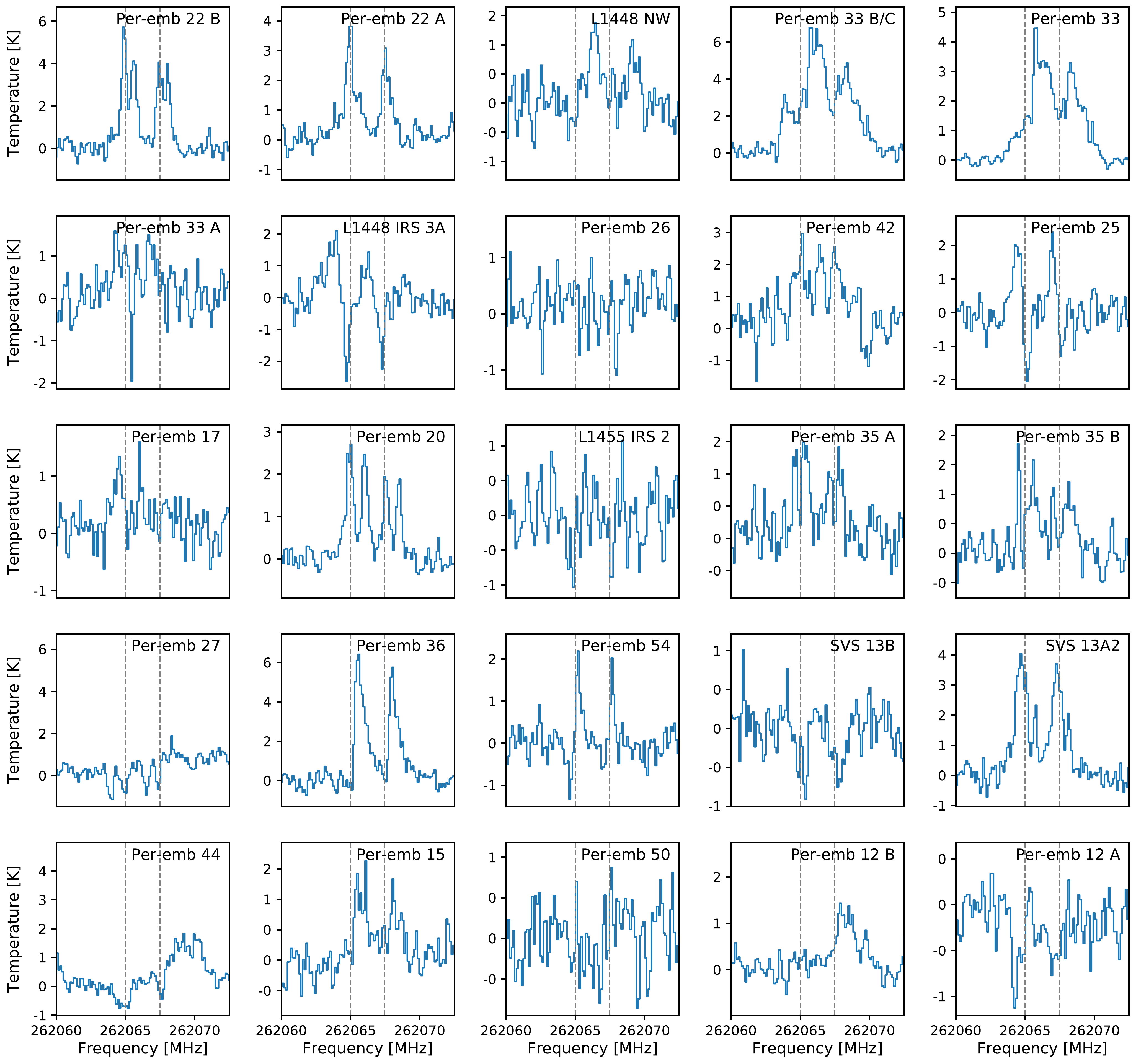}
  \caption{The \cch\ spectra of all PEACHES sources extracted from the region of the continuum source.  The vertical dashed lines indicate the rest frequencies of \cch\ transitions.}
  \label{fig:all_cch}
\end{figure*}

\renewcommand{\thetable}{F\arabic{table}}
\label{sec:cch}
The \cch\ spectra toward the continuum emission have irregular line profiles.  Some spectra have strong self-absorption, while some spectra only show the blue-shifted emission.  Due to the absorption and irregular line profile, the \textsc{xclass} fitting routine often fails to faithfully reproduce the observed \cch\ spectra.  \cch\ can easily form in the outflow cavity wall due to the abundant CH$_{4}$ sublimated from dust grains as well as C$^+$ ionized by the UV radiation.  Thus, the \cch\ spectra can have a broad line width and multiple components.  Furthermore, the morphology of the \cch\ emission traces the outflows, making our extraction from the continuum emission not ideal for representing the nature of the \cch\ emission.  Figure\,\ref{fig:all_cch} shows the spectra of \cch\ toward the PEACHES sample.

\renewcommand{\thefigure}{\arabic{figure} (Cont.)}
\addtocounter{figure}{-1}
\begin{figure*}[htbp!]
  \centering
  \includegraphics[width=\textwidth]{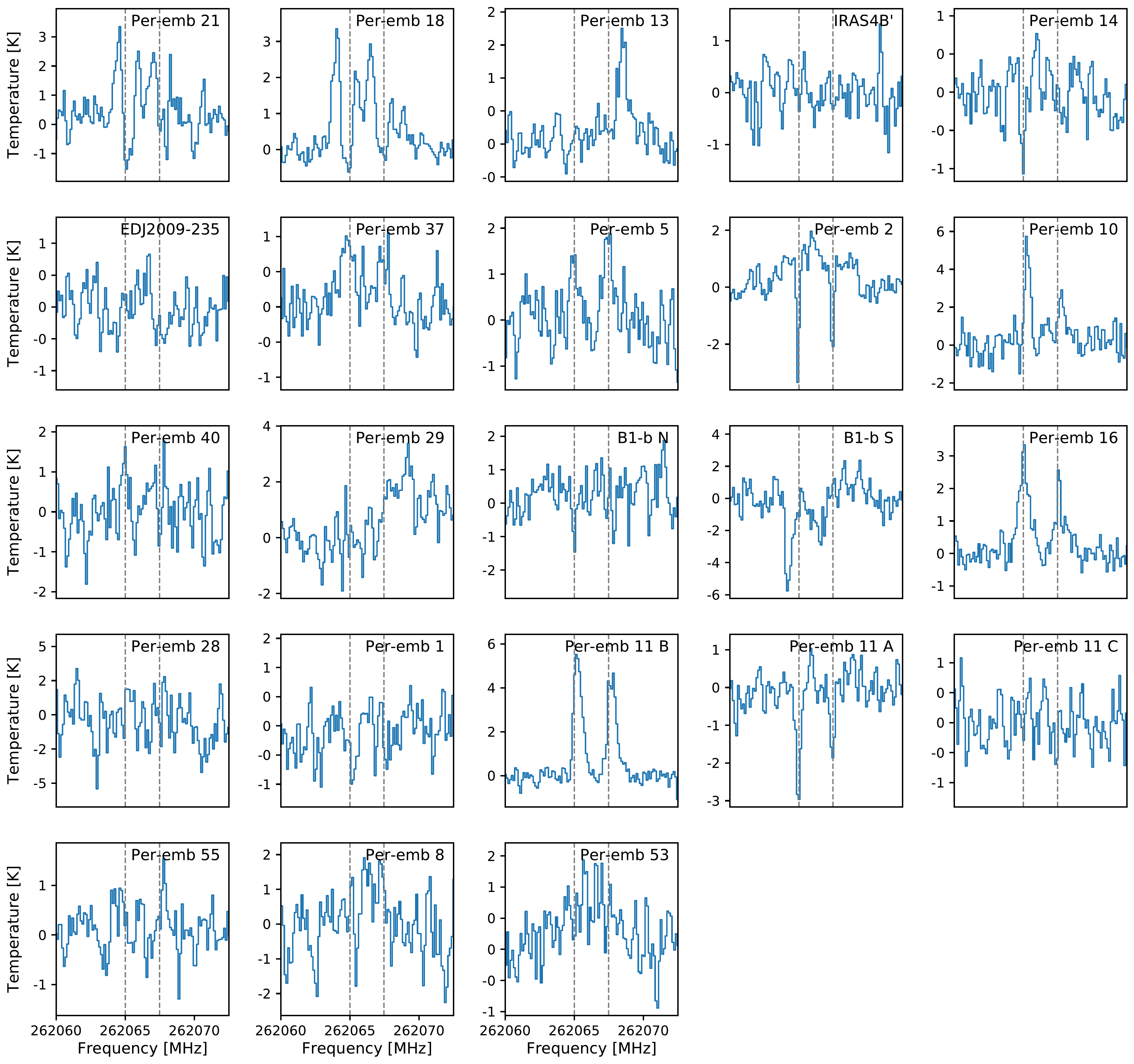}
  \caption{}
\end{figure*}
\renewcommand{\thefigure}{\arabic{figure}}

\section{The Effect of Source Size and Beam Dilution on Column Densities}
\label{sec:beam_dilution}
\begin{figure*}[htbp!]
  \centering
  \includegraphics[width=0.8\textwidth]{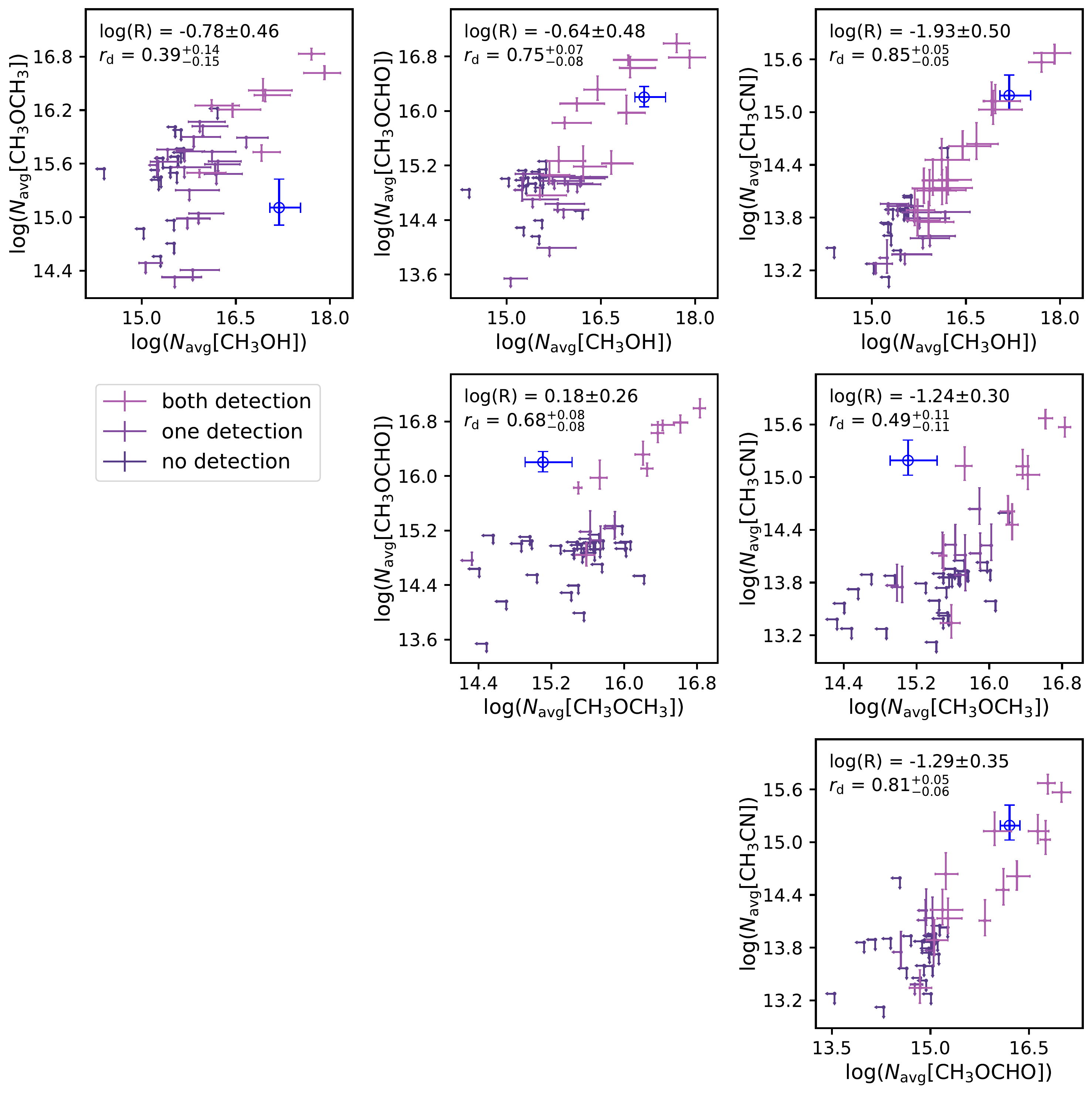}
  \caption{Corner plot of the correlations of the averaged column densities between \methanol, \methylcyanide, \methylformate, and \dimethylether.  The legends and the color code follow the legends in Figure\,\ref{fig:ch3oh_ch3cn} and \ref{fig:corner}.}
  \label{fig:corner_beam_avg}
\end{figure*}

\begin{figure*}[htbp!]
  \centering
  \includegraphics[width=0.8\textwidth]{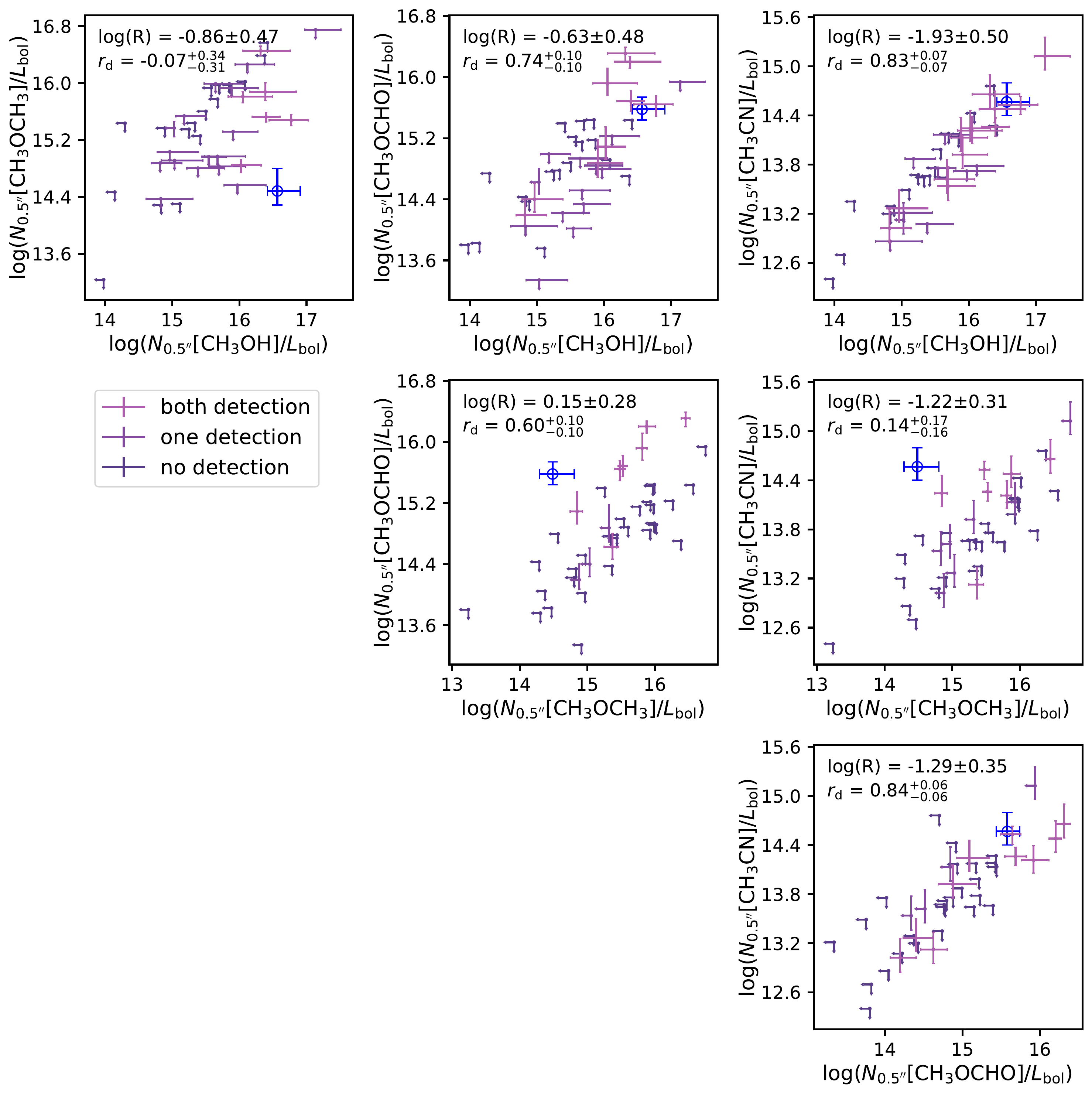}
  \caption{Corner plot of the correlations of the column densities normalized by \lbol\ between \methanol, \methylcyanide, \methylformate, and \dimethylether.  The legends and the color code follow the legends in Figure\,\ref{fig:ch3oh_ch3cn} and \ref{fig:corner}.}
  \label{fig:corner_Lbol}
\end{figure*}

Our modeling assumes a fixed 0\farcs{5} source size to estimate the column densities because most of the emission of COMs is unresolved or marginally resolved in our observations.  To study the impact of this assumption on the resulting correlation and scatter in the derived column densities, we tested whether the correlations shown in Section\,\ref{sec:correlations} still persist in two extreme cases, the column density averaged over the size of the continuum source, and assuming that COMs strictly emit from the $T=$100 K radius.  For the case of the averaged column density, we simply multiplied the modeled column density ($N_{0\farcs{5}}$) by $\Omega_{0\farcs{5}}/\Omega_{\rm cont.}$, where $\Omega$ is the solid angle.  Figure\,\ref{fig:corner_beam_avg} shows the averaged column densities of the four most frequently detected COMs.  The scatter in the averaged column densities is about 1--3 orders of magnitudes, similar to that in the modeled column densities assuming a size of 0\farcs{5}.  When we assume that COMs emit from the $T=$100 K radius where water ice desorbs, we need to multiply the modeled column density by $\Omega_{0\farcs{5}}/\Omega_{\rm 100\,K}$.  Using the equation for hot cores in \citet{2007AA...465..913B}, we have
\begin{equation}
  R_{\rm T=100\,K} \approx 15.4 \sqrt{\frac{L}{L_{\odot}}} \text{AU}.
\end{equation}
Thus, $\Omega_{0\farcs{5}}/\Omega_{\rm 100\,K}$ is proportional to $L^{-1}$.  Figure\,\ref{fig:corner_Lbol} shows the modeled column densities of the four most frequently detected COMs normalized by \lbol.  The scatter of the normalized column densities is still about 1--3 orders of magnitude, suggesting that the intrinsic variation in the number of COMs dominates the scatter in the column density.  We also note that the \lbol\ may not represent the protostellar luminosity due to the bias of the inclination of the protostars.  

% \bibliography{research,fixed}

\end{document}